\documentclass[10pt,aps,pra,twocolumn,export,superscriptaddress,nofootinbib,longbibliography]{revtex4-2}
\usepackage[colorlinks=true,linkcolor=blue,citecolor=blue,urlcolor=blue]{hyperref}

\usepackage{amsmath, mathtools, amssymb}
\usepackage{comment}
\usepackage{braket}
\usepackage{suffix}
\usepackage{booktabs}
\usepackage{physics}
\usepackage[export]{adjustbox} % valign
\usepackage{microtype}
\usepackage[dvipsnames]{xcolor}

\usepackage{newtxtext}
\usepackage[smallerops]{newtxmath}

\usepackage{bbold}

\DeclareMathAlphabet{\mathcal}{OMS}{cmsy}{m}{n}
\DeclareMathAlphabet{\mathsf}{OT1}{cmss}{m}{n}
\usepackage{courierten} % texttt font replace

\usepackage{tikz}
\usetikzlibrary{shapes.geometric}
\usetikzlibrary{decorations.markings}
\usetikzlibrary{arrows.meta}
\usetikzlibrary{calc}
\usetikzlibrary{3d}
\usetikzlibrary{knots}

\usepackage[inline]{enumitem}  % inline lists
\usepackage{anyfontsize}

\newcommand{\eq}[1]{\begin{align}\begin{split} #1 \end{split}\end{align}}
\newcommand{\tta}{\mathtt{a}}
\newcommand{\ttb}{\mathtt{b}}
\newcommand{\ttc}{\mathtt{c}}
\newcommand{\ttd}{\mathtt{d}}
\newcommand{\tte}{\mathtt{e}}

\newcommand{\ttg}{\mathtt{g}}
\newcommand{\tth}{\mathtt{h}}

\newcommand{\ttm}{\mathtt{m}}

\newcommand{\tto}{\mathtt{o}}

\newcommand{\ttr}{\mathtt{r}}

\newcommand{\ttx}{\mathtt{x}}
\newcommand{\tty}{\mathtt{y}}
\newcommand{\ttz}{\mathtt{z}}

\newcommand{\zz}{\mathbb{Z}}

\def\Z#1{\mathbb{Z}_{#1}}
\def\U#1{\mathrm{U}({#1})}
\def\SU#1{\mathrm{SU}({#1})}

\newcommand{\dehn}{{\sf Dehn}}
\newcommand{\poly}{{\sf poly}}
\newcommand{\mo}{\mathcal{O}}
\newcommand{\bs}{{\sf BS}}
\newcommand{\el}{{\sf EL}}
\newcommand{\Ker}{{\sf Ker}}
\newcommand{\red}{{\sf Red}}
\newcommand{\Span}{{\sf span}}

\DeclareMathOperator{\growth}{\mathsf{Growth}}

\renewcommand{\vec}[1]{\mathbf{#1}}
\renewcommand{\ketbra}[2]{\ket*{#1}\!\!\bra*{#2}}

\DeclareMathOperator{\GL}{GL}
\DeclareMathOperator{\SL}{SL}

\definecolor{agen}{RGB}{255,0,0}
\definecolor{bgen}{RGB}{0,84,212}
\definecolor{cgen}{RGB}{169,212,0}
\definecolor{dgen}{RGB}{255,221,85}

\tikzset{
kb1/.style={postaction={decorate,
   decoration={markings,mark=at position .6 with {\arrow[scale=1.3]{stealth};}}}
   }}

\makeatletter
\tikzset{nomorepreaction/.code=\let\tikz@preactions\pgfutil@empty}
\makeatother

\makeatletter
\def\l@subsubsection#1#2{}
\makeatother

\begin{document}

\title{Robust Hilbert space fragmentation in group-valued loop models}

\date{\today}

\author{Alexey Khudorozhkov}
\email{alehud@bu.edu}
\affiliation{Department of Physics, Boston University, Boston, Massachusetts 02215, USA}

\author{Charles Stahl}
\affiliation{Department of Physics and Center for Theory of Quantum Matter, University of Colorado Boulder, Boulder, Colorado 80309 USA}

\author{Oliver Hart}
\affiliation{Department of Physics and Center for Theory of Quantum Matter, University of Colorado Boulder, Boulder, Colorado 80309 USA}

\author{Rahul Nandkishore}
\affiliation{Department of Physics and Center for Theory of Quantum Matter, University of Colorado Boulder, Boulder, Colorado 80309 USA}

\begin{abstract}
We introduce a large class of models exhibiting robust ergodicity breaking in quantum dynamics. Our work is inspired by recent discussions of ``topologically robust Hilbert space fragmentation,'' but massively generalizes in two directions: first, from states describable as ``loop-soups'' to a broader class of states reminiscent of string-nets and sponges and, second, from models restricted to square or cubic lattices, to models defined on arbitrary lattices and graphs, including those without translation invariance. Our constructions leverage a 
recently proposed group-theory framework~[\href{https://doi.org/10.1103/PhysRevX.14.021034}{PRX \textbf{14}, 021034 (2024)}], and identify a host of new phenomena arising from the interplay of ``group-model dynamics'' and lattice structure. We make crisp connections to gauge theories, and our construction generalizes Kitaev's quantum double to infinite groups.
\end{abstract}

\maketitle

\tableofcontents

\section{Introduction}
\label{sec:intro}

One of the paramount questions in the field of quantum dynamics is whether an isolated many-body quantum system eventually thermalizes under its own dynamics, meaning that it reaches a state where the system acts as a thermal bath for any of its small subsystems, while local observables reach their equilibrium values predicted by statistical mechanics \cite{MBLARCMP, MBLRMP}. ``Generic'' quantum systems are believed to thermalize, and to obey the eigenstate thermalization hypothesis (ETH) \cite{Deutsch, Srednicki, Rigol}. There are several known classes of isolated quantum systems that fail to thermalize:
\begin{enumerate*}[label={(\roman*)}]
    \item integrable systems, where the number of local conserved quantities scales with the number of degrees of freedom, and the time evolution is set by these integrals of motion (see Ref.~\cite{Baxter} for a review); 
    \item many-body localized (MBL) systems, where thermalization is arrested by strong disorder (see Refs.~\cite{MBLARCMP, MBLRMP} for reviews); 
    \item systems with quantum many-body scars (QMBS) (see Ref.~\cite{scarsarcmp} for a review) -- non-thermal eigenstates in the middle of an energy spectrum that otherwise appears thermal; 
    \item systems exhibiting Hilbert space fragmentation (HSF) (aka Hilbert space shattering), where local kinetic constraints split the Hilbert space into dynamically disconnected components (Krylov sectors), with conserved quantities associated to these sectors being highly non-local (see Ref.~\cite{moudgalya2022quantum} for a review).
\end{enumerate*}

How robust are these routes to ``ergodicity breaking''?  For integrability, QMBS, and the majority of HSF instances there is no known argument for stability. That is, it is entirely possible that generic local perturbations to the Hamiltonian produce thermalization, on a timescale that scales as a low-order polynomial in inverse perturbation strength. MBL is more robust, in the sense that the corresponding timescale is generically superpolynomially long in perturbation strength \cite{deRoeckHuveneers, GopalakrishnanHuse}, and for the special case of one-dimensional spin chains may even be infinite \cite{imbrie}, for spatially local perturbations below some critical strength, although this last claim is contested \cite{selspolkovnikov}.\footnote{We are restricting to finite-dimensional systems with a notion of geometric locality, otherwise one can also get infinite timescales in models inspired by locally testable codes \cite{Chao}} Meanwhile, HSF arising due to the presence of a multipole symmetry~\cite{PPN, KHN, Sala2020, MPNRB} (historically, the first example of HSF) is robust (infinite timescale) to spatially local \emph{symmetric} perturbations~\cite{KHN}, but not robust to non-local or asymmetric perturbations. Still more recently, models with {\it topologically} robust HSF have been proposed ~\cite{stephen2022ergodicity,stahl2023topologically}, where for the first time spatial locality is no longer a requirement. The ergodicity breaking originates from an emergent higher-form symmetry~\cite{Nussinov2007, Gaiotto2015,  McGreevy2022} , and is robust to arbitrary $k$-local perturbations (which do not need to be geometrically local), with the ``thermalization timescale'' being infinite if the perturbations respect a single constraint, and exponential in perturbation strength if they do not. The ``topologically robust HSF'' models are thus arguably the most robust examples of ergodicity breaking known to date, in finite dimensional systems. However, so far topologically robust HSF has only been demonstrated on square and cubic lattices, in a restricted class of models where the Hilbert space is spanned by a basis of non-intersecting closed loops.

In this paper, we generalize topologically robust HSF to a much broader class of systems. Our generalization is twofold: first, we allow for a wider class of local degrees of freedom, such that states satisfying the imposed constraints are not necessarily configurations of closed colored loops, but rather webs of merging colored strings, or sponges of merging colored membranes. Second, we generalize the construction to arbitrary lattices (or even arbitrary graphs without translation invariance or geometric locality). The generalization leverages the picture of ``group models'' recently proposed in Ref.~\cite{balasubramanian2023glassy}. This generalization establishes topologically robust HSF as a general phenomenon, which can arise in generic settings, and also helps connect to gauge theories, which are perhaps the most famous examples in physics of constrained quantum dynamics. Our generalizations also open the door to the realization of new phenomena, not present in the square (or cubic) lattice models of Refs.~\cite{stephen2022ergodicity,stahl2023topologically}. 

This paper is structured as follows. In Sec.~\ref{sec:1D_stuff} we discuss group-model dynamics in one dimension. This section reviews results from Ref.~\cite{balasubramanian2023glassy}, establishes essential vocabulary and technology, explains how the discussion of Ref.~\cite{balasubramanian2023glassy} may be extended to systems with periodic boundary conditions, and identifies global symmetries of the group models. In Sec.~\ref{sec:2d} we demonstrate how ``group-model dynamics'' may be defined on arbitrary two-dimensional crystalline lattices, and how topologically robust HSF may emerge as a result. We also discuss the connection to gauge theories. In Sec.~\ref{sec:lattice} we discuss new phenomena arising from the interplay of topologically robust HSF and lattice structure. In Sec.~\ref{sec: 3d} we discuss the extension to three space dimensions, and new possibilities that arise therein.
Finally, in Sec.~\ref{sec:arbitrary-graphs} we discuss extensions to arbitrary graphs (without even translation invariance), and comment on implications. We conclude in Sec.~\ref{sec:conclusion}.

\section{Review of group dynamics in 1D}
\label{sec:1D_stuff}
We warm up by discussing the group-dynamics formalism. In Secs.~\ref{subsec:presentation} and \ref{subsec:word_problem} we review the formalism and results of Ref.~\cite{balasubramanian2023glassy}, referring the reader to that reference for in-depth discussion.\footnote{Although in Ref.~\cite{balasubramanian2023glassy}, many of the results are formulated for the more general case of \emph{semi}-groups, here we restrict our attention to groups only.} We then introduce new concepts related to the formalism. In Sec.~\ref{sec:boundary-conditions} we discuss the role of periodic boundary conditions, and in Sec.~\ref{sub:conservation} we discuss global symmetries.

\subsection{Local Hilbert space and the presentation of a group}
\label{subsec:presentation}

Consider a discrete, finitely presentable group $G$ and choose one of its presentations,
\eq{
G = \braket{ \mathcal{S}\,}{\,R},
}
where $\mathcal{S}$ is a set of characters (e.g., $\mathcal{S} = \{ \tta, \ttb, \ttc\}$) and $R$ is a set of relations, explained below. A \emph{word} is a sequence of elements from the \emph{symmetric alphabet} $\mathcal{A} \equiv \mathcal{S} \cup \mathcal{S}^{-1} \cup \{\tte\}$, where $\mathcal{S}^{-1}$ is the set of inverses of elements from $\mathcal{S}$ (e.g., $\mathcal{S}^{-1} = \{\tta^{-1}, \ttb^{-1}, \ttc^{-1}\}$) and $\tte$ is the identity character.  We introduce a binary operation of \emph{concatenation} ($\cdot$) on the space of all words, $\mathcal{A}^*$ (where $^*$ is the Kleene star\footnote{$\{ \tta, \ttb, \ttc \}^* \equiv \{ \varepsilon, \tta, \ttb, \ttc, \tta\tta, \tta\ttb, \tta\ttc, \ttb\tta, \ttb\ttb, \ttb\ttc, \ttc\tta, \ttc\ttb, \ttc\ttc, \tta\tta\tta, \tta\tta\ttb, \dots \}$, where $\varepsilon$ is an empty string.}), which appends the second word to the end of the first one (e.g., $\tta\ttr\ttg\tth\ttm \cdot \ttg\tto\ttg = \tta\ttr\ttg\tth\ttm\ttg\tto\ttg$). 

We can define a homomorphism $\varphi$ from $\mathcal{A}^*$ to the group $G$, $\varphi: \mathcal{A}^* \to G$, that multiplies the characters in a word $w \in \mathcal{A}^*$ to obtain a group element $\varphi(w) \in G$. This homomorphism preserves the binary operation, meaning $\varphi(w_1 \cdot w_2) = \varphi(w_1) \varphi(w_2)$ with group multiplication on the right. 
The relations in $R$ are equations, saying that two words map to the same group element. For example, the relation $\tta^4 = \ttb^2$ means that $\varphi(\tta^4) = \varphi(\ttb^2)$. Alternatively, any relation can be rewritten as some sequence of generators equals identity, e.g., $\tta^4 \ttb^{-2} = \tte$. The corresponding words (e.g., $\tta^4 \ttb^{-2}$) are called \emph{relators}, and $R$ is often given as a list of relators. Note that due to the existence of words with $\varphi(w) = \tte$ (relators being some of them), the map $\varphi$ is surjective but not injective, i.e., there are many words that are mapped to the same group element.

Some simple examples of group presentations include:
\begin{subequations}
\begin{gather}
\zz  = \braket{ \ttx \,}{\,\, } = \braket{ \tta, \ttb \,}{\, \tta^3 = \ttb^2,\, \tta \ttb = \ttb \tta },
\\
\zz^2 \equiv \zz \times \zz  = \braket{ \ttx, \tty \,}{\, \ttx\tty = \tty\ttx },
\\
\mathbb{Z}_N^{*3} \equiv \mathbb{Z}_N * \mathbb{Z}_N * \mathbb{Z}_N  = \braket{ \ttr , \ttg , \ttb  \,}{\, \ttr^N = \ttg^N = \ttb^N = \tte }
    \, ,
\\
\begin{aligned}
    S_3 &= \braket{ \tta, \ttb, \ttc \,}{\, \tta^2 = \ttb^2 = \ttc^3 = \tta\ttb\ttc = \tte } \\
    &= \braket{ \tta, \ttb \,}{\, \tta^2 = \ttb^2 = (\tta\ttb)^3 = \tte },
    \end{aligned}
\label{eq:presentations_example}
\end{gather}
\end{subequations}
where $\zz$ is the group of integers under addition, $\zz_N$ is the cyclic group of order $N$, $\times$ is the direct product, $*$ denotes the free product, and $S_N$ is the symmetric group of degree $N$. Importantly, we do not restrict ourselves to finite groups. Note that a group can admit several (in fact, infinitely many) different presentations. 

To turn the group presentation into a physical system, consider a 1D chain of length $L$, where the local Hilbert space at every site is spanned by states $\ket*{s}$ with $s \in \mathcal{A}$. Product states $\ket*{w}$ in the computational basis are words in $\mathcal{A}^*$, e.g. $\ket*{w} = \ket*{\tta \cdot \ttr \cdot \ttg \cdot \tth} = \ket*{\tta} \ket*{\ttr} \ket*{\ttg} \ket*{\tth}$. We introduce generic local dynamics that updates $w$ while preserving $\varphi(w)$, so that $\ket*{w_1}$ can never mix with $\ket*{w_2}$ unless $\varphi(w_1)=\varphi(w_2)$. This is a manifestation of Hilbert space fragmentation, which we call \emph{intrinsic fragmentation}, as opposed to \emph{fragile fragmentation} discussed below. Define the set of all words that multiply to the group element $g \in G$,
\eq{
K_g = \{ w \in \mathcal{A}^* \,|\, \varphi(w) = g \in G \},
}
as well as the subset of $K_g$ containing only words of a certain length $L$,
\eq{
K_{g, L} = \big\{ w \in K_g \,\big| \,\abs{w} = L \big\},
}
where $\abs{w}$ is the length of the word. Then the Krylov sector labeled by a group element $g \in G$ consists of all states $\ket*{w}$ such that $w \in K_{g,L}$. We call these \emph{intrinsic Krylov sectors}.\footnote{For certain groups $G$, all intrinsic Krylov sectors can be associated to quantum numbers of a global symmetry. We discuss this in detail in Sec.~\ref{sub:conservation}} The number of intrinsic Krylov sectors is, then, the number of group elements that can be written as words of length $L$. As long as we include the $\tte$ character, this is the same as the number of group elements that can be written as words of length $\leq L$. This quantity is called the \emph{growth rate} of the group, 
\begin{align}
\growth(L) = \big| \big\{ g \in G \,\big\vert\, g = \varphi(w), \abs{w} \leq L \big\} \big|,
\end{align}
and is also called the \emph{group volume}. 
Although $\growth(L)$ depends on the presentation as written, it is independent of the presentation up to polynomial-in-$L$ corrections. 

Group-preserving dynamics can be introduced as classical stochastic dynamics, quantum  evolution under a spatially local Hamiltonian $H$, or random-unitary dynamics. All three cases consist of local fundamental blocks (classical update, Hamiltonian term, or random unitary) of a certain length. The minimal ``interaction range'' needed to capture the essential physics is determined by the length of relations in $R$. Every relation written as $r_\text{left} = r_\text{right}$ can be rewritten in the form $r \equiv r_\text{left} r_{\smash{\text{right}}}^{\smash -1} = \tte$. Then the minimal interaction range we require is
\eq{
l_R = \max_{r \in R} \left\lceil \frac{\abs{r}}{2} \right\rceil,
\label{eq:int_range}
}
where $\lceil \cdots \hspace{-0.1pt} \rceil$ denotes rounding up to the nearest integer. With such interaction range, we can always split $r$ into $r = \tilde{r}_\text{left} \tilde{r}_{\smash{\text{right}}}^{\smash -1}$, where
$\abs*{\tilde{r}_\text{left}}$ and $\abs*{\tilde{r}_\text{right}}$ differ by at most one,
and we can flip a local configuration $\ket*{\tilde{r}_\mathrm{left}}$ to $\ket*{\tilde{r}_\mathrm{right}}$, padding the shorter of the two words with an identity character $\tte$ if necessary.

In the case of the Hamiltonian dynamics, the Hamiltonian can be written as
\begin{multline}
H_G = \sum_i \Bigg(
\sum_{s\in\mathcal{A}} \big( \lambda_{1,i} \ket*{s s^{-1}}\bra{\tte\tte}_i +  \lambda_{2,i} \ket{s\tte}\bra{\tte s}_i \big)
 + \\[-5pt]
 \sum_{r \in R} \lambda_{3,r,i} \ket*{r_\mathrm{left}} \bra*{r_\mathrm{right}}_i
\Bigg) + \mathrm{h.c.}
\label{eq:H_G}
\end{multline}
Here, $\lambda_{1,i}, \lambda_{2,i}, \lambda_{3,r,i}$ are arbitrary constants and $i$ denotes a spatial site.
The first term implements creation or annihilation of a character and its inverse (known as \emph{free reduction}), the second term allows for a commutation with an identity $\tte$, while the last term implements relations in $R$. 
We discuss how implementing different boundary conditions on~\eqref{eq:H_G} affects the dynamics in Sec.~\ref{sec:boundary-conditions}.\footnote{While~\eqref{eq:H_G} encompasses a large family of fragmented models, not all local quantum dynamics exhibiting fragmentation can be understood as \emph{group} dynamics; one notable exception is in models with dipole conservation in 1D~\cite{PPN, KHN, Sala2020}, which can instead be formulated as \emph{semigroup} dynamics~\cite{balasubramanian2023glassy}.}

In the case of random-unitary dynamics, each random unitary $U$ with support on $l$ sites can act non-trivially only within each sector $K_{g,l}$, where $l$ must be $\geq l_R$. In other words, for any two words $w, w'$ with length $\abs*{w} = \abs*{w'} = l$, $\mel*{w'}{U}{w}$ can only be nonzero if $\varphi(w) = \varphi(w')$. Other than that, the $U$'s can be randomly drawn from the appropriate ensemble.

If we instead consider dynamics with a shorter interaction range than $l_R$, then we will end up implementing the wrong group.
Indeed, to properly implement the group-associated dynamics, we must have at least one substitution rule derived from every relation in $R$. In Appendix~\ref{app:pair-flip}, we discuss how this phenomenon leads to fragmentation in the one-dimensional pair-flip model of Refs.~\cite{CahaNagaj, moudgalya2022hilbert}.

\subsection{The word problem}
\label{subsec:word_problem}

The task of determining whether $\varphi(w) = \varphi(w')$ for any two given words $w, w' $ constitutes the \emph{group word problem}. For certain classes of groups, this problem can be solved in polynomial (in the length of the word) time. However, there exist groups with simple presentations for which the word problem is very difficult, or even undecidable \cite{markov1948impossibility, novikov1955algorithmic, tseitin1958associative, boone1959word}.

We focus our attention on a specific way of solving the word problem, the so-called \emph{Dehn proof system}. Within this system, one proves that $\varphi(w) = \varphi(w')$ by presenting a \emph{derivation} from $w$ to $w'$,
\eq{
D(w \rightsquigarrow w') = w \to v_1 \to v_2 \to \dots \to v_n \to w',
\label{eq:derivation}
}
where one sequentially deforms $w$ into $w'$ by applying a single local substitution rule at each step. Allowed substitution rules include:
\begin{enumerate*}[label={(\roman*)}]
    \item rules derived from relations in $R$ (e.g., if a relation is $\ttx\tty = \tty\ttx$, then the substitution rules derived from it are $\ttx\tty \leftrightarrow \tty\ttx$, $\tty^{-1}\ttx\tty \leftrightarrow \ttx$, $\ttx\tty\ttx^{-1}\tty^{-1} \leftrightarrow \varepsilon$, etc., where $\varepsilon$ is an empty string); 
    \item free reductions (e.g., $\ttx\ttx^{-1} \leftrightarrow \varepsilon$, etc.); 
    \item rules involving the identity character $\tte$ (e.g., $\tte \leftrightarrow \varepsilon$, $\ttx\tte \leftrightarrow \tte\ttx$, etc.).
\end{enumerate*}
These substitution rules are like the ones from the previous subsection except that they are allowed to change the length of the word.

The temporal complexity of the word problem involves the length of the shortest derivation from $w$ to $w'$.
This length is given by the \emph{Dehn function} \cite{clay2017office, brady2007geometry, bridson2002geometry}, defined as \eq{
\dehn(w, w') = \min_D \abs{D(w \rightsquigarrow w')},
}
where $\abs{D}$ is the number of words in derivation $D$. Up to polynomial-in-$L$ corrections, $\dehn(w, w')$ is equal to the number of substitutions (i) only. Similarly, up to $\poly(L)$ corrections, in the language of circuits, $\dehn(w,w')$ corresponds to the minimal number of local gates obeying group constraints required to connect the two words.
To characterize the temporal complexity of the word problem for the whole group, we can define 
\eq{
\dehn(L) \equiv \max_{w \in K_{\tte,L}} \dehn(w, \tte^L),
}
the maximal Dehn function between $\tte^L$ and all other words of length $L$ that are equivalent to the identity. Reference~\cite{balasubramanian2023glassy} shows that this function bounds all Dehn functions, and depends only on the group (not the presentation), up to contributions linear in $L$.

As discussed in detail in Ref.~\cite{balasubramanian2023glassy}, the Dehn function completely governs thermalization timescales in group-dynamics models. It is therefore expected that models where $\dehn(L)$ scales exponentially (or faster) with $L$ exhibit glassy behavior, in that there would exist product states that would take at least an exponentially long (in $L$) time to thermalize.\footnote{Since the Dehn function is defined as the ``worst-case'' instance of temporal complexity, these bounds only indicate the existence of \emph{some} ``glassy'' words. On the other hand, to figure out whether \emph{typical} product states would take a long time to thermalize, one needs to introduce the \emph{typical Dehn function}, which characterizes the shortest derivation length between two typical words. Such a function can also grow faster than $\poly(L)$, and similar bounds for thermalization time exist~\cite{balasubramanian2023glassy}.}

For different groups, $\dehn(L)$ can scale with $L$ as nearly any function. The ``easiest'' groups are hyperbolic groups (which include all finite groups), whose Dehn function scales linearly with $L$, $\dehn(L) \sim L$.\footnote{We write $f(L) \sim g(L)$ if $f(L/c) \leq g(L) \leq f(cL)$ for all $L \geq L_0$, for some $c > 0$ and $L_0 > 0$. Correspondingly, $f(L) \lesssim g(L)$ if $f(cL) \leq g(L)$ for all $L \geq L_0$, for some $c > 0$ and $L_0 > 0$.} For infinite Abelian groups, $\dehn(L) \sim L^2$, and for automatic groups, $\dehn(L) \lesssim L^2$. For infinite non-Abelian groups (with finite presentations), $\dehn(L)$ can scale polynomially $\sim L^\alpha$ 
~\cite{brady2000there, sapir2002isoperimetric}, 
exponentially $\sim \exp(L)$, faster than $\exp(L)$, or even faster than any recursive function, rendering the word problem undecidable~\cite{novikov1955algorithmic, boone1959word}.\footnote{Note that $\dehn(L)$ indicates the (non-deterministic) complexity of the word problem within the Dehn proof system. For some groups, the word problem can be solved faster by  other methods. For example, for the Baumslag-Solitar group $\bs(1,2)$ introduced below, $\dehn(L) \sim \exp(L)$, while the word problem can be solved in $\sim L$ time by employing the 2-dimensional linear representation of the group.}

An example of a ``hard'' group with a simple presentation is the Baumslag-Solitar group $\bs(1,2)$ \cite{baumslag1969non} (for brevity, henceforth denoted $\bs$). Its simplest presentation is
\eq{
\bs = \braket{ \tta, \ttb \,}{\, \tta\ttb = \ttb \tta^2 },
}
and its Dehn function  scales as $\dehn(L) \sim 2^L$.
A slightly modified group, called the iterated Baumslag-Solitar group,
\eq{
\bs^{(2)} = \braket{ \tta, \ttb, \ttc \,}{\, \tta\ttb = \ttb \tta^2, \ttb\ttc = \ttc\ttb^2 },
}
exhibits an even more rapid Dehn function growth, scaling as $\dehn(L) \sim 2^{2^L}$ \cite{balasubramanian2023glassy}.

Having asked the question ``How much time do we need to solve the word problem?'' it is then logical to also ask ``How much space do we need to solve the word problem?'' Indeed, when we perform a derivation \eqref{eq:derivation}, every subsequent word need not be shorter than the previous one. 
Let us define the \emph{expansion length}\footnote{Called the \emph{filling length} function in the math literature.} between two words $w, w'$ as
\eq{
\el(w, w') = \min_{D(w \rightsquigarrow w')} \max_{u_i \in D(w \rightsquigarrow w')} \abs{u_i},
}
the maximal length of an intermediate word minimized among all possible derivations from $w$ to $w'$. In other words, one must grow the length of the word to at least $\el(w, w')$ in order to deform $w$ into $w'$.\footnote{A derivation that is optimal in terms of the extra space required need not be optimal in terms of the number of time steps required.} We can introduce
\eq{
\el(L) \equiv \max_{w \in K_{\tte, L}} \el(w, \tte^L)
}
as the measure of spatial complexity of the group word problem.
For finite groups, $\el(L) \leq L + C$ for some constant $C$, while for Abelian groups, $\el(L) \sim L$. While the $\bs$ group has $\el(L) \sim L$, the $\bs^{(2)}$ group has $\el(L) \sim \exp(L)$. In general, the expansion length is lower bounded by $\log_{\abs{\mathcal{A}}} \dehn(L)$~\cite{gersten2002filling, balasubramanian2023glassy}.

By design, the group dynamics fragments the Hilbert space into Krylov sectors $K_{g,L}$ associated with group elements $g \in G$. However, if the expansion length is larger than the system size, $\el(L) > L$, then some $K_{g,L}$ will be further fragmented into smaller sectors because connecting some product states with support of size $L$ requires an intermediate state with support of size $>L$.
In this case, ergodicity within $K_{g,L}$ is broken, but can be restored by temporarily padding the system with sites in the $\ket{\tte}$ state, allowing the dynamics to happen within this extended system, and then removing the extra sites, having brought them back to the $\ket{\tte}$ state. For this reason, the authors of \cite{balasubramanian2023glassy} call this phenomenon "fragile" Hilbert space fragmentation. 

\subsection{Boundary conditions}
\label{sec:boundary-conditions}
Above, we always implicitly assumed open boundary conditions (OBC), which is a natural choice in the math setting, since a word has a beginning and an end. However, in a physical setting, we may also be interested in periodic boundary conditions (PBC).
Imposing PBC affects both intrinsic and fragile fragmentation.

Let us first consider the consequences for the intrinsic fragmentation. Since with PBC, the choice of origin is arbitrary, words are defined only up to a cyclic shift. This leads to intrinsic Krylov sectors being labeled by conjugacy classes of the group $G$, instead of its group elements. Indeed, consider a chain of length $L$ and arbitrarily fix an origin. Suppose that starting from that site, the degrees of freedom read a word $w \equiv w_1 \cdot w_2$, with $\varphi(w) = g \in G$, and $\varphi(w_{1,2}) = g_{1,2} \in G$, respectively. Through a cyclic shift of characters, one can obtain the word $w' \equiv w_2 \cdot w_1$ with $\varphi(w') = g' \in G$. Therefore, $g' = g_2 g g_2^{-1}$, i.e., $g'$ is in the same conjugacy class as $g$. In fact, intrinsic Krylov sectors for \emph{any} two group elements in the same conjugacy class would merge. For any group element $\tilde{g} \in G$, we can pad the word $w$ with $\tilde{w} \cdot \tilde{w}^{-1}$ (where $\tilde{w}$ is some word with $\varphi(\tilde{w}) = \tilde{g}$) without changing its group element, since $\varphi(w) = \varphi(w \cdot \tilde{w} \cdot \tilde{w}^{-1}) = g$. Then, through a cyclic shift, the word can be changed into $\tilde{w}^{-1} \cdot w \cdot \tilde{w}$, with $\varphi(\tilde{w}^{-1} \cdot w \cdot \tilde{w}) = \tilde{g}^{-1} g \tilde{g}$. Thus, with PBC, for any $g, \tilde{g} \in G$,  intrinsic Krylov sectors $K_{g,L}$ and $K_{\tilde{g}^{-1} g \tilde{g},L}$ merge. We denote the intrinsic Krylov sectors for the system of size $L$ with PBC as $K_{\left[g\right], L}$, where $\left[g\right]$ is the conjugacy class of $g \in G$.
Note that choosing a different origin for a chain with PBC is equivalent to performing a conjugation by some group element and does not relabel the intrinsic Krylov sectors.

In addition to changing the structure of the intrinsic Krylov sectors, introducing PBC will merge some of the fragile Krylov sectors. For example, if $w = w_1 \cdot w_2$, $w' = w_2 \cdot w_1$ and $\varphi(w) = \varphi(w')$, but $\el(w,w') > \abs{w}+1 = L$, then states $\ket{w \cdot \tte}$ and $\ket*{w' \cdot \tte}$ would be dynamically disconnected in a system of size $L$ with OBC. With PBC, however, these states would become connected through a cyclic shift of characters. Note here that the extra $\tte$ character ensures that it is possible to perform the cyclic shift through simply commuting every character with $\tte$. Otherwise, the cyclic shift might be only obtainable through a global update of the whole word (unless $w$ already contains an $\tte$, or the cyclic shift can be performed by non-trivial local updates), in which case $\ket{w}$ and $\ket*{w'}$ would still remain disconnected. The inability to perform the cyclic shift with local updates can give rise to additional fragmentation since certain words belonging to the same conjugacy class remain dynamically disconnected. However, these words would become connected in systems of larger size where they are padded with at least one identity character.

\subsection{Conserved quantities}
\label{sub:conservation}

Here, we first establish to what extent the Krylov sector structure can be attributed to the presence of simple global symmetries, then explicitly construct the complete set of conserved quantities for groups with faithful linear representations.

\subsubsection{Global symmetries}

Appendix D in Ref.~\cite{balasubramanian2023glassy} shows that if $G$ is non-Abelian, then no simple global symmetry can be employed to distinguish \emph{all} intrinsic Krylov sectors: completely resolving all $K_{g,L}$ sectors requires unitaries that are not locality-preserving.
This does not, however, rule out the possibility that \emph{some} $K_{g,L}$ sectors can be distinguished by global symmetries.
In Appendix~\ref{app:general_charges} we show which intrinsic Krylov sectors can be distinguished using global symmetries and which cannot by using the \emph{commutator subgroup}. The commutator subgroup of $G$, denoted $\comm{G}{G}$, is a subgroup generated by all commutators $\comm{g}{h} \equiv ghg^{-1}h^{-1}$ of group elements $g,h \in G$. 
The commutator subgroup satisfies $\comm{G}{G} \vartriangleleft G$, so that $\comm{G}{G}$ splits $G$ into cosets. As shown in Appendix \ref{app:general_charges}, the sector $K_{g_1,L}$ can be distinguished from $K_{g_2,L}$ by a global symmetry only if $g_1$ and $g_2$ belong to different cosets, while distinguishing $g_1$ and $g_2$ from the same coset requires non-local operators.

Thus, for an Abelian group (with $\comm{G}{G}$ trivial), all sectors can be distinguished by global symmetry charges, which we construct explicitly in Appendix~\ref{app:abelian_charges}. In particular, if $G=\zz_N$ then the global symmetry is $\zz_N$, and if $G=\zz$, then the symmetry group is $\U{1}$. On the other hand, a group that satisfies $\comm{G}{G} = G$ is known as a \emph{perfect group} and possesses no intrinsic Krylov sectors that can be distinguished using a simple global symmetry. Such perfect groups allow for the possibility of Hilbert space fragmentation without a concomitant protecting global symmetry.\footnote{An example of a finitely generated~\cite{Conder1992}, infinite, perfect group is $\SL(3, \zz)$.} In general, the full global symmetry for a (possibly non-Abelian) group $G$ is completely determined by the \emph{Abelianization} of $G$, called $G^\text{ab} = G / [G,G]$, which is an Abelian group and hence can always be written as 
\begin{equation}
G^\mathrm{ab} = \zz_{p_1} \times \dots \times \zz_{p_{n-l}} \times \zz^l .
\end{equation}
In this case, the global symmetry group is $\zz_{p_1} \times \dots \times \zz_{p_{n-l}} \times \mathrm{U}(1)^l$. We construct the symmetry charges for this group in Appendix~\ref{app:abelian_charges}.

As an example, consider the group $\Z{2}^{*3}$, which is related to the three-colored pair-flip model~\cite{CahaNagaj, moudgalya2022hilbert} studied in the literature, with presentation
\eq{
\Z{2} * \Z{2} * \Z{2} = \braket{ \ttr , \ttg , \ttb  \, }{ \, \ttr^2 = \ttg^2 = \ttb^2 = \tte }
\, .
\label{eqn:pair-flip-group}
}
To systematically identify discrete global symmetries of the model, consider the elements of the commutator subgroup $\comm{G}{G}$. 
Since any element of $[G, G]$ can be decomposed in terms of group elements $g=g_1 \cdots g_n$ that can be rearranged to give the identity,
$\comm{G}{G}$ consists of products of colors, $\ttr$, $\ttg$, $\ttb$, containing an \emph{even} number of each color. The associated cosets are then described by boolean variables $n_\ttr$, $n_\ttg$, and $n_\ttb$, which describe the parity of each color [e.g., $n_\ttr = 0$ ($1$) for an even (odd) number of $\ttr$'s, and similarly for the other colors].
Hence, the model's Krylov structure can be partially attributed to a $\Z2^3$ global symmetry associated to conservation of the parity of each color separately.

Interestingly, the pair-flip model in the literature, defined by the same group as in Eq.~\eqref{eqn:pair-flip-group}, but without an onsite identity character (i.e., $\mathcal{A} = \{\ttr, \ttg, \ttb \}$)
exhibits a global $\U{1}^2$ symmetry~\cite{moudgalya2022hilbert} (see also Appendix~\ref{app:pair-flip}). Thus, the symmetry is a function of the group alone only when the local Hilbert space is spanned by a minimal set of generators, their inverses, and the identity. Restricting $\mathcal{A}$ leads to additional constraints and larger symmetries. On the other hand, some of the fragmentation phenomena we describe can be removed by enlarging $\mathcal{A}$. 
 
As another example, consider the group 
\eq{
    \bs = \braket{ \tta, \ttb \, }{ \, \tta\ttb = \ttb \tta^2 }.
}
Once again, the symmetries of the model are identified by modding out by the commutator subgroup $[\bs, \bs]$. Any word can always be mapped to the standard form $w_{kn\ell} = \ttb^k \tta^{n} \ttb^{-\ell}$ for integers $k$, $n$, $\ell$, with $k, \ell \geq 0$~\cite{burillo2015metric,balasubramanian2023glassy}, where $n$ can only be even if either $k$ or $\ell$ is zero. These words are then in one-to-one correspondence with group elements of $\bs$. 
Elements of $[\bs, \bs]$ are of the form $\ttb^{\ell} \tta^{n} \ttb^{-\ell}$, and the associated cosets are described by the number of $\ttb$'s, i.e., consist of words of the form $\ttb^{\ell + n_\ttb} \tta^{n} \ttb^{-\ell}$ for fixed $n_\ttb$. This is leads to a global $\U1$ symmetry associated with the conserved quantity
\eq{
    n_\ttb = \sum_{i} \left( \ket{\ttb}\! \bra{\ttb}_i -\ket*{\ttb^{-1}}\! \bra*{\ttb^{-1}}_i \right)
    \, ,
}
which is the symmetry considered in Ref.~\cite{balasubramanian2023glassy}.

The global symmetry group is enlarged at special points in parameter space. For example, with special choices of the parameters, the pair-flip model~\eqref{eqn:pair-flip-group} becomes the Temperley-Lieb model, with a global $\SU{3}$ symmetry~\cite{moudgalya2022hilbert}. This enlarged symmetry is beyond the scope of this paper.

\subsubsection{Non-local conserved quantities}

Suppose that $G$ admits a finite-dimensional, faithful, linear representation $\rho$ in vector space $V$. That is, $\rho : G \to \GL(V)$ is a group homomorphism, $\rho(g_1g_2) = \rho(g_1)\rho(g_2)$, that is also an isomorphism. The existence of such a finite-dimensional representation allows the group word problem to be solved in a time polynomially large in the length of the word: one simply multiplies the matrices $\rho(\varphi(s))$ for $s \in \mathcal{A}$ to find the group element to which the word corresponds. One can also use this fact to explicitly construct conserved quantities associated to dynamics of the form~\eqref{eq:H_G}. Specifically,
\eq{
\hat{O} = \bigotimes_{i=1}^{L} 
\sum_{s \in \mathcal{A}} \rho(\varphi(s))  \ket{s} \bra{s}_i = \sum_{w} \rho(\varphi(w)) \ket{w} \bra{w}
\, ,
\label{eq:matrix-conserved-q}
}
where $\bigotimes$ implements a tensor product on the quantum states and matrix multiplication on $\rho$. The matrix-valued operator $\hat{O}$ is ``diagonal'' in the computational basis, meaning that $\mel*{w}{\hat{O}}{w'}$ is a non-zero (invertible) matrix if $w = w'$, and a zero-matrix otherwise. The matrix satisfies $\mel{w}{\hat{O}}{w} = \rho(g)$ for any $\ket{w} \in K_{g,L}$. While this operator distinguishes all Krylov sectors, it is generically not locality preserving and cannot be interpreted as a global symmetry operator. If, however, $G$ is cyclic, then it admits a faithful complex 1-dimensional representation. In this case, $\rho(g)$ are $1 \times 1$ complex matrices (i.e., scalars) and, as a consequence, $\hat{O}$ becomes a global symmetry.

\section{Constructing group models in 2D} \label{sec:2d}
We now generalize group models to 2D. A coarse-grained, ``continuous'' approach to 2D group models has been introduced in Ref.~\cite{balasubramanian2023glassy}. In this section and the next, we focus on microscopic realizations on two-dimensional crystalline lattices. Specifically, we describe the interplay between the lattice and the group, and discuss the concomitant phenomena that can arise.

In this section, we will define the 2D lattice group model and see that all of the phenomenology from the 1D model can be found in 2D. This includes intrinsic fragmentation, as expected from the continuum picture, and also fragile fragmentation.  Furthermore, both types of fragmentation become topologically robust. In the next section we will see how microscopic lattice details lead to new ways that the dynamics can be even less ergodic.

\subsection{The flatness condition}

Consider a translationally invariant 2D lattice where every edge, $i$, is assigned one of the two orientations, $q_i \in \{ \pm 1\}$, pictorially denoted as an arrow $\Vec{q}_i$ along the edge (e.g., the black arrows in Fig.~\ref{fig:loops}). The orientations can be assigned in an arbitrary way, but we consider them fixed. The degrees of freedom are situated on the edges. Similar to the 1D case, the local Hilbert space is spanned by states $\ket{s}$, with $s \in \mathcal{A} \equiv \mathcal{S} \cup \mathcal{S}^{-1} \cup \{\tte\}$, where $\mathcal{S}$ is a set of generators of a group $G$ with presentation $G = \braket{ \mathcal{S}\,}{\,R}$. Any oriented continuous path on the lattice that hosts consecutive generators $s_1 s_2 \cdots s_L$ constitutes a word $w = s_1^{q_1} s_2^{q_2} \cdots s_L^{q_L}$. If the path is anti-aligned with the edge, then the corresponding generator is read as its inverse.

We can view every closed loop as a 1D group model with PBC and impose the \emph{flatness condition}:\footnote{This flatness condition looks like a generalized Gauss law in a discrete gauge theory. Indeed, for Abelian gauge group $G$, Gauss laws and flatness conditions are dual to each other. However, for non-Abelian $G$, violations of Gauss laws are valued in representations of a group while violations of flatness are valued in conjugacy classes of $G$~\cite{Kitaev2003}.} we require that, around every elementary plaquette $p$,
\begin{equation}
    \varphi(w_p) = \tte
    \, ,
    \label{eqn:flatness}
\end{equation}
where $w_p$ is the word formed around $p$.
First, note that this constraint is independent of where we choose the origin for every loop, since the identity group element is its own conjugacy class, so that $\varphi(w_1 \cdot w_2) = \tte$ implies $\varphi(w_2 \cdot w_1) = \tte$. It also follows that the orientation of each loop does not matter, since if $\varphi(w) = \tte$, then $\varphi(w^{-1}) = \tte$. Furthermore, note that imposing the flatness condition locally, on every elementary plaquette of the 2D lattice, implies that the condition is satisfied on \emph{every} contractible loop. Indeed, suppose two loops are in the states $\ket{w_1 \cdot w}$ and $\ket*{w_2 \cdot w^{-1}}$, and that they share a common path in the state $\ket{w}$. Then, the loop obtained by concatenation of the two smaller loops hosts the state $\ket{w_1 \cdot w_2}$. If $\varphi(w_1 \cdot w) = \varphi(w_2 \cdot w^{-1}) = \tte$, then we immediately get that $\varphi(w_1 \cdot w_2) = \tte$ as well. This argument shows that the flatness condition~\eqref{eqn:flatness} implies $\varphi(w_{\partial S})=\tte$ for any oriented path $\partial S$ corresponding to the boundary of a region $S$ constructed from elementary plaquettes. 

On the other hand, the flatness condition leaves open the possibility of a nontrivial $w_\gamma$ for noncontractible paths $\gamma$. Furthermore, two paths that are homotopically deformable into each other will host words that map to the same group element (for open boundary conditions as discussed below) or the same conjugacy class (for periodic boundary conditions).
These group elements or conjugacy classes will label intrinsic Krylov sectors.

\subsection{Allowed product states: loops and nets}
\label{sec:tiles}

The flatness condition~\eqref{eqn:flatness} defines ``allowed'' configurations of characters around each elementary plaquette. These allowed configurations derive from both
\begin{enumerate*}[label={(\roman*)}]
    \item \label{item:free-reduction} free reduction, and
    \item \label{item:relations} the group $G$'s generating relations $R$.
\end{enumerate*}
The existence of \ref{item:free-reduction} can be interpreted as closed loop segments on the dual lattice. For example, around a hexagonal plaquette $p$, one can have such product states as
\begin{equation} \label{eqn:loop-segments}
    \scalebox{0.75}{
    \begin{tikzpicture}[baseline={(0,-0.1)}]
    \begin{scope}[very thick,decoration={markings,
    mark=at position 0.7 with {\arrow[scale=1]{Classical TikZ Rightarrow}}}
    ] 
    \path
    node[regular polygon, regular polygon sides=6, inner sep=14pt] (hex) {};
    \draw [thick, ->, >=stealth] (hex.corner 3) to (hex.corner 2);
    \draw [thick, ->, >=stealth] (hex.corner 2) to (hex.corner 1);
    \draw [thick, ->, >=stealth] (hex.corner 1) to (hex.corner 6);
    \draw [thick, ->, >=stealth] (hex.corner 3) to (hex.corner 4);
    \draw [thick, ->, >=stealth] (hex.corner 4) to (hex.corner 5);
    \draw [thick, ->, >=stealth] (hex.corner 5) to (hex.corner 6);
    \fill [agen] (hex.side 1) circle[radius=2pt];
    \fill [agen] (hex.side 3) circle[radius=2pt];
    \fill [bgen] (hex.side 4) circle[radius=2pt];
    \fill [bgen] (hex.side 5) circle[radius=2pt];
    \draw [line width=2, agen, bend left=30pt, postaction={decorate}] (hex.side 1) to (hex.side 3);
    \draw [line width=2, bgen, bend left=30pt, postaction={decorate}] (hex.side 4) to (hex.side 5);
    \end{scope}
    \end{tikzpicture}
    }
    \Longleftrightarrow 
    \scalebox{0.75}{
    \begin{tikzpicture}[baseline={(0,-0.1)}]
    \begin{scope}[very thick,decoration={markings,
    mark=at position 0.7 with {\arrow{Classical TikZ Rightarrow}}}
    ] 
    \path
    node[regular polygon, regular polygon sides=6, inner sep=14pt] (hex) {};
    \draw [thick, ->, >=stealth] (hex.corner 3) to (hex.corner 2);
    \draw [thick, ->, >=stealth] (hex.corner 2) to (hex.corner 1);
    \draw [thick, ->, >=stealth] (hex.corner 1) to (hex.corner 6);
    \draw [thick, ->, >=stealth] (hex.corner 3) to (hex.corner 4);
    \draw [thick, ->, >=stealth] (hex.corner 4) to (hex.corner 5);
    \draw [thick, ->, >=stealth] (hex.corner 5) to (hex.corner 6);
    \node [circle, agen, fill=agen!20!white, inner sep=0, minimum size=3.75mm, draw, thick, font=\footnotesize] at (hex.side 1) {$\tta$};
    \node [circle, agen, fill=agen!20!white, inner sep=0, minimum size=3.75mm, draw, thick, font=\footnotesize] at (hex.side 3) {$\tta$};
    \node [circle, bgen, fill=bgen!20!white, inner sep=0, minimum size=3.75mm, draw, thick, font=\footnotesize] at (hex.side 4) {$\bar{\ttb}$};
    \node [circle, bgen, fill=bgen!20!white, inner sep=0, minimum size=3.75mm, draw, thick, font=\footnotesize] at (hex.side 5) {$\ttb$};
    \end{scope}
    \end{tikzpicture}
    }
    \quad
    \text{or}
    \quad\:
    \scalebox{0.75}{
    \begin{tikzpicture}[baseline={(0,-0.1)}]
    \begin{scope}[very thick,decoration={markings,
    mark=at position 0.3 with {\arrow{Classical TikZ Rightarrow}},
    mark=at position 0.85 with {\arrow{Classical TikZ Rightarrow}}}
    ] 
    \path
    node[regular polygon, regular polygon sides=6, inner sep=14pt] (hex) {};
    \draw [thick, ->, >=stealth] (hex.corner 3) to (hex.corner 2);
    \draw [thick, ->, >=stealth] (hex.corner 2) to (hex.corner 1);
    \draw [thick, ->, >=stealth] (hex.corner 1) to (hex.corner 6);
    \draw [thick, ->, >=stealth] (hex.corner 3) to (hex.corner 4);
    \draw [thick, ->, >=stealth] (hex.corner 4) to (hex.corner 5);
    \draw [thick, ->, >=stealth] (hex.corner 5) to (hex.corner 6);
    \fill [agen] (hex.side 1) circle[radius=2pt];
    \fill [agen] (hex.side 3) circle[radius=2pt];
    \fill [agen] (hex.side 4) circle[radius=2pt];
    \fill [agen] (hex.side 5) circle[radius=2pt];
    \draw [line width=2, agen, postaction={decorate}] (hex.side 1) to (0,0) to (hex.side 3);
    \draw [line width=2, agen, postaction={decorate}] (hex.side 4) to (0,0) to (hex.side 5);
    \end{scope}
    \end{tikzpicture}
    }
    \Longleftrightarrow 
    \scalebox{0.75}{
    \begin{tikzpicture}[baseline={(0,-0.1)}]
    \begin{scope}[very thick,decoration={markings,
    mark=at position 0.7 with {\arrow{Classical TikZ Rightarrow}}}
    ] 
    \path
    node[regular polygon, regular polygon sides=6, inner sep=14pt] (hex) {};
    \draw [thick, ->, >=stealth] (hex.corner 3) to (hex.corner 2);
    \draw [thick, ->, >=stealth] (hex.corner 2) to (hex.corner 1);
    \draw [thick, ->, >=stealth] (hex.corner 1) to (hex.corner 6);
    \draw [thick, ->, >=stealth] (hex.corner 3) to (hex.corner 4);
    \draw [thick, ->, >=stealth] (hex.corner 4) to (hex.corner 5);
    \draw [thick, ->, >=stealth] (hex.corner 5) to (hex.corner 6);
    \node [circle, agen, fill=agen!20!white, inner sep=0, minimum size=3.75mm, draw, thick, font=\footnotesize] at (hex.side 1) {$\tta$};
    \node [circle, agen, fill=agen!20!white, inner sep=0, minimum size=3.75mm, draw, thick, font=\footnotesize] at (hex.side 3) {$\tta$};
    \node [circle, agen, fill=agen!20!white, inner sep=0, minimum size=3.75mm, draw, thick, font=\footnotesize] at (hex.side 4) {$\bar{\tta}$};
    \node [circle, agen, fill=agen!20!white, inner sep=0, minimum size=3.75mm, draw, thick, font=\footnotesize] at (hex.side 5) {$\tta$};
    \end{scope}
    \end{tikzpicture}
    }
    \enspace,
\end{equation}
where an edge without a colored dot is assumed to be in the $\ket{\tte}$ state and a bar above a character denotes its inverse.
When the characters are concatenated around $p$ to read a word (taking into account the orientation of the edges), we have $\varphi(w_p) = \tte$. Note that reversing the orientation of one of the arrows on the primary lattice is equivalent to a unitary transformation that sends $\ket{s} \leftrightarrow \ket*{s^{-1}}$ $\forall s \in \mathcal{A}$ on the corresponding edge~\cite{Kitaev2003}. Similarly, reversing the direction of a colored dual loop represents the state with $\ket{s} \leftrightarrow \ket*{s^{-1}}$ on all edges this loop intersects, since, in our graphical notation, whether an edge hosts a character or its inverse is determined by whether the cross product $\vec{d}_i \times \vec{q}_i$ points into or out of the page, where $\vec{d}_i$ denotes the local orientation of the dual loop.
To make the graphical notation in one-to-one correspondence with the configuration of generators, we draw a crossing of loops if there exists an ambiguity in how the generators are paired [as on the right-hand side of~\eqref{eqn:loop-segments}]. 
The relations \ref{item:relations} give rise to allowed configurations such as 
\begin{equation} \label{eqn:basepoints-example}
    \scalebox{0.75}{
    \begin{tikzpicture}[baseline={(0,-0.1)}]
    \begin{scope}[very thick,decoration={markings,
    mark=at position 0.9 with {\arrow{Classical TikZ Rightarrow}},
    mark=at position 0.275 with {\arrow{Classical TikZ Rightarrow}}}
    ] 
    \path
    node[regular polygon, regular polygon sides=6, inner sep=14pt] (hex) {};
    \draw [thick, ->, >=stealth] (hex.corner 3) to (hex.corner 2);
    \draw [thick, ->, >=stealth] (hex.corner 2) to (hex.corner 1);
    \draw [thick, ->, >=stealth] (hex.corner 1) to (hex.corner 6);
    \draw [thick, ->, >=stealth] (hex.corner 3) to (hex.corner 4);
    \draw [thick, ->, >=stealth] (hex.corner 4) to (hex.corner 5);
    \draw [thick, ->, >=stealth] (hex.corner 5) to (hex.corner 6);
    \fill [bgen] (hex.side 2) circle[radius=2pt];
    \fill [cgen] (hex.side 3) circle[radius=2pt];
    \fill [bgen] (hex.side 5) circle[radius=2pt];
    \fill [cgen] (hex.side 6) circle[radius=2pt];
    \draw [line width=2, bgen, postaction={decorate}] (hex.side 2) to (hex.side 5);
    \draw [line width=2, cgen, postaction={decorate}] (hex.side 6) to (hex.side 3);
    \draw [black, fill=white] (-2pt, -2pt) rectangle (2pt, 2pt);
    \end{scope}
    \end{tikzpicture}
    }
    \Longleftrightarrow 
    \scalebox{0.75}{
    \begin{tikzpicture}[baseline={(0,-0.1)}]
    \begin{scope}[very thick,decoration={markings,
    mark=at position 0.9 with {\arrow{Classical TikZ Rightarrow}},
    mark=at position 0.275 with {\arrow{Classical TikZ Rightarrow}}}
    ] 
    \path
    node[regular polygon, regular polygon sides=6, inner sep=14pt] (hex) {};
    \draw [thick, ->, >=stealth] (hex.corner 3) to (hex.corner 2);
    \draw [thick, ->, >=stealth] (hex.corner 2) to (hex.corner 1);
    \draw [thick, ->, >=stealth] (hex.corner 1) to (hex.corner 6);
    \draw [thick, ->, >=stealth] (hex.corner 3) to (hex.corner 4);
    \draw [thick, ->, >=stealth] (hex.corner 4) to (hex.corner 5);
    \draw [thick, ->, >=stealth] (hex.corner 5) to (hex.corner 6);
    \node [circle, bgen, fill=bgen!20!white, inner sep=0, minimum size=3.75mm, draw, thick, font=\footnotesize] at (hex.side 2) {$\ttb$};
    \node [circle, cgen, fill=cgen!20!white, inner sep=0, minimum size=3.75mm, draw, thick, font=\footnotesize] at (hex.side 3) {$\ttc$};
    \node [circle, bgen, fill=bgen!20!white, inner sep=0, minimum size=3.75mm, draw, thick, font=\footnotesize] at (hex.side 5) {$\ttb$};
    \node [circle, cgen, fill=cgen!20!white, inner sep=0, minimum size=3.75mm, draw, thick, font=\footnotesize] at (hex.side 6) {$\ttc$};
    \end{scope}
    \end{tikzpicture}
    }
    \quad
    \text{or}
    \quad\:
    \scalebox{0.75}{
    \begin{tikzpicture}[baseline={(0,-0.1)}]
    \begin{scope}[very thick,decoration={markings,
    mark=at position 0.7 with {\arrow{Classical TikZ Rightarrow}}}
    ] 
    \path
    node[regular polygon, regular polygon sides=6, inner sep=14pt] (hex) {};
    \draw [thick, ->, >=stealth] (hex.corner 3) to (hex.corner 2);
    \draw [thick, ->, >=stealth] (hex.corner 2) to (hex.corner 1);
    \draw [thick, ->, >=stealth] (hex.corner 1) to (hex.corner 6);
    \draw [thick, ->, >=stealth] (hex.corner 3) to (hex.corner 4);
    \draw [thick, ->, >=stealth] (hex.corner 4) to (hex.corner 5);
    \draw [thick, ->, >=stealth] (hex.corner 5) to (hex.corner 6);
    \node [inner sep=0pt] (C) at (0, 0) {};
    \fill [agen] (hex.side 1) circle[radius=2pt];
    \draw [line width=2, agen, postaction={decorate}] (hex.side 1) to (C);
    \fill [agen] (hex.side 2) circle[radius=2pt];
    \draw [line width=2, agen, postaction={decorate}] (hex.side 2) to (C);
    \fill [agen] (hex.side 3) circle[radius=2pt];
    \draw [line width=2, agen, postaction={decorate}] (C) to (hex.side 3);
    \fill [dgen] (hex.side 4) circle[radius=2pt];
    \draw [line width=2, dgen, postaction={decorate}] (hex.side 4) to (C);
    \fill [cgen] (hex.side 5) circle[radius=2pt];
    \draw [line width=2, cgen, postaction={decorate}] (hex.side 5) to (C);
    \fill [bgen] (hex.side 6) circle[radius=2pt];
    \draw [line width=2, bgen, postaction={decorate}] (hex.side 6) to (C);
    \draw [black, fill=white] (-2pt, -2pt) rectangle (2pt, 2pt);
    \end{scope}
    \end{tikzpicture}
    }
    \Longleftrightarrow
    \scalebox{0.75}{
    \begin{tikzpicture}[baseline={(0,-0.1)}]
    \begin{scope}[very thick,decoration={markings,
    mark=at position 0.9 with {\arrow{Classical TikZ Rightarrow}},
    mark=at position 0.275 with {\arrow{Classical TikZ Rightarrow}}}
    ] 
    \path
    node[regular polygon, regular polygon sides=6, inner sep=14pt] (hex) {};
    \draw [thick, ->, >=stealth] (hex.corner 3) to (hex.corner 2);
    \draw [thick, ->, >=stealth] (hex.corner 2) to (hex.corner 1);
    \draw [thick, ->, >=stealth] (hex.corner 1) to (hex.corner 6);
    \draw [thick, ->, >=stealth] (hex.corner 3) to (hex.corner 4);
    \draw [thick, ->, >=stealth] (hex.corner 4) to (hex.corner 5);
    \draw [thick, ->, >=stealth] (hex.corner 5) to (hex.corner 6);
    \node [circle, agen, fill=agen!20!white, inner sep=0, minimum size=3.75mm, draw, thick, font=\footnotesize] at (hex.side 1) {$\tta$};
    \node [circle, agen, fill=agen!20!white, inner sep=0, minimum size=3.75mm, draw, thick, font=\footnotesize] at (hex.side 2) {$\tta$};
    \node [circle, agen, fill=agen!20!white, inner sep=0, minimum size=3.75mm, draw, thick, font=\footnotesize] at (hex.side 3) {$\tta$};
    \node [circle, dgen, fill=dgen!20!white, inner sep=0, minimum size=3.75mm, draw, thick, font=\footnotesize] at (hex.side 4) {$\bar{\ttd}$};
    \node [circle, cgen, fill=cgen!20!white, inner sep=0, minimum size=3.75mm, draw, thick, font=\footnotesize] at (hex.side 5) {$\bar{\ttc}$};
    \node [circle, bgen, fill=bgen!20!white, inner sep=0, minimum size=3.75mm, draw, thick, font=\footnotesize] at (hex.side 6) {$\ttb$};
    \end{scope}
    \end{tikzpicture}
    }
    \enspace ,
\end{equation}
where the relations $\ttb\ttc=\ttc\ttb$ and $\tta\ttb\ttc\ttd = \tte$ have been assumed. 
The latter example combines \ref{item:free-reduction} and \ref{item:relations}.
Plaquettes that host a word connected to $\tte$ via the application of at least one generating relation are denoted by a white square and will be referred to as \emph{basepoints}. 
Having made this choice, our graphical notation does not distinguish between generating relations and derived relations, i.e., the combination of multiple generating relations to form a word $w$ satisfying $\varphi(w)=\tte$.

\begin{figure}[t]
\centering
  \includegraphics[scale=0.65]{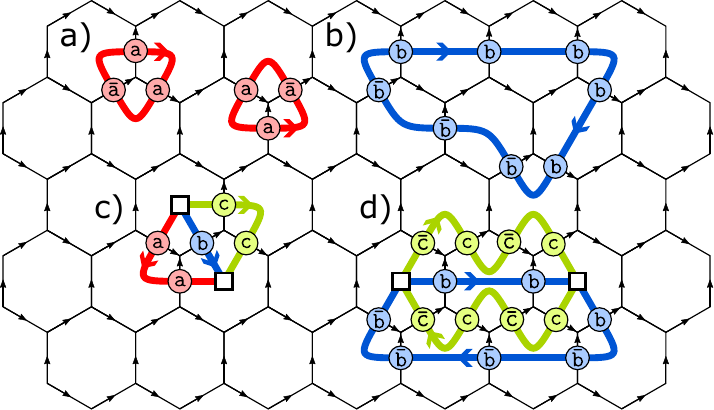}
  \caption{An example of a product state that satisfies the flatness condition. Bubbles of different colors denote different generators, with the bar above the character denoting the inverse. Every edge without a bubble is in the $\ket{\tte}$ state. Product states satisfying the flatness condition consist of the following objects: (a,b) Closed (non-intersecting) loops on the dual lattice, each corresponding to a single generator type. Each loop can be given one of the two orientations, which, together with the orientations of the lattice edges, determines whether to put a generator or its inverse on every given edge. 
  (c,d) Nets associated with words from the identity sector $K_{\tte, L}$ (e.g., relators from the group's presentation). Net (c) assumes that $\tta\ttb\ttc = \tte$, while net (d) is associated with the relation $\ttb\ttc = \ttc\ttb$. Every net has two basepoints (denoted using white squares).}
  \label{fig:loops}
\end{figure}

The allowed configurations of characters around plaquettes define a set of locally flat ``tiles,'' which may be tiled to generate a two-dimensional configuration satisfying~\eqref{eqn:flatness}. Not all tiles can be laid next to one another; two adjacent tiles that share an edge must agree on the character hosted by that edge.
This consistency condition ensures that a configuration satisfying the flatness condition consists of closed loops that can only terminate at basepoints.
We refer to a collection of several strings (generically corresponding to different generators) that connect two basepoints as a \emph{net}. Examples of simple closed-loop configurations and nets are shown in Figs.~\ref{fig:loops}\,(a) and (b) and Figs. (c) and (d), respectively.
Compatible collections of tiles define a Hilbert space of flat configurations that generically does not have tensor-product structure and grows exponentially with system volume $L^2$.

We can define two different boundary conditions, in analogy with the rough and smooth boundaries of the toric code~\cite{Bravyi1998Boundary}. The first is the rough boundary, or reflecting boundary. Here, the lattice terminates so that there is a dangling edge on every vertex. These dangling edges, along with the edges that connect them, define an ``open'' plaquette; impose flatness around this plaquette just like any other. Loops cannot end on this boundary, but the path on which we read words may terminate. Furthermore, the flatness condition on open plaquettes means that paths that begin and end on the same boundary read trivial words, while two paths that end at different points on the same boundary read the same word. The other boundary is the smooth or absorbing boundary. On this boundary, there are no dangling edges on any vertex, and the flatness condition is still enforced on every boundary plaquette. The paths on which we read the words cannot consistently terminate on this boundary, but the colored loops can.

\subsection{Dynamics} \label{sub:dynamics}

We now discuss the most general local dynamics compatible with the flatness condition. The operators that we describe nucleate and rearrange the strings and nets introduced in Sec.~\ref{sec:tiles}.
Define the diagonal projection operators
\begin{equation} \label{eqn:T_op}
    T^{h}_+ \equiv \ket{h}\bra{h}\, ,  \qquad
    T^{h}_- \equiv \ket*{h^{-1}}\bra*{h^{-1}} 
    \, ,
\end{equation}
for characters $h \in \mathcal{A}$ belonging to the set of generators.
From~\eqref{eqn:T_op}, we are able to define the projection operators onto flat configurations around $p$
\begin{equation} \label{eqn:flatness-operator}
    B_p = \sum_{ h_{e_1} \cdots h_{e_k} = \tte} \: \prod_{e \in \gamma_p} T^{h_e}(e, p)
    \, , 
\end{equation}
where the path $\gamma_p$ traverses $\partial p$ counterclockwise, crossing the ordered set of edges $\{ e_1, \dots, e_k \}$. The operator $T^{h}(e, p)$ evaluates to $T^{h}_+$ ($T^{h}_-$) if the edge $e$ is oriented with (against) the path $\gamma_p$. The operator $B_p$ acts as the identity on configurations such as~\eqref{eqn:loop-segments} and \eqref{eqn:basepoints-example} and annihilates configurations that violate Eq.~\eqref{eqn:flatness}.
We now construct the minimal dynamics that commutes with $B_p$ for all plaquettes, thereby preserving flatness.

First, we introduce the operators~\cite{Kitaev2003}
\begin{equation} \label{eqn:L_op}
    L^{g}_+ \equiv \sum_{s \in G} \ket{g s}\bra{s}\,, \qquad
    L^{g}_- \equiv \sum_{s \in G} \ket*{s g^{-1}}\bra{s}
    \, ,
\end{equation}
which multiply an onsite state by $g \in G$ on the left and by $g^{-1}$ on the right, respectively (i.e., $gs$ is interpreted as group multiplication rather than character concatenation).
Note that these operators may involve group elements that do not belong to the chosen onsite Hilbert space if $\mathcal{A}$ does not include the full group.
We therefore introduce the projector $\Pi = \sum_{s\in\mathcal{A}} \ket{s}\bra{s} = \sum_{s\in\mathcal{A}}T^s_+$ to select only the transitions between states belonging to the local Hilbert space and define $\tilde{L}_\pm^g \equiv \Pi L^g_\pm \Pi$.
From these projected operators we construct
\begin{equation} \label{eqn:generalized-Av}
    \tilde{A}_v(g) = \prod_{e \in \partial^\dagger v} \tilde{L}^g(e, v)
    \, ,
\end{equation}
where $\tilde{L}^g(e, v)$ acts as $\tilde{L}^g_+$ ($\tilde{L}^g_-$) if the edge $e$ points out of (into) vertex $v$ of the primary lattice, and $\partial^\dagger v$ is the dual boundary of the vertex $v$. While there are an infinite number of $\tilde{A}_v(g)$ operators for an infinite group, the only operators that can act nontrivially are those with $g \in \varphi(\mathcal{A} \times \mathcal{A})$, the set of group elements that can be written using two letters (which trivially includes elements that can be written using a single letter). See Sec.~\ref{sub:range} for an example of an operator that must be written with two letters.

To see that this operator preserves $B_p$, consider the following example. The action of $\tilde{A}_v(g)$ on a vertex $v$ shared between three plaquettes is:
\begin{equation}
\label{eq:action_of_Av}
    \scalebox{0.75}{
    \begin{tikzpicture}[baseline={(0,0)}]
        \draw [thick, ->, >=stealth] (0:0) to (0:1);
        \draw [thick, <-, >=stealth] (0:0) to (120:1);
        \draw [thick, ->, >=stealth] (0:0) to (240:1);
        \node [above] at (0:0.5) {$\ttb$};
        \node [left] at (120:0.5) {$\tta$};
        \node [right] at (240:0.5) {$\ttc$};
    \end{tikzpicture}
    }
    \quad
    \xrightarrow{\tilde{A}_v(g)}
    \quad
    \scalebox{0.75}{
    \begin{tikzpicture}[baseline={(0,0)}]
        \draw [thick, ->, >=stealth] (0:0) to (0:1);
        \draw [thick, <-, >=stealth] (0:0) to (120:1);
        \draw [thick, ->, >=stealth] (0:0) to (240:1);
        \node [above] at (0:0.5) {$g \ttb$};
        \node [left] at (120:0.5) {$\tta g^{-1}$};
        \node [right] at (240:0.5) {$g \ttc$};
    \end{tikzpicture}
    }
\end{equation}
The three loop segments on the primary lattice evaluate to $(\tta g^{-1})(g \ttb) = \tta\ttb$, $(g \ttb)^{-1}(g \ttc) = \ttb^{-1}\ttc$, and $(g \ttc)^{-1}(\tta g^{-1})^{-1} =\ttc^{-1}\tta^{-1}$. Hence, if the initial configuration was flat, then the state that results after applying $\tilde{A}_v(g)$ will be, too. When acting on the vacuum, the operator~\eqref{eqn:generalized-Av} never acts trivially and nucleates a closed, directed loop:
\begin{equation}
\label{eq:Av_creates_a_loop}
    \scalebox{0.75}{
    \begin{tikzpicture}[baseline={(0,0)}]
        \draw [thick, ->, >=stealth] (0:0) to (0:1);
        \draw [thick, <-, >=stealth] (0:0) to (120:1);
        \draw [thick, ->, >=stealth] (0:0) to (240:1);
    \end{tikzpicture}
    }
    \quad
    \xrightarrow{\tilde{A}_v(\tta)}
    \quad
    \scalebox{0.75}{
    \begin{tikzpicture}[baseline={(0,0)}]
        \begin{scope}[very thick,decoration={markings,
    mark=at position 0.6 with {\arrow[agen,scale=1.25]{Classical TikZ Rightarrow}}}
    ] 
        \draw [white, postaction={decorate}] (120:0.685) to (0:0.685);
        \draw [white, postaction={decorate}] (0:0.685) to (240:0.685);
        \draw [white, postaction={decorate}] (240:0.685) to (120:0.685);
        \draw [thick, ->, >=stealth] (0:0) to (0:1);
        \draw [thick, <-, >=stealth] (0:0) to (120:1);
        \draw [thick, ->, >=stealth] (0:0) to (240:1);
        \fill [agen] (0:0.5) circle[radius=2.5pt];
        \fill [agen] (120:0.5) circle[radius=2.5pt];
        \fill [agen] (240:0.5) circle[radius=2.5pt];
        \draw [line width=2, agen, rounded corners=10pt] (0:0.685) to (120:0.685) to (240:0.685) --cycle;
        \node [circle, agen, fill=agen!20!white, inner sep=1.05pt, draw, thick, font=\small] at (120:0.51) {$\bar{\tta}$};
        \node [circle, agen, fill=agen!20!white, inner sep=1.5pt, draw, thick, font=\small] at (240:0.51) {$\tta$};
        \node [circle, agen, fill=agen!20!white, inner sep=1.5pt, draw, thick, font=\small] at (0:0.51) {$\tta$};
        \end{scope}
    \end{tikzpicture}
    }
    \, .
\end{equation}
When acting on pre-existing loop configurations, the operators~\eqref{eqn:generalized-Av} can freely deform the strings locally (provided that they don't intersect). Furthermore, acting with $\prod_{v \in S}\tilde{A}_v(g)$ for some region $S$ will nucleate a loop on the boundary $\partial S$ since each interior edge will be sent to $\ket{e} \to \ket*{geg^{-1}} = \ket{e}$.
Also, if there exist relations of the form $\tta\ttb=\ttc^{-1}$, then simple basepoints associated to this relation can be generated:
\begin{equation}
\label{eq:abc_net}
    \scalebox{0.75}{
    \begin{tikzpicture}[baseline={(0,0)}]
        \coordinate (R) at (1, 0) {}; 
        \coordinate (L1) at (120:0.66) {}; 
        \coordinate (L2) at (240:0.66) {}; 
        \coordinate (R1) at ($ (R)+(60:0.66) $) {}; 
        \coordinate (R2) at ($ (R)+(300:0.66) $) {}; 
        \coordinate (C1) at ($(L1)!0.5!(R1)$);
        \coordinate (C2) at ($(L2)!0.5!(R2)$);
        \begin{scope}[very thick,decoration={markings,mark=at position 0.6 with {\arrow[agen, scale=1.25]{Classical TikZ Rightarrow}}}] 
            \draw [white, postaction={decorate}] (120:0.685) to (240:0.66);
        \end{scope}
        \begin{scope}[very thick,decoration={markings,mark=at position 0.6 with {\arrow[cgen, scale=1.25]{Classical TikZ Rightarrow}}}] 
            \draw [white, postaction={decorate}] (R1) to (R2);
        \end{scope}
        \begin{scope}[very thick,decoration={markings,mark=at position 0.275 with {\arrow[bgen, scale=1.25]{Classical TikZ Rightarrow}},mark=at position 0.875 with {\arrow[bgen, scale=1.25]{Classical TikZ Rightarrow}}}] 
            \draw [white, postaction={decorate}] (C1) to (C2);
        \end{scope}
        \draw [thick, ->, >=stealth] (0:0) to (0:1);
        \draw [thick, <-, >=stealth] (0:0) to (120:1);
        \draw [thick, ->, >=stealth] (0:0) to (240:1);
        \draw [thick, ->, >=stealth] (R) to ($ (R)+(60:1) $);
        \draw [thick, ->, >=stealth] (R) to ($ (R)+(300:1) $);
        \fill [cgen] ($ (R)+(60:0.51) $) circle[radius=2.5pt];
        \fill [cgen] ($ (R)+(300:0.51) $) circle[radius=2.5pt];
        \draw [line width=2, agen, rounded corners=10pt] (C1) to (120:0.66) to (240:0.66) to (C2);
        \draw [line width=2, cgen, rounded corners=10pt] (C1) to ($ (R)+(60:0.66) $) to ($ (R)+(300:0.66) $) to (C2);
        \draw [line width=2, bgen, rounded corners=10pt] (C1) to (C2);
        \draw [very thick, black, fill=white] (C1) +(-2.5pt, -2.5pt) rectangle +(2.5pt, 2.5pt);
        \draw [very thick, black, fill=white] (C2) +(-2.5pt, -2.5pt) rectangle +(2.5pt, 2.5pt);
        % labels
        \node [circle, agen, fill=agen!20!white, inner sep=1.5pt, draw, thick, font=\small] at (120:0.51) {$\tta$};
        \node [circle, agen, fill=agen!20!white, inner sep=1.05pt, draw, thick, font=\small] at (240:0.51) {$\bar{\tta}$};
        \node [circle, cgen, fill=cgen!20!white, inner sep=1.5pt, draw, thick, font=\small] at ($ (R)+(60:0.51) $) {$\ttc$};
        \node [circle, cgen, fill=cgen!20!white, inner sep=1.5pt, draw, thick, font=\small] at ($ (R)+(300:0.51) $) {$\ttc$};
        \node [circle, bgen, fill=bgen!20!white, inner sep=1.pt, draw, thick, font=\small] at (0:0.5) {$\ttb$};
    \end{tikzpicture}
    }
    \, .
\end{equation}
Note that any non-self-retracing path through this simple net  reads one of the words $\tta\tta^{-1}$, $\ttc^{-1}\ttc$,  $\tta\ttb\ttc$ or $\ttc^{-1}\ttb^{-1}\tta^{-1}$, as required by flatness. 

More generally, the operators $\tilde{A}_v(g)$ preserve the conjugacy class of any closed loop. Any closed loop goes into and out of vertex $v$ the same number of times. If $v$ is not the starting point of the path, then the group element is preserved because $\tilde{A}_v(g)$ only inserts $g g^{-1}$ or $g^{-1} g$ pairs into the middle of the group element. If $v$ is the starting point of the loop (and therefore also the endpoint) then $\tilde{A}_v(g)$ sends the group element $\varphi(w_\gamma)$ to another member of its conjugacy class, either $g \varphi(w_\gamma)g^{-1}$ or $g^{-1} \varphi(w_\gamma) g$.

\begin{figure}
    \centering
    \includegraphics[width=\linewidth]{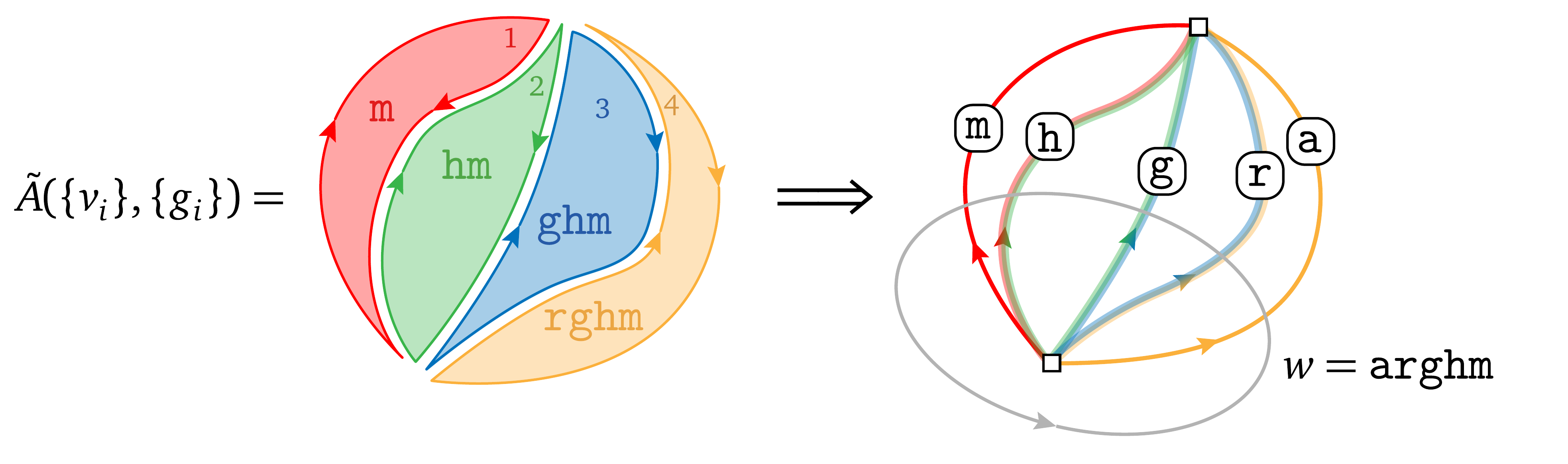}
    \caption{Procedure for creating an operator $\tilde{A}(\{ v_i \}, \{ g_i \})$ capable of nucleating a ``net'' from the vacuum. Illustrated is the operator associated with the relation $w = \tta \ttr \ttg \tth \ttm = \tte$. Each shaded region is associated with the application of an operator $A_{v}(g)$. The operators are applied sequentially from left to right, i.e., $A_{v_1}(\ttm)$ is applied first and $A_{v_4}(\ttr\ttg\tth\ttm)$ last. Any path that passes between the two basepoints reads either the word $\tta \ttr \ttg \tth \ttm$ or its inverse.}
    \label{fig:net_creation}
\end{figure}

However, to create basepoints associated to more complex generating relations, we need to consider a larger class of dynamics. For instance, the configuration shown in Fig.~\ref{fig:net_creation} generically cannot be nucleated from the vacuum using the $\tilde{A}_v(g)$ operators in Eq.~\eqref{eqn:generalized-Av}. To generate such configurations, introduce a variant of~\eqref{eqn:generalized-Av} absent the projection operators:
\begin{equation}
    A_v(g) = \prod_{e \in \partial^\dagger v} L^g(e, v)
    \, .
\end{equation}
We may then consider operators that generate an intermediate configuration outside of the onsite Hilbert space before being projected back in:
\begin{equation} \label{eqn:larger-Av}
    \tilde{A}(\{v_i\}, \{g_i\}) = \Pi \left[ \prod_{i=1}^n A_{v_i}(g_i) \right] \Pi
    \, ,
\end{equation}
where the collection of vertices $\{v_i\}$ define a subgraph consisting of a single connected component. Crucially, the generalized family \eqref{eqn:larger-Av} includes operators that \emph{do not factorize into a product of $\tilde{A}_v(g)$ operators}. For instance, consider an operator that nucleates a net associated to a word $w$ of length $n$ with $w \in K_{\tte, n}$, and $w=\tta_1 \cdots \tta_n$. Such a net can be created by sequentially applying $A_v(g)$ operators corresponding to group elements $\varphi(\tta_n), \varphi(\tta_{n-1}\tta_n), \dots, \varphi(\tta_2 \cdots \tta_{n-1}\tta_n)$. This operator will have a nontrivial projection under $\Pi$ if the vertices acted upon by the $A_v(g)$ are arranged as in Fig.~\ref{fig:net_creation}. Note that the operators~\eqref{eqn:larger-Av} would appear in perturbation theory if we were to take Kitaev's quantum double construction from Sec.~\ref{sub:quantum-double} and energetically penalize the generators that do not belong to our onsite Hilbert space.  If the interaction range of the operators in Eq.~\eqref{eqn:larger-Av} is truncated then additional fragmentation may occur, a phenomenon that we discuss in further detail in Sec.~\ref{sec:lattice}.

\subsection{Intrinsic Krylov sectors}\label{subsec:intrinsic_krylov}

In order to support nontrivial group elements on open boundary conditions, we consider lattices with smooth boundary conditions on the top and bottom boundaries and rough boundary conditions on left and right boundaries. Then, flat configurations can support strings that stretch from the top boundary to the bottom boundary and therefore nontrivial words on paths that stretch from the left boundary to the right boundary. Flatness (in the bulk and on the rough boundaries) implies that the word read on any such path will map to the same group element. Furthermore, the dynamics discussed in Sec.~\ref{sub:dynamics} will preserve this group element. It follows that intrinsic Krylov sectors are labeled by group elements. This label structure is the closest analog of open boundary conditions in 1D.

As in 1D, periodic boundary conditions allow the Krylov sectors corresponding to two group elements in the same conjugacy class to merge. 
Flat configurations consist of both contractible and non-contractible colored loops and nets on the dual lattice,\footnote{A non-contractible net can always be smoothly deformed into non-contractible loops using the operators $\tilde{A}(\{v_i\}, \{g_i\})$ defined in Eq.~\eqref{eqn:larger-Av}.} as shown in Fig.~\ref{fig:loops_2}. 
Contractible loops on the primary lattice must support trivial words, whereas noncontractible loops (i.e., the loops that are not the boundary of any finite region) may support words that map to nontrivial group elements $g\in G$.
Flatness implies that the words read on any two homotopically equivalent loops that share the same starting point will map to the same group element. Homotopic deformations that move the starting point of a loop may change the group element, but will not change the conjugacy class. Furthermore, the dynamics described in Sec.~\ref{sub:dynamics} will preserve this conjugacy class. It follows that intrinsic Krylov sectors are labeled by conjugacy classes in the presence of periodic boundary conditions.

For simplicity, we focus the rest of our discussion on periodic boundary conditions, although everything that follows is easily generalizable to open boundary conditions, or a mix of open and periodic boundary conditions.

\subsubsection{Compatibility of intrinsic Krylov sectors}
\label{sec:label-compatibility}
A 2D torus has two non-trivial cycles supporting non-contractible loops. Therefore, an intrinsic Krylov sector must be labeled by two conjugacy classes, $K_{([g_x], [g_y]), (L_x, L_y)}$, where $L_x \times L_y$ is the size of the lattice. However, not all conjugacy classes are compatible with each other. In particular, the compatibility condition can be written as
\eq{
\comm*{g_x}{g_y} \equiv g_x g_y g_x^{-1} g_y^{-1} = \tte.
\label{eq:compatibility}
}
This can be shown by considering a contractible ``rectangular'' loop of size $L_x \times L_y$. The word along it would be $w_x w_y w_x^{-1} w_y^{-1}$, with $\varphi(w_x) = g_x$, $\varphi(w_y) = g_y$. But, due to the flatness condition, $\varphi(w_x w_y w_x^{-1} w_y^{-1}) = \tte$, which gives Eq.~\eqref{eq:compatibility}. If this condition is satisfied for any two representative members of the conjugacy classes, then it is also satisfied over the entire conjugacy classes. In Sec.~\ref{sec:arbitrary-graphs}, we construct constraints between the labels of intrinsic Krylov sector, such as Eq.~\eqref{eq:compatibility}, in all generality.

\subsubsection{Higher-form symmetries} 

As in 1D, it is possible to derive global symmetries for the 2D group models directly from the input group $G$. These global symmetries are capable of distinguishing \emph{some} Krylov sectors from others, but not all if the input group $G$ is non-Abelian. Say we have $G = \braket*{ h_1, \dots, h_m \,}{\, R }$. As in Sec.~\ref{sub:conservation}, we define the Abelianization $G^{\text{ab}}$ of $G$,
\begin{gather}
G^\text{ab} = \braket*{ h_1, \dots, h_m \,}{\, R \cup C_h }, \\ C_h = \left\{ [h_j, h_k] = \tte \right\}_{j,k=1}^m , \notag
\end{gather}
where elements of $C_h$ permit commutation between all characters. This definition is equivalent to the one $G^\text{ab} = G/[G, G]$ introduced in Sec.~\ref{sub:conservation}. Via transformations of the presentation (known as \emph{Tietze transformations}), we can construct an equivalent presentation,
\begin{subequations}
\begin{align}
    G^\text{ab} &= \braket*{ s_1, \dots, s_n, h_1, \dots, h_m \,}{\, C \cup P \cup E } \\
    &= \zz_{p_1} \times \dots \times \zz_{p_{n-l}} \times \zz^l,
\end{align}
\end{subequations}
where $s_j$ are the ``essential'' generators, $C$ contains commutators between all characters, $P$ consists of $s_1^{p_1} = \dots = s_{n-l}^{p_{n-l}} = \tte$, and $E$ contains relations of the form
\begin{equation}
h_j = s_1^{d_{j1}} s_2^{d_{j2}} \cdots s_n^{d_{jn}},\quad j = 1,\dots, m, 
\end{equation}
which allow us to write the original generators in terms of the essential generators. See Appendices~\ref{app:abelian_charges} and~\ref{app:general_charges} for details. 

Given any closed oriented loop $\gamma = \{e_1,\dots, e_\lambda\}$
on the (primary) lattice, we can construct the operators 
\begin{equation}
n_k = \sum_{e \in \gamma} \sum_{g \in \mathcal{A}} d_{g,k} T^{g}(e,\gamma), \quad k = 1, \dots, n,
\end{equation}
where $T^g(e, \gamma)$ evaluates to $T^g_+$ ($T^g_{-}$)~\eqref{eqn:T_op} if the edge $e$ is oriented with (against) the path $\gamma$. The value of this operator is conserved mod $p_k$ for $k\le n-l$ and conserved exactly otherwise, for any $\gamma$. The weights are 
\begin{equation}
d_{g,k} = \begin{cases}
d_{jk},  & g = h_j \\
-d_{jk}, & g = h_j^{-1} \\
0,       & g = \tte.
\end{cases}
\end{equation}
The flatness condition~\eqref{eqn:flatness} implies that the charge around any plaquette must vanish, and that deforming the loop around a plaquette does not change the charge. Together, these qualities imply that the charge around any contractible loop must vanish.

As in the case of the Krylov sector labels, this leaves open the possibility of nontrivial charge measured on noncontractible loops. The flatness condition implies that the charge measured on two loops that are homotopically deformable into each other must be the same. All taken together, these statements imply that the 2D group model possesses a higher-form symmetry~\cite{Nussinov2007, Gaiotto2015, McGreevy2022}. Since the symmetry operators live on $(d-1)$-dimensional sublattices, the symmetry is a 1-form symmetry. In particular, it is a $\zz_{p_1}\times \cdots \times \zz_{p_{n-l}} \times \U{1}^l$ symmetry. Furthermore, higher-form symmetries must be Abelian, so the 2D group models are able to display any possible 1-form symmetry.

\begin{figure}[t]
\centering
  \includegraphics[scale=0.25]{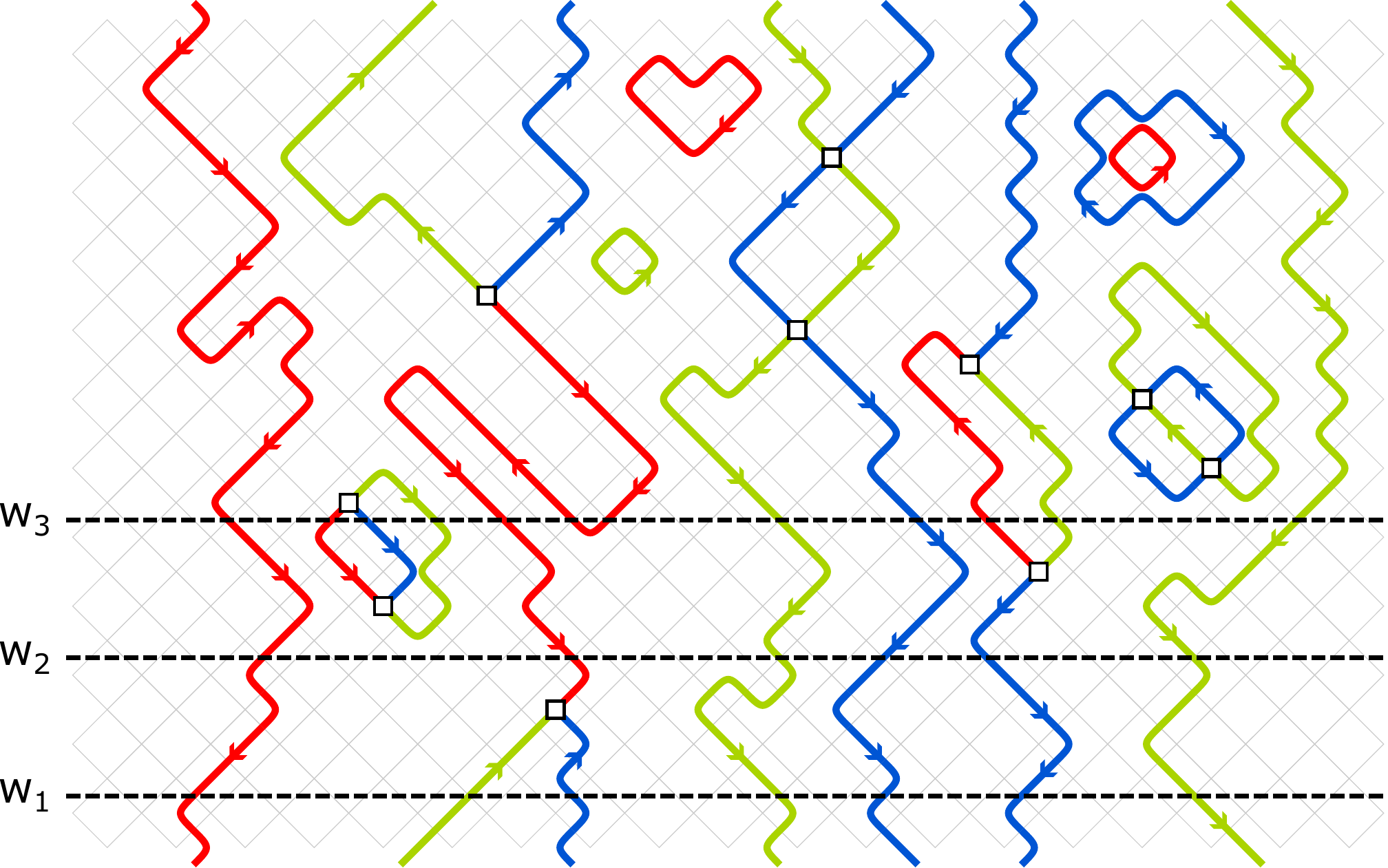}
  \caption{A product state obeying the flatness condition on the square lattice with PBC (imposed at 45\textdegree\! relative to the lattice directions). Similarly to Fig.~\ref{fig:loops}, we consider generators $\tta, \ttb, \ttc$ and relations $\tta\ttb\ttc = \tte$, $\ttb\ttc = \ttc\ttb$. Here, we omit explicitly writing the generators, since they are uniquely determined by the orientations of the colored strings and the arbitrarily fixed orientations of the primary lattice edges (hence, there is a gauge freedom --- reversing an orientation of a primary lattice edge inverts the generator on that edge). The product state consists of both contractible colored loops and nets, as well as non-contractible ones. The latter determine the label of the intrinsic Krylov sector, which can be read from a word along any of the non-contractible loops on the primary lattice. For example, in this case, going from left to right along the dashed lines, $\varphi(w_1) = \varphi(\tta\ttc^{-1}\ttb^{-1}\ttc\ttb^2\ttc) = \varphi(w_2) = \varphi(\tta^2\ttc\ttb^2\ttc) = \varphi(w_3) = \varphi(\tta^2\ttb\ttc\tta\tta^{-1}\tta\ttc\ttb\tta^{-1}\ttc^{-1}\ttc) = \tte$. Any vertical non-contractible loop on the primary lattice would also spell a word $w$ with $\varphi(w) = \tte$. Therefore, this particular product state belongs to the $K_{(\tte, \tte), (L_x, L_y)}$ sector (i.e., all non-contractible colored loops can be removed using local group dynamics).}
  \label{fig:loops_2}
\end{figure}

\subsection{2D models as ``time evolving'' 1D models} \label{sub:timeslice}
Due to the continuous nature of strings, one can view a 2D product state as a classical time evolution of 1D product states.
Consider a set of $T$ non-contractible and non-intersecting primary loops of equal length, $\{\gamma_t\}_{t=1}^T$, (e.g., an example is shown as horizontal black dashed lines in Fig.~\ref{fig:loops_2}). Words supported on the loops from this set, $\{w_t\}_{t=1}^{T}$, can be viewed as instantaneous product states of a 1D group model on a chain of length $\abs{w_1}=\hdots=\abs{w_T}$ at different times during a classical Markovian time evolution. For example, in Fig.~\ref{fig:loops_2}, if we assume that the ``time'' is running from bottom to top of the picture, then words $w_1, w_2, w_3$ would correspond to classical states at times $t_1 < t_2 < t_3$, respectively.

Since all these loops are homotopically equivalent, the associated words multiply to group elements in the same conjugacy class of $G$. Product states on two consecutive loops, $\gamma_t$ and $\gamma_{t+1}$, are different by a number of local updates, determined by the behavior of the colored strings in the region between $\gamma_t$ and $\gamma_{t+1}$. For example, if this region contains one of the two basepoints of a net associated with relator $r\in R$, then an update associated with $r$ has been performed between $w_t$ and $w_{t+1}$. At the same time, free reduction and commutation with the $\tte$ character is performed by kinks in the colored loops and contractible colored loops crossing $\gamma_t$ or $\gamma_{t+1}$.

\subsubsection{Fragile fragmentation}
Consequently, the time evolution picture provides a natural proof of the following fact: 1D product states on any two homotopically equivalent non-contractible loops of equal length, $\gamma_t$ and $\gamma_{t'}$, must belong to the same fragile Krylov sector of the corresponding 1D group model. Indeed, since $\ket{w_{t'}}$ is the time-evolved $\ket{w_t}$, both states must belong to the same dynamical sector of the 1D group model. This observation allows us to define fragile Krylov sectors of the 2D group model --- by associating each of them with a specific fragile Krylov sector of the respective 1D group model defined on a chain of length $\abs{w_t}$.

\subsubsection{Equal-time correlators}
In addition, the time evolution picture makes it apparent that the Dehn function of the associated 1D group model affects equal-time correlators of the 2D model, $\expval*{P_t P_{t'}}$, where $P_t = \ket{w_t} \! \bra{w_t}_{\gamma_t}$ is a projector onto the product state $\ket{w_t}$ on the loop $\gamma_t$. In particular, for a set of densely packed loops $\{\gamma_t\}$, the correlator vanishes, $\expval*{P_t P_{t'}} = 0$, if $\dehn(w_t, w_{t'}) > L_x \abs*{t' - t}$.
Indeed, the number of accessible ``time steps'' along, say, $y$-direction of the lattice is only $\sim L_y$. Therefore, if $L_x \sim L_y \sim L$, then a Dehn function that scales faster than linear with $L$ would prohibit large numbers of 1D product states on $\gamma_t$'s from simultaneously coexisting with each other, even though these states belong to the same fragile (and intrinsic) Krylov sectors of the 1D group model. 

\subsection{Topological stability of fragmentation}
Instead of implementing the flatness condition as a hard constraint, which leads to a Hilbert space without tensor product structure, we could consider a Hamiltonian that enforces flatness energetically:
\eq{
H = \Delta \sum_p 
(1-B_p)
- h \sum_i \sum_{s=1}^{\abs{\mathcal{A}}} \sum_{k=1}^{\abs{\mathcal{A}}} t_i^{sk} \ket{s}\bra{k}_i,
\label{eqn:perturbation}
} 
where $\Delta > 0$.
The first summation goes over all plaquettes $p$ and assigns an energy cost $\Delta$ to every violation of the flatness condition. The operator $1-B_p$~\eqref{eqn:flatness-operator} projects onto local configurations violating the flatness condition.
Thus, violations of the flatness condition are permitted, although they are energetically costly.
The second, non-diagonal, term allows for generic flips between any two basis states ($s, k \in \mathcal{A} \equiv \mathcal{S} \cup \mathcal{S}^{-1} \cup \{\tte\}$) on lattice edge $i$, with arbitrary complex matrix coefficients $t_i^{sk}$.

Assuming $\Delta \gg h$ and treating the non-diagonal term within perturbation theory, the effective low-energy Hamiltonian will consist exactly of all possible loop and net nucleation operators. Although different loop and net nucleation operators will appear at different orders of perturbation theory depending on the support of these operators, any such order will be $\mo(1)$ with respect to the lattice size.
Any such operator will necessarily preserve the intrinsic Krylov sector structure, since a non-contractible loop that defines the intrinsic Krylov sector can always be deformed to avoid the operator's support (this argument is analogous to the cleaning lemma from the theory of quantum codes \cite{bravyi2009no, kalachev2022linear}).
Conversely, operators that change the intrinsic Krylov sector require support on $\sim L$ sites. 
The intrinsic Krylov structure is therefore \emph{perturbatively} stable, i.e., to all finite orders in perturbation theory when the lattice is thermodynamically large. For further details, see Ref.~\cite{stahl2023topologically}.

The first few orders in perturbation theory result in the Hamiltonian
\begin{equation}\label{eqn:BplusA}
H = -\Delta \sum_p B_p - \sum_{v,g} \lambda_v^g \tilde{A}_v(g) - \cdots ,
\end{equation} 
where $\lambda_v^g$ is a function of $h$ and $t_i^{sk}$ depending on the order of perturbation theory, and the $\cdots$ represents loop and net nucleation terms found at higher orders of perturbation theory. 

This construction preserves the intrinsic Krylov structure at every order in perturbation theory up to order $L$, suggesting that the sectors might have a chance of being absolutely stable (as opposed to just perturbatively stable). This expectation proves correct in the prethermal regime: Under dynamics generated by the full Hamiltonian~\eqref{eqn:perturbation}, local correlation functions are well approximated by the optimally truncated perturbative Hamiltonian up to times exponentially long in $\Delta/h$ \cite{ADHH, ChaoLucas}. Furthermore, states from different Krylov sectors remain dynamically disconnected up to the same timescale.

The Krylov sectors clearly must break down in some cases. For example, consider $G=\zz_2$, so that we are constructing a $\zz_2$ gauge theory. Flatness in our construction is equivalent to a flux-free condition in the gauge theory, so $B_p$ penalizes magnetic fluxes and the $h$ term in \eqref{eqn:perturbation} gives them dynamics. The allowed states are free of magnetic fluxes but do have electric charges present. The Krylov sectors are labeled by the magnetic quantum numbers of the gauge theory. Unfortunately, generic states from any Krylov sector have finite energy density with respect to the full gauge theory Hamiltonian~\eqref{eqn:BplusA} because of the electric charges, and we know that the magnetic quantum numbers cease to be good at nonzero energy density~\cite{FradkinBook}. The problem is that the system is able to transfer energy from the electric sector to the magnetic sector, creating fluxes that are able to traverse the system and change the quantum numbers. Since the problem here is proliferation of constraint violations, it is worth considering these violations in more detail.

\subsection{Local excitations}
\label{sec:excitations}
Here, we discuss elementary excitations that take the system out of the constrained subspace. According to Eq.~\eqref{eqn:perturbation}, such an excitation arises from a violation of the flatness condition on a plaquette --- call it $p_0$ --- and costs energy $\Delta$. Now the word around $p_0$ can multiply to a nonidentity group element, $g_0 \in G$.
Changing the starting point of the word may change the group element but it will not change the conjugacy class $[g_0]$, so we refer to $[g_0]$ as the \emph{charge} of the excitation. 

\begin{figure}[t]
\centering
  \includegraphics[scale=0.5]{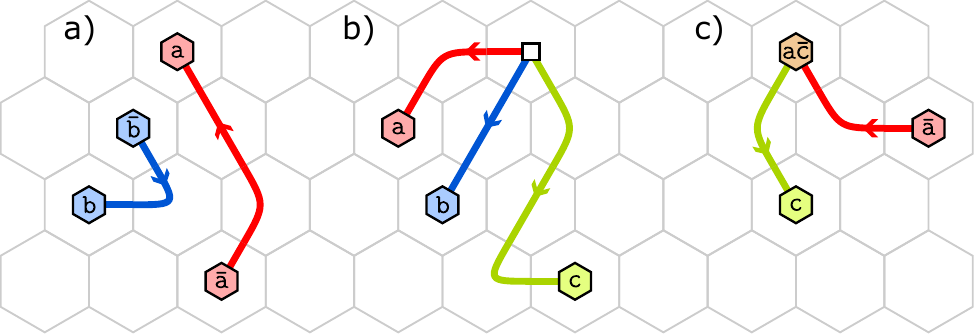}
  \caption{Local excitations that take the system out of the low-energy subspace, in which the flatness condition is satisfied on every plaquette. The charge of an excitation is the conjugacy class $[g]$ of the group element $g = \varphi(w) \in G$ that corresponds to the word $w$ read by going clockwise around the plaquette. Excitations can appear either (a) in pairs with opposite charges, or (b) in multiplets corresponding to a relator $r \in R$ (in the current example, the relation is $\tta\ttb\ttc = \tte$). A single plaquette can act as a source/sink of several colored strings (c) --- the charge is then obtained by the product of the corresponding generators.} 
  \label{fig:anyons}
\end{figure}

Such excitations, however, cannot appear in isolation. Note that the plaquette $p_0$ acts as a source/sink of colored strings, and as a result, the elementary excitations must come either in pairs (with the opposite charges, $[g_0]$ and $[g_0]^{-1}$), as shown in Fig.~\ref{fig:anyons}(a), or in multiplets corresponding to a relator $r \in R$ (call them $r$-multiplets), such that their charges multiply to $r$, as shown in Fig.~\ref{fig:anyons}(b). A single plaquette can act as a source/sink of multiple colored strings, as shown in Fig.~\ref{fig:anyons}(c) --- the charge is then determined by the corresponding product of generators.

By sequentially applying pair-creation operators (as in Fig.~\ref{fig:anyons}), one can move charges across the lattice.\footnote{Note, however, that generically colored strings cannot intersect and therefore the charge propagation is not completely free.} In turn, by applying operators that create $r$-multiplets, one can split a charge into multiple other charges or merge several charges, according to the group-$G$ multiplication rules. The conjugacy class $[\varphi(w)]$ for any contractible loop $\gamma$ supporting the word $w$ is determined solely by the charges (and their relative position) inside of the region bounded by $\gamma$. If the charge creation operator is fully contained within this region, then $\varphi(w) = \tte$. 

Creation and subsequent annihilation of several excitations is equivalent to applying a closed loop-creation or a net-creation operator. If such an operator is local (such as the ones shown in Fig.~\ref{fig:loops}), then the system remains in the same (topological) intrinsic and fragile Krylov sectors. If, on the other hand, the excitations proliferate and are annihilated after winding around the torus, the topological Krylov sector is changed. In this sense, the excitations are similar to anyons in topologically ordered systems, which change the ground state of the system when created, brought around a non-contractible loop, and annihilated.

This similarity prompts us to introduce error correction in group models. Assume dynamics induced by the Hamiltonian $H$ from Eq.~\eqref{eqn:BplusA} and suppose that the system is initialized in a state $\ket{\psi}$, that is not necessarily an eigenstate of $H$, but is such that $B_p \ket{\psi} = \ket{\psi}$ for every plaquette $p$. If the system is isolated, the state will time-evolve as $\ket{\psi(t)} = \exp(-iHt) \ket{\psi}$. Now, introduce noise that applies local operators that are non-diagonal in the computational basis (for simplicity, assume that the operators are on-site ``transverse field'' operators, which flip one generator to another). Assume that between times $t$ and $t+\dd{t}$, such an operator is applied to each site with probability $P$. This will lead to creation of excitations. One can then measure local operators $B_p$ on every plaquette to detect the excitations. Next, one can run a decoder\footnote{The existence of an efficient decoder for any given group $G$ is an open question. In fact, it is likely that for certain groups, such a decoder cannot exist, due to the hardness of the word problem.} that determines an optimal way to locally annihilate the excitations and apply the corresponding operators that perform this task. In this way, one can correct for flip-errors in a time-evolving system, provided that the probability $P$ is lower than some threshold value $P_\mathrm{cr}$. However, since the group models generically do not have a structure of a stabilizer or a subsystem code, there is no obvious way to correct for phase errors, i.e., ones induced by operators diagonal in the computational basis. Hence, the proposed error correction is classical in nature.

\subsection{Reproducing Kitaev's Quantum Double}
\label{sub:quantum-double}
We have already named our operators to suggest a connection to the quantum double models of Ref.~\cite{Kitaev2003}. In fact, our construction reduces to the construction in that paper when $G$ is a finite group and the alphabet is the ``faithful'' presentation $\mathcal{A}_G$, which has one character for every element of $G$. Using this larger alphabet means that the projectors introduced below Eq.~\eqref{eqn:L_op} are always satisfied and $A_v(g) = \tilde{A}_v(g)$.

In that case, the Hamiltonian in Eq.~\eqref{eqn:BplusA} becomes
\begin{align}
H_\text{QD} = -\sum_p B_p - \sum_{v,g} A_v(g), \label{eqn:KitaevQD}
\end{align}
which is equivalent to Eq.~(13) of Ref.~\cite{Kitaev2003} up to constant shifts and rescalings. 
It is known that the quantum double has a finite number of ground states corresponding to the finite number of Krylov sectors, and that each ground state is absolutely stable to arbitrary local perturbations at zero temperature.

This connection means that for the setting in which our model reduces to the quantum double construction, namely a finite group $G$ and the faithful presentation $\mathcal{A}_G$, the Krylov sectors are absolutely stable when considering zero-temperature dynamics. The constraint violations of Sec.~\ref{sec:excitations} do not proliferate, and no error correction is necessary. Whether the same holds with a non-faithful presentation or with infinite groups remains a question for future research.

\section{Lattice-induced phenomena} \label{sec:lattice}
The product states and dynamics of the previous section can be conveniently understood in a ``continuous'' picture of 2D group models, as introduced in Sec. VII of Ref.~\cite{balasubramanian2023glassy}. In this picture, the lattice is (nearly) disregarded and the allowed states consist of continuous colored strings that can be freely deformed. The only effect of the lattice is through fragile fragmentation, which depends only on the lattice size $L$.  Any two states with the same intrinsic and fragile Krylov labels are connected by loop-nucleation operators and sufficiently large net-nucleation operators.
However, such a continuous description misses some effects relevant to the lattice regularization of the model. 

In this section, we elaborate on how different lattices might lead to further fragmentation, and the support that net-nucleation operators must have in order to be fully ergodic within Krylov sectors. All of the effects in this section depend on microscopic details of the lattice. As such, while practically all effects discussed in Sec.~\ref{sec:2d} do not require translation invariance, effects discussed in this section will locally depend on each elementary plaquette of the considered graph. 

\subsection{Compatibility of the group presentation and the lattice} \label{sec:group-lattice-compatibility}

The first purely lattice effect arises from the fact that every plaquette has a finite number of edges. As a consequence, certain lattices might be incompatible with specific group presentations.
An incompatibility arises when a plaquette has fewer edges than the length of the longest relator in the presentation. In this case, such a plaquette cannot host a basepoint of the net associated with this relator. For example, if the presentation of a group $G$ contains a relator $r$ of length at least 4, then a triangular lattice would not be able to support the corresponding net. The allowed product states (and the allowed dynamics) instead represent a bigger group, $G'$.\footnote{$G$ is the quotient of $G'$ by $\Ker(f)$, where $f$ is a homomorphism from $G'$ to $G$ that maps the group element $\varphi(r) \in G'$ (and its conjugates) to the identity $\tte \in G$.} 

We now discuss an explicit construction that illustrates this incompatibility. 
Consider a two-dimensional triangular lattice with periodic boundary conditions imposed.
We will attempt to look at group dynamics that derive from the group $G = \mathbb{Z}_2^3$ with the presentation 
\begin{equation}
    G = \braket{ \ttr , \ttg , \ttb  \, }{ \, \ttr^2 = \ttg^2 = \ttb^2 = \tte, \ttr\ttg=\ttg\ttr, \ttr\ttb=\ttb\ttr, \ttg\ttb=\ttb\ttg }
    \, .
    \label{eqn:triangular-Z2}
\end{equation}
Since the group is Abelian, its intrinsic Krylov sectors can be labeled by discrete 1-form symmetry operators (Appendix~\ref{app:abelian_charges}). However, note that the relators that impose commutation of the group elements require at least four edges around a plaquette to implement. Consequently, the triangular plaquettes are unable to implement commutation of the group elements. As mentioned in Sec.~\ref{subsec:presentation} and discussed in further detail in Appendix~\ref{app:pair-flip}, this implies that the allowed states (dynamics) instead derive from (respect) the enlarged group $G'$ with the appropriate relations removed
\begin{equation}
    G' = \braket{ \ttr , \ttg , \ttb  \, }{ \, \ttr^2 = \ttg^2 = \ttb^2 = \tte } = \mathbb{Z}_2^{*3}
    \, .
\end{equation}
That is, the triangular lattice is fundamentally incompatible with the desired presentation~\eqref{eqn:triangular-Z2}, and attempting to implement symmetry-respecting dynamics ends up giving rise to an exponential number of Krylov sectors labeled by $G'$. This therefore represents an example of \emph{lattice-induced fragmentation}.
We note that this effect is also responsible for the fragmentation observed in Ref.~\cite{stahl2023topologically}, in which imposing a 1-form $\U1^n$ symmetry (which in our language corresponds to a $\mathbb{Z}^{n}$ group) gave rise to fragmentation on the square lattice.

This result can also be understood at the level of the graphical depictions we have been utilizing.
Since $\ttr$, $\ttg$, and $\ttb$ are their own inverses, we may neglect the orientation of edges and strings. Allowed configurations around plaquettes are of the form
\begin{equation}
    \begin{tikzpicture}[x=2.4ex, y=2.4ex, baseline={(0, 0)}]
        \draw [thick] (0, 0) to (2, 0) to (1, {2*sin(60)}) --cycle;
    \end{tikzpicture}\,,
    \quad
    \begin{tikzpicture}[x=2.4ex, y=2.4ex, baseline={(0, 0)}]
        \draw [thick] (0, 0) to (2, 0) to (1, {2*sin(60)}) --cycle;
        \draw [ultra thick, red, rounded corners=4pt] (2, {1/cos(30)}) to (1, {tan(30)}) to (1, {-tan(30)});
    \end{tikzpicture}\,,
    \quad
    \begin{tikzpicture}[x=2.4ex, y=2.4ex, baseline={(0, 0)}]
        \draw [thick] (0, 0) to (2, 0) to (1, {2*sin(60)}) --cycle;
        \draw [ultra thick, red, rounded corners=4pt] (0, {1/cos(30)}) to (1, {tan(30)}) to (2, {1/cos(30)});
    \end{tikzpicture}\,,
    \quad
    \begin{tikzpicture}[x=2.4ex, y=2.4ex, baseline={(0, 0)}]
        \draw [thick] (0, 0) to (2, 0) to (1, {2*sin(60)}) --cycle;
        \draw [ultra thick, red, rounded corners=4pt] (0, {1/cos(30)}) to (1, {tan(30)}) to (1, {-tan(30)});
    \end{tikzpicture}\,,
\end{equation}
with analogous diagrams permitted for the other two colors.
The graphical notation makes it clear that loops are forbidden from intersecting on faces, which prevents loops from passing through one another and leads to additional conserved quantities associated to the ordering of the loops.
If the model is defined on a torus, the labels in the horizontal and vertical directions must obey certain compatibility relations, as discussed in Sec.~\ref{sec:label-compatibility}. 

Although we have just shown that certain group \emph{presentations} may be incompatible with certain lattices, any \emph{group} can be compatible with any lattice by choosing the right presentation. Every face in a lattice has at least three edges around it, and every group admits a presentation with relations of length at most three~\cite{balasubramanian2023glassy}. Thus, this minimal presentation is compatible with any lattice we might want to choose.

\subsection{Lattice defects and absence of translation invariance}

Whether a specific plaquette $p$ can host a net associated with a relator $r \in R$ is determined solely based on how $\abs{r}$ compares with the number of edges around $p$ --- no information about other plaquettes is needed. Therefore, for lattices consisting of several types of plaquettes, some plaquettes might be able to host certain nets, while other plaquettes might not.

In addition, lattice defects that alter the number of edges of a plaquette (be they topological defects or not), might allow or prohibit certain nets originating on them. Again, consider the example of the triangular lattice and attempt to implement group-$G$ dynamics with $G$ given in Eq.~\eqref{eqn:triangular-Z2}. While crossing of the differently colored loops is prohibited on the translationally invariant lattice, removing a lattice link creates a plaquette with four edges that can support such a crossing. Nets that are allowed to originate only on lattice defects can be protected from annihilation if the lattice defects serving as the basepoints are separated by the distance larger than the interaction range of the considered dynamics. In this case, additional fragmentation appears with sectors labeled by the local presence or absence of these nets.

\subsection{Finite interaction range} \label{sub:range}

Interaction range plays a crucial role on the lattice, since implementing the group dynamics might require operators with support on a significant number of edges. If the group dynamics arises in a perturbative regime, such operators will be of a high order in perturbation theory. If the flatness condition is strictly enforced, then limiting interaction range might rule out all operators associated with a specific relator $r$, and the dynamics would only reflect a bigger group, $G'$, as above.

In order to fully implement the group dynamics we must be able to at minimum nucleate and move loops and nets. The loop-nucleation operators $\tilde{A}_v(g)$, $g \in \varphi(\mathcal{A})$ are always sufficient to nucleate and move loops. In order to nucleate and move nets that correspond to relators of length at most 4, the operators $\tilde{A}_v(g)$, $g \in \varphi(\mathcal{A} \times \mathcal{A})$ are sufficient. To see this, consider the series of configurations
\begin{equation}
\label{eq:4_letter_net_nucleation}
\vcenter{\hbox{\includegraphics[scale = .5]{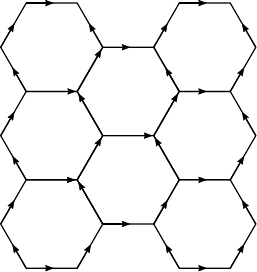}}} \longleftrightarrow \vcenter{\hbox{\includegraphics[scale = .5]{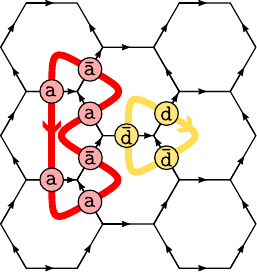}}} \longleftrightarrow \vcenter{\hbox{\includegraphics[scale = .5]{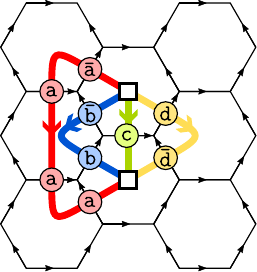}}},
\end{equation}
for a group with the relation $\tta \ttb \ttc \ttd = \tte$. The second configuration can be reached from the first by applying two $\tilde{A}_v(\tta^{-1})$ operators and one $\tilde{A}_v(\ttd)$ operator. To get from the second to the third, apply an $\tilde{A}_v(\ttb^{-1} \tta^{-1} )$ operator to the central vertex. This operator replaces $\tta$ with $\ttb^{-1} \tta^{-1} \tta = \ttb^{-1}$, replaces $\tta^{-1}$ with $\tta^{-1} \tta \ttb = \ttb$ (due to the inward pointing arrow), and also replaces $\ttd^{-1}$ with $\ttb^{-1} \tta^{-1} \ttd^{-1} = \ttc$ (due to the relation).

For longer relations, it is instead necessary to work with the generalized $\tilde{A}(\{v_i\}, \{g_i\})$ operators from~\eqref{eqn:larger-Av}. For example, consider the configurations
\begin{equation}
\label{eq:5_letter_net_nucleation}
\vcenter{\hbox{\includegraphics[scale = .5]{pics/abcd1.pdf}}} \longleftrightarrow \vcenter{\hbox{\includegraphics[scale = .5]{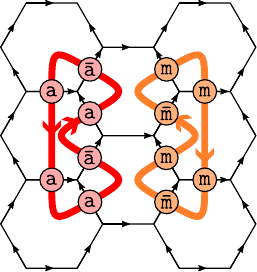}}}\longleftrightarrow \vcenter{\hbox{\includegraphics[scale = .5]{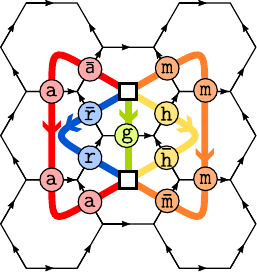}}},
\end{equation}
with the relation $\tta\ttr \ttg\tth \ttm = \tte$. It is once again possible to go between the first two relations using $A_v(g)$ operators, but not between the second and third. Instead, it is necessary to act with an $\tilde{A}(\{v_i\}, \{g_i\})$ operator on the central two vertices, which affects the central five edges.
Among all the necessary $\tilde{A}(\{v_i\}, \{g_i\})$ for a given group, the one with the maximal support is determined by $l_R$ introduced in Eq.~\eqref{eq:int_range}. This support is equal to the number of edges that must be crossed by $(l_R-2)$ strings connecting two basepoints, and the necessary $(l_R-2)$ strings might not be contiguous as they are in the above example.
We refer to the dynamics that consists of all $\tilde{A}_v(g)$ operators and the $\tilde{A}(\{v_i\}, \{g_i\})$ required by the relations as the ``minimal group-$G$ dynamics.''

\subsection{Pinning of basepoints}

It is possible that the minimal group-$G$ dynamics are not able to implement the ``full'' group-$G$ dynamics, in the sense that the interaction range is not large enough to support operators that connect all states within the same intrinsic or fragile Krylov sector.
Consider a plaquette with $\ell$ edges that hosts a word $w$, with $\varphi(w) = \tte$ and expansion length $\el(w) > \ell$. Then, the size of the plaquette is not enough to transform $\ket{w}$ into $\ket*{\tte^\ell}$ by performing local changes on parts of the plaquette. This is a local analog of fragile fragmentation. The only way to switch between $\ket{w}$ and $\ket*{\tte^\ell}$ is to apply an operator that acts simultaneously on the whole plaquette, nucleating a net associated with the word $w$. However, the minimal group-$G$ dynamics might not include such an operator. Without the operator in question, the basepoints of the net associated with $w$ will remain pinned to their original plaquettes during the dynamics. This creates additional Krylov sectors labeled by the positions of such basepoints.

\begin{figure}[t]
\centering
  \includegraphics[scale=0.55]{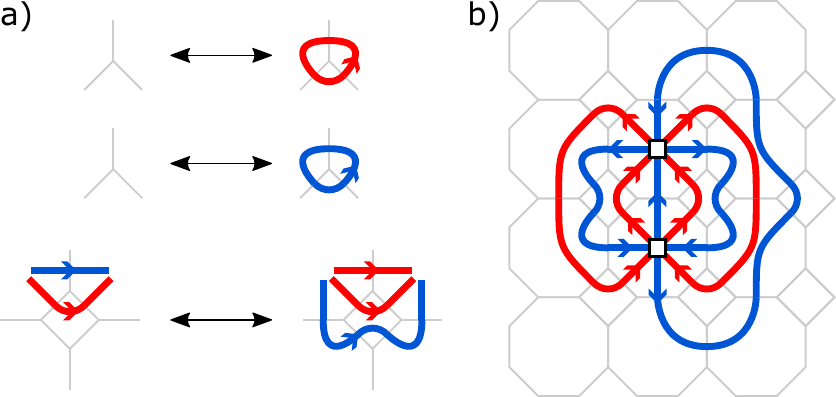}
  \caption{(a) Operators sufficient to implement $\bs$ group dynamics on the square-octagon lattice. The top two operators are associated with free reduction, i.e., $\tte = \tta\tta^{-1}$, $\tte = \ttb\ttb^{-1}$. The bottom operator is associated with the relation $\tta\ttb = \ttb\tta\tta$. (b) Minimal-support net associated with the word $w = \tta^{-1} \ttb^{-1} \tta^{-1} \ttb \tta \ttb^{-1} \tta \ttb$, with $\varphi(w) = \tte$. Such a net (or any net of this type) cannot be created or annihilated under the dynamics from (a) and therefore the basepoints are pinned in space.}
  \label{fig:bs_loops}
\end{figure}

To be more concrete, consider the following example: take the Baumslag-Solitar group, $G = \bs = \braket{ \tta, \ttb \,}{\, \tta\ttb = \ttb \tta^2 }$, on the square-octagon lattice. The minimal group-$\bs$ dynamics consists of just three operators shown in Fig.~\ref{fig:bs_loops}(a) (along with their Hermitian conjugates and operators related via lattice symmetries): two operators associated with the free reduction and a single operator associated with the only generating relation. Now, consider a word $w = \tta^{-1} \ttb^{-1} \tta^{-1} \ttb \tta \ttb^{-1} \tta \ttb$. This word multiplies to identity, $\varphi(w) = \tte$, but the expansion length is larger than the length of the word, $\el(w) > \abs{w}$ (see Ref.~\cite{balasubramanian2023glassy} for a thorough analysis of the $\bs$ group dynamics in 1D). Since $\abs{w} = 8$, we can write this word around an octagon. A minimal associated net is depicted in Fig.~\ref{fig:bs_loops}(b). However, it is not possible to perform a series of local updates (using operators from Fig.~\ref{fig:bs_loops}(a)) to transform $w$ into $\tte^8$ on the octagon, since $\el(w) > 8$. The basepoints of the net therefore remain pinned to particular octagons. The only way to unpin the basepoints is to include operators that nucleate nets from Fig.~\ref{fig:bs_loops}(b) into the dynamics. Such an operator must act simultaneously on 31 spins!

\section{Three-dimensional lattices}
\label{sec: 3d}

The construction presented in Sec.~\ref{sec:2d} can also be generalized to three spatial dimensions. The extra spatial dimension affords additional robustness: As we show below, charged operators that modify the model's robust Krylov sectors now act on $\sim L^2$ spins, and must surmount a large energy barrier, in contrast with the 2D models of Sec.~\ref{sec:2d}. For simplicity, the discussion in this section is presented for cubic lattices, but it may be generalized {\it mutatis mutandis} to arbitrary crystalline lattices in three dimensions. 

\subsection{Flatness and allowed product states: membranes and sponges}

For simplicity, consider a 3D cubic lattice with degrees of freedom on edges. Around every face of the cubic lattice we impose the flatness condition~\eqref{eqn:flatness}.
By composing adjacent faces, we deduce that $\varphi(w_{\partial S})=\tte$ for any contractible surface $S$ composed of faces on the primary cubic lattice. Note that the allowed configurations around faces are identical to 2D, but the constraints on how these configurations can be placed next to one another are modified, since each edge is now adjacent to \emph{four} faces. As in 2D, the word read around every elementary face can be connected to the identity word either via free reduction or via the application of a relation, or some combination thereof.

\begin{figure}[t]
    \centering
    \includegraphics[scale=0.8]{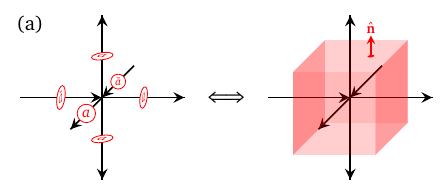}\\
    \includegraphics[scale=0.8]{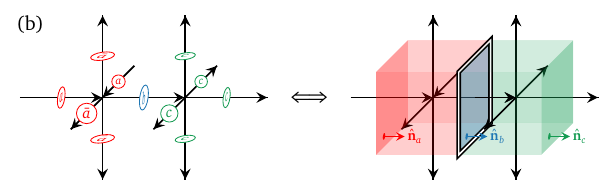}\\
    \caption{Examples of product states on the edges of a three-dimensional cubic lattice satisfying the flatness condition around faces. (a) A closed (non-intersecting) surface on the dual lattice that consists of a single type of generator. The surface can be given two orientations, which, together with the orientation of edges, determine whether the generator ($a$, if parallel) or its inverse ($\bar{a} \equiv a^{-1}$, if antiparallel) is placed on the corresponding edge. Any closed, directed loop on the primary lattice that passes through the pictured vertex necessarily passes through the surface twice and hence reads one of the words $w = \tta \tta^{-1}, \tta^{-1}\tta$. (b) A simple membrane configuration (i.e., sponge) associated with the relation $\tta \ttb \ttc = \tte$. A line of faces on the primary lattice host a basepoint. Any path that passes once through the sponge reads words $\tta\tta^{-1}$, $\ttc \ttc^{-1}$, $\tta \ttb \ttc$, or their inverses.}
    \label{fig:3D-states}
\end{figure}

As in Sec.~\ref{sec:tiles}, it is convenient to introduce a graphical representation of allowed product states. The directed strings on the dual lattice are generalized to oriented membranes composed of dual faces in 3D. If an edge on the primary lattice is parallel (respectively, antiparallel) to the orientation of the dual face associated to the generator $g$, it is understood that the edge hosts the generator $g$ (respectively, the inverse generator $g^{-1}$). Examples of simple product-state configurations satisfying the flatness condition are shown in Fig.~\ref{fig:3D-states}. Generically, states satisfying the flatness condition consist of closed surfaces, possibly terminating along one-dimensional lines of basepoints, which we dub \emph{baselines}. Such baselines may branch and fuse, subject to a kind of ``basepoint-flux conservation,'' described below. The resulting states, made up of membranes joined along baselines, form ``sponge'' configurations that span the Hilbert space. 

One way to motivate the allowed configurations in 3D is to utilize the spacetime mapping from Sec.~\ref{sub:timeslice}. A time-evolving string sweeps out a membrane in spacetime. Similarly, a time-evolving basepoint will lead to a one-dimensional baseline in spacetime, and the splitting and merging of basepoints in time will lead to branching and fusion of baselines in the corresponding 3D configuration.

\begin{figure}[t]
    \centering
    \includegraphics[scale=1]{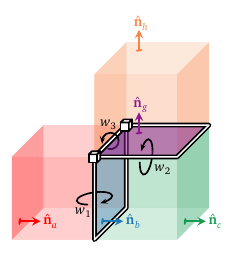}
    \caption{Splitting and merging of baselines in 3D. Pictured is a group model with the relations $\tta \ttb \ttc = \tte $ and $ \ttg \tth = \ttc$, where the membranes shown live on the dual lattice. The primary lattice is omitted for clarity. Paths through this sponge on the primary lattice read either $w_1 = \tta \ttb \ttc$, $w_2 = \ttc^{-1} \ttg \tth$, the composite word $w_3 = \tta \ttb \ttg \tth$, or their inverses, depending on which of the baselines is encircled. The splitting of baselines is denoted by a white cube.}
    \label{fig:baseline-split}
\end{figure}

More formally, we can derive a microscopic basepoint-flux conservation condition.
Consider the words $w_i$ with $i \in \{1, \dots, 6\}$ read around the six faces of a cube on the primary lattice (with a particular starting position and orientation, defined below). Let us also introduce the equivalence relation $\sim_f$. Two words are equivalent under $\sim_f$ if they can be transformed into one another using only free reduction and cyclic permutations of characters, e.g., $\tta \ttb \tta^{-1} \sim_f \ttb \ttc^{-1}\ttc$. Consider the following definition of $w_1$, $w_2$, and $w_3$:
\begin{equation}
\label{eq:3d_loops_concat}
    \begin{tikzpicture}[baseline={(0, 0.25)},every node/.append style={transform shape}]
        \begin{scope}[canvas is xy plane at z=-0.1]
            \draw [thick, ->, >=stealth] (1, 1) to (0, 1) to (0, 0) to (1, 0) to (1, 0.9);
            \node at (0.5, 0.5) {$w_1$};
            \fill (1, 1) circle [radius=0.35ex];
        \end{scope}
        \begin{scope}[canvas is zy plane at x=1.32]
            \draw [thick, ->, >=stealth]  (0, 1) to (0, 0) to (1, 0) to (1, 1) to (0.15, 1);
            \node [xscale=-1, scale=1.3] at (0.35, 0.5) {$w_2$};
            \fill (0, 1) circle [radius=0.4ex];
        \end{scope}
        \begin{scope}[canvas is xz plane at y=1.32]
            \draw [thick, ->, >=stealth] (1, 0) to (1, 1) to (0, 1) to (0, 0) to (0.9, 0);
            \node [rotate=180,xscale=-1, scale=1.3] at (0.5, 0.35) {$w_3$};
            \fill (1, 0) circle [radius=0.4ex];
        \end{scope}
    \end{tikzpicture}
    \:
    \overset{w_1 \cdot w_2 \cdot w_3}{\longrightarrow}
    \:
    \begin{tikzpicture}[baseline={(0, 0.25)}]
        \draw [thick, rounded corners=1pt] (0.9, 1, 0) to (0, 1, 0) to (0, 0, 0) to (0.95, 0, 0) to (0.95, 0.85, 0) to (1.05, 0.85, 0) to (1.05, 0, 0) to (1.05, 0, 1) to (1.05, 1, 1) to (1.05, 1, 0.2) to (0.95, 1, 0.15) to (0.95, 1, 1) to  (0, 1, 1) to (0, 1, 0.15) to (0.9, 1, 0.15)--cycle;
        \draw [kb1] (0, 0.75, 0) to (0, 0.15, 0);
        \draw [kb1] (0.15, 0, 0) to (0.75, 0, 0);
        \begin{scope}[canvas is zy plane at x=1.05]
            \draw [kb1] (0.25, 0) to (0.75, 0);
            \draw [kb1] (1, 0.25) to (1, 0.85);
        \end{scope}
        \begin{scope}[canvas is xz plane at y=1]
            \draw [kb1] (0.85, 1) to (0.25, 1);
            \draw [kb1] (0, 0.75) to (0, 0.25);
        \end{scope}
    \end{tikzpicture}
    \:
    \overset{\sim_f}{\longrightarrow}
    \:
    \begin{tikzpicture}[baseline={(0, 0.25)}]
        \fill [gray, opacity=0.2] (0, 0) rectangle (1, 1);
        \begin{scope}[canvas is zy plane at x=1]
            \fill [gray, opacity=0.2] (0, 0) rectangle (1, 1);
            \draw [kb1] (0.25, 0) to (0.75, 0);
            \draw [kb1] (1, 0.25) to (1, 0.85);
        \end{scope}
        \begin{scope}[canvas is xz plane at y=1]
            \fill [gray, opacity=0.2] (0, 0) rectangle (1, 1);
            \draw [kb1] (0.85, 1) to (0.25, 1);
            \draw [kb1] (0, 0.75) to (0, 0.25);
        \end{scope}
        \draw [kb1] (0, 0.75, 0) to (0, 0.15, 0);
        \draw [kb1] (0.15, 0, 0) to (0.75, 0, 0);
        \draw [thick, lightgray] (1, 1, 0) to (1, 1, 1);
        \draw [thick, lightgray] (1, 1, 0) to (0, 1, 0);
        \draw [thick, lightgray] (1, 1, 0) to (1, 0, 0);
        \draw [thick] (0, 0, 0)--++(1, 0, 0)--++(0, 0, 1)--++(0, 1, 0)--++(-1, 0, 0)--++(0, 0, -1)--++(0, -1, 0);
    \end{tikzpicture}
\end{equation}
When these words $w_i$ are concatenated the resulting path involves self-retracing steps, which can be removed using free reduction.
Since we could have performed the same construction using the words $w_4$, $w_5$, and $w_6$ read around the complementary faces (starting on the diametrically opposite vertex),
we deduce that
\begin{equation}
    w_1 \cdot w_2 \cdot w_3 \sim_f w_4 \cdot w_5 \cdot w_6
    \, .
    \label{eqn:basepoint-conservation}
\end{equation}
If one side of~\eqref{eqn:basepoint-conservation} belongs to $[\tte]$ (the equivalence class under $\sim_f$ containing $\tte$), then the other side contains either no basepoints or a simple net that can be created in isolation from the vacuum. If one side of~\eqref{eqn:basepoint-conservation} belongs to an equivalence class distinct from $[\tte]$, then at least one of the relations $r \in R$ needs to be applied to connect the word to $\tte$ (i.e., at least one of the constituent faces hosts a basepoint). Hence, in this case, the complementary faces must also host at least one basepoint. An example configuration involving the splitting and merging of baselines is shown in Fig.~\ref{fig:baseline-split}.

\subsection{Dynamics}

The most general dynamics preserving flatness follow from the operators defined in Sec.~\ref{sub:dynamics}. The projector onto flat configurations around plaquettes~\eqref{eqn:flatness-operator} is identical to 2D. The operator with smallest range that implements nontrivial dynamics while preserving flatness is notationally identical to 2D: 
\begin{equation}
    \tilde{A}_v(g) = \prod_{e \in \partial^\dagger v} \tilde{L}^g(e, v)
    \, ,
    \label{eqn:Av-3D}
\end{equation}
but now acts on the dual closed surface around the vertex $v$. We remind the reader that $\tilde{L}_\pm^g \equiv \Pi L^g_\pm \Pi$ with $\Pi$ the projector onto the onsite alphabet $\mathcal{A}$. When acting on the vacuum, this operator nucleates a minimal closed surface associated to the character $g$, as depicted in Fig.~\ref{fig:3D-states}(a). When acting on a configuration that contains $g$ membranes, Eq.~\eqref{eqn:Av-3D} is capable of locally fluctuating the shape of the membrane. More complicated structures can also be nucleated or fluctuated using an analogous generalization of Eq.~\eqref{eqn:larger-Av} to 3D.

\subsection{Robust fragmentation}

The flatness condition implies that $\varphi(w_{\partial S}) = \tte$ for the boundary of any contractible surface $S$.
However, the dynamics also preserve $\varphi(w_\gamma)$ for noncontractible curves $\gamma$. Since these curves cannot be constructed from elementary plaquettes, the group element $\varphi(w_\gamma)$ is not constrained to be the identity. The model therefore supports intrinsic Krylov sectors labeled by group elements (with appropriately defined OBC) or conjugacy classes (with PBC) along noncontractible curves. The flatness condition ensures that any two homotopically equivalent curves multiply to group elements in the same conjugacy class.
In contrast to the 2D models from Sec.~\ref{sec:2d}, operators that modify the group element must act simultaneously on $\mo(L^2)$ degrees of freedom (constituting a system-spanning dual membrane) to prevent violations of the flatness condition. In contrast, any sequence of local moves that aims to change the intrinsic Krylov sector must involve intermediate states that violate flatness on at least $\mo(L)$ faces \cite{stahl2023topologically}. Absent diagonal contributions to the Hamiltonian, this creates an energy barrier that diverges with system size. This diverging energy barrier could endow group dynamics in three dimensions with even greater stability --- a possibility that we will discuss elsewhere.  

A generic diagonal contribution will energetically prefer a small number of configurations. In that case, states from other Krylov sectors will evolve out of their sectors, in analogy to false-vacuum decay~\cite{Langer1969}. As a result, the fragmentation will be only prethermal. However, the size of the ``bubble'' that needs to be nucleated itself grows as the strength of the perturbation decreases, so that the energy barrier increases with decreasing perturbation strength. Consequently, the prethermal timescale will diverge parametrically faster than in two dimensions.  

\subsection{Fragmentation from knots}
\label{subsec:knots}

The above construction is based on a 3D system with a 2-form symmetry, which implies that flat configurations consist of closed membranes, perhaps splitting or merging at baselines. Here, we argue how a different type of fragmentation arising from knotting of strings could occur in 3D in group models possessing a 1-form symmetry.

Starting with an Abelian group, such models can be constructed as follows. Consider a 3D cubic lattice with degrees of freedom on faces. The 1-form flatness condition then requires that the group element constructed by multiplying generators on faces around a cube equals the identity. This presents a potential ordering ambiguity that is resolved by considering only Abelian groups $G$ for which the order of multiplication does not affect the resulting group element. 
In such a model flat configurations consist of closed strings of dual edges, which are allowed to intersect at cube centers.\footnote{In general, it is also possible for flat configurations to involve branching and merging of these dual loops, which will give rise to additional fragmentation structure.} 
If we consider dynamics that respect $G$, then the model will exhibit Krylov sectors whose structure can be understood entirely by the global 1-form symmetry of the model.
To achieve fragmentation, we require one additional ingredient. If the intersection of any two species of string is forbidden as a hard local constraint then it is possible to achieve fragmentation due to nontrivially linked or knotted loops.
In particular, this occurs when the system is endowed with dynamics that allows loops to locally fluctuate but does not allow loops to tunnel through one another (therefore effectively implementing the \emph{Reidemeister moves} from knot theory~\cite{Kauffman1991}).
For example, the states represented by
\begin{equation}
    \begin{tikzpicture}[baseline={(0, -0.1)}]
        \pgfmathsetmacro{\r}{0.6};
        \draw [thick, line cap=round, preaction={draw, line width=5, white}] ({0+\r*cos(0)},{0+\r*sin(0)}) arc (0:180:{\r});
        \draw [thick, line cap=round, preaction={draw, line width=5, white}] ({\r+\r*cos(0)},{0+\r*sin(0)}) arc (0:180:{\r});
        \draw [thick, line cap=round, preaction={draw, line width=5, white}] ({\r+\r*cos(0)},{0+\r*sin(0)}) arc (0:-180:{\r});
        \draw [thick, line cap=round, preaction={draw, line width=5, white}] ({0+\r*cos(0)},{0+\r*sin(0)}) arc (0:-180:{\r});
    \end{tikzpicture}
    \quad \text{and} \quad
    \begin{tikzpicture}[baseline={(0, -0.1)}]
        \pgfmathsetmacro{\r}{0.6};
        \draw [thick, line cap=round, preaction={draw, line width=5, white}] ({0+\r*cos(0)},{0+\r*sin(0)}) arc (0:180:{\r});
        \draw [thick, line cap=round, preaction={draw, line width=5, white}] ({\r+\r*cos(0)},{0+\r*sin(0)}) arc (0:180:{\r});
        \draw [thick, line cap=round, preaction={draw, line width=5, white}] ({0+\r*cos(0)},{0+\r*sin(0)}) arc (0:-180:{\r});
        \draw [thick, line cap=round, preaction={draw, line width=5, white}] ({\r+\r*cos(0)},{0+\r*sin(0)}) arc (0:-180:{\r});
    \end{tikzpicture}
\end{equation}
would remain disconnected under the dynamics and hence these configurations must belong to distinct Krylov sectors. Similarly, a single loop can be knotted with itself giving rise to configurations such as
\begin{equation}
    \scalebox{0.7}{
    \begin{tikzpicture}[baseline={(0, -0.1)}]
        \begin{knot}[
            clip width=5,
            consider self intersections
        ]
            \strand[line width=1.2] (90:1.2)
                \foreach \x in {1,2,3} {to [bend left=120,looseness=1.75] ({90+120*\x}:1.2)};
            \flipcrossings{1,3}
        \end{knot}
    \end{tikzpicture}}
    \quad \text{and} \quad
    \scalebox{0.7}{
    \begin{tikzpicture}[baseline={(0, -0.1)}]
        \begin{knot}[
            clip width=5,
            consider self intersections
        ]
            \strand[line width=1.2] (90:1.2)
                \foreach \x in {1,2,3} {to [bend left=120,looseness=1.75] ({90+120*\x}:1.2)};
            \flipcrossings{3}
        \end{knot}
    \end{tikzpicture}}
    \hspace{-3mm} ,
\end{equation}
which are dynamically disconnected.

\begin{figure}
    \centering
    \includegraphics[scale=0.135]{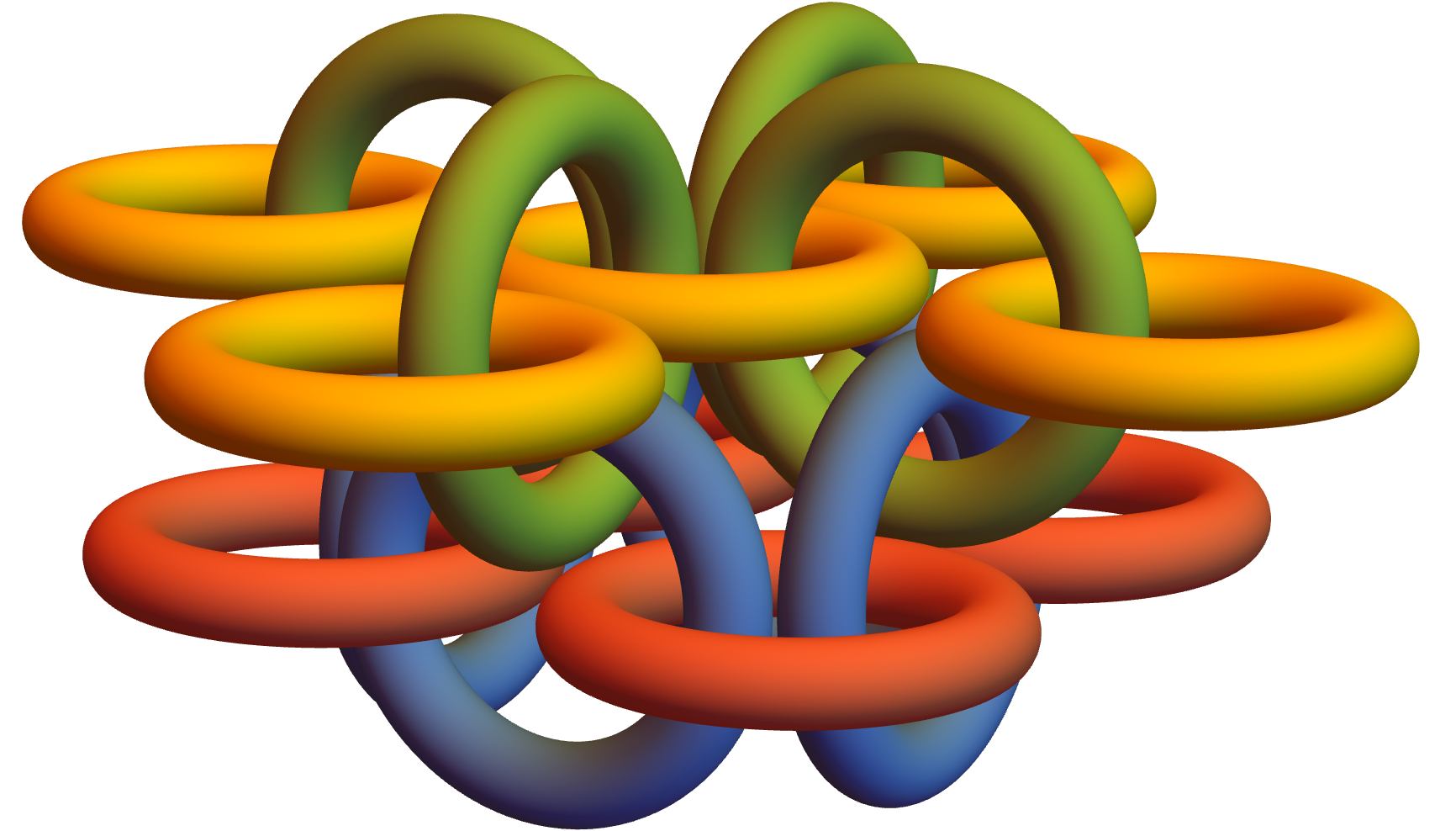}
    \hspace{0.5cm}
    \includegraphics[scale=0.135]{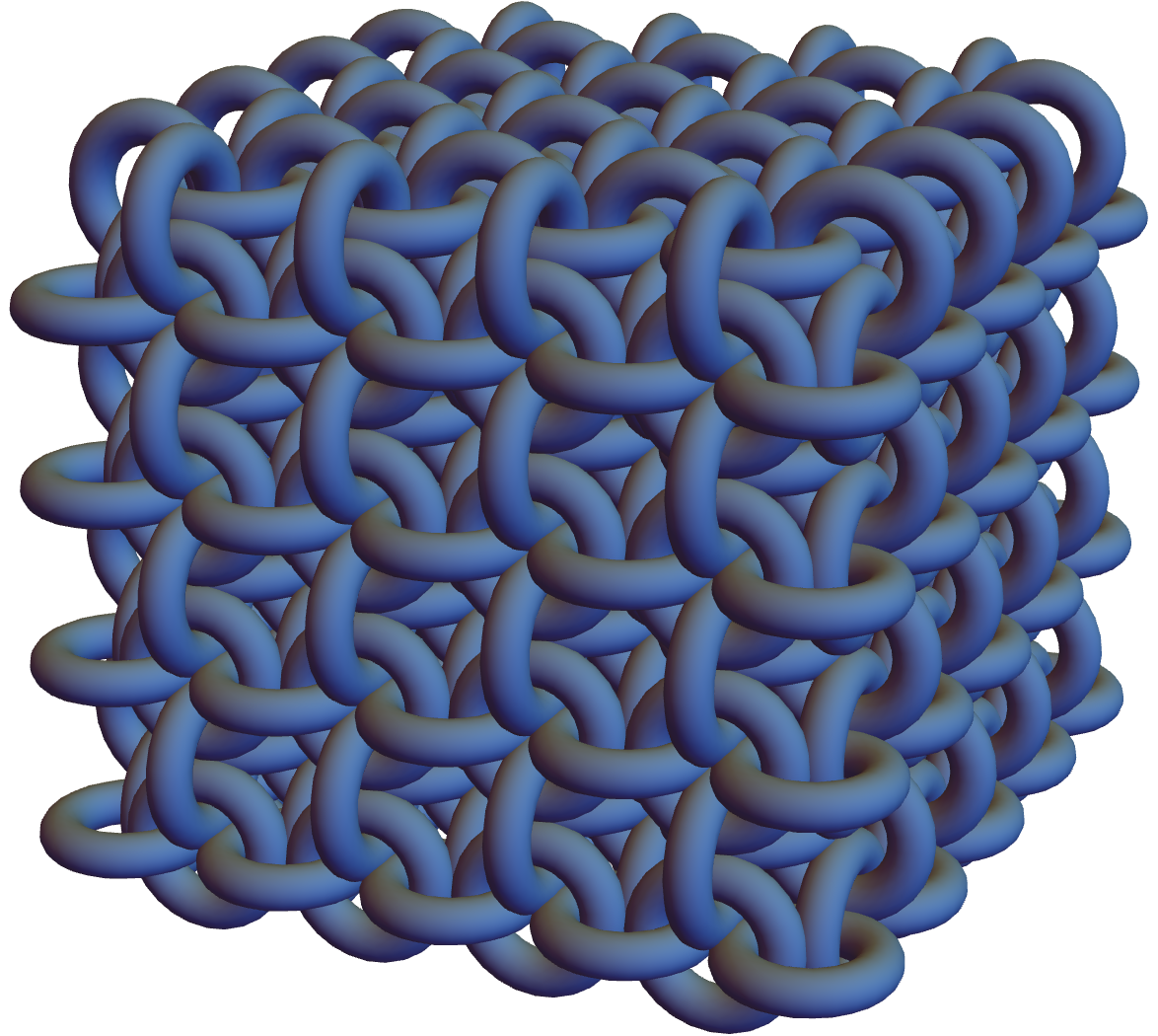}
    \caption{Left: 3D ``chainmail'' configuration of interlocked rings that is capable of repeating periodically throughout space. The colors help to distinguish the relative positioning of the rings as opposed to denoting different species of loop. Right: A larger chainmail configuration. Removing any one ring does not unlink any of the other rings.}
    \label{fig:3D-chainmail}
\end{figure}

Such a construction will give rise to an exponential (in system volume) number of Krylov sectors.
This may be seen by constructing the periodic, densely interlocked configuration of rings shown in Fig.~\ref{fig:3D-chainmail}, for example.
For an Abelian group $G$ with $m$ generators, each ring can be colored in $m$ ways, giving rise to an exponential number of dynamically disconnected configurations.
Similarly, even if there is only a single species of ring present, we can choose a subset of rings with some minimum separation (e.g., the yellow rings in the left panel of Fig.~\ref{fig:3D-chainmail}). Each ring belonging to this subset may be either present or absent, again giving rise to an exponential-in-volume number of sectors.

Another interesting feature of models exhibiting fragmentation due to knots relates to the
computational complexity of questions regarding thermalization. That is, there exist a number of decision problems in knot theory that have implications for the dynamics of the systems described above.
One famous question asks whether two knots can be deformed into one another, sometimes known as the ``equivalence problem'' for knots and links\footnote{A knot is an embedding of the circle $S^1$ in $\mathbb{R}^3$, while a link is an embedding of at least one copy of $S^1$ in $\mathbb{R}^3$.}~\cite{lackenby2016knottheory}, whose complexity class remains unknown. Physically, this is equivalent to asking whether two product-state configurations belong to the same Krylov sector.
A simpler question asks whether a given configuration of, e.g., a single loop, is equivalent to the trivial knot (the \emph{unknot}). It is not known whether there exists a polynomial-time algorithm to recognize the trivial knot~\cite{lackenby2016knottheory}, but there do exist upper bounds on the number of Reidemeister moves needed to transform a knot with $n$ crossings into the unknot~\cite{hass2001number}, $\leq 2^{cn}$ with $c=10^{11}$. This is the analog of a large Dehn function from Sec.~\ref{subsec:word_problem} and hints that the thermalization time in such knotted models could be incredibly slow.

\section{Group models on arbitrary graphs}
\label{sec:arbitrary-graphs}

While previously we restricted ourselves to crystalline lattices in 1D, 2D, and 3D (or, at the very most, geometrically local graphs embeddable in 2D manifolds), in this section we generalize group dynamics to models on completely arbitrary graphs (which include non-planar graphs and graphs not embeddable in 2D manifolds, as well as graphs without geometric locality). First, we introduce some basic concepts from graph theory relevant for further discussion. Next, we describe the emergent structure of intrinsic Krylov sectors upon imposing the flatness condition on an arbitrary set of closed walks on a graph. Further, we discuss the allowed product states and introduce the analogs of loops and nets on an arbitrary graph (assuming the flatness condition is imposed on every closed walk of the graph). Finally, we introduce the group dynamics on arbitrary graphs.

\subsection{Relevant graph theory concepts}
\label{sec:graph_basics}

A directed graph is defined as $\Gamma = (V, E, \delta)$, where $V$ is a set of vertices, $E$ is a set of edges, and $\delta: E \to V \times V$ is an incidence function that maps an edge $e \in E$ to an ordered pair of vertices, $(v, u)$, $v,u \in V$. We say that vertices $v,u$ are \emph{endpoints} of edge $e$ --- the edge leaves its \emph{source} vertex $v$ and enters its \emph{target} vertex $u$. We allow $E$ to contain edges with the source and target vertices being the same, i.e., $\delta(e) = (v,v)$, as well as multiple distinct edges with the same endpoints, i.e., $\delta(e) = \delta(e') = (v, u)$ for $e \neq e'$.  

We define a \emph{walk}, $\gamma$, as a finite sequence of edges, $\gamma = (e_1, e_2, \dots, e_{n-1})$, together with a sequence of vertices $(v_1, v_2, \dots, v_n)$ such that the endpoints of $e_i$ are $v_i$ and $v_{i+1}$, for $i=1,2,\dots,n-1$. Note that in our definition, a walk is allowed to traverse against the direction of an edge. For an edge $e$ with $\delta(e) = (v,v)$, we must distinguish between traversing along and against the direction of the edge --- therefore, such an edge will appear in the walk in one of the two orientations, $e^+$ or $e^-$. A graph is said to be \emph{connected} if there exists a walk between any two vertices. From now on, for simplicity, we will only consider connected graphs. A \emph{closed walk} is a walk with $v_1 = v_n$.

Next, we introduce a procedure of \emph{free reduction} on a walk $\gamma$: (i) if the walk traverses the same edge twice in a row (i.e., if $e_i = e_{i+1}$),\footnote{For edges with the same source and target vertices, free reduction is only performed when two consecutive edges have opposite signs.} remove these two edges and the corresponding vertex $v_{i+1}$ from the walk --- the resulting sequence of edges and vertices still defines a valid walk; (ii) iteratively repeat step (i) until no edges can be deleted anymore. We denote the resulting walk as $\red(\gamma)$. Consequently, a walk where no two consecutive edges are the same (i.e., $\red(\gamma) = \gamma$) is called \emph{freely reduced}. Intuitively, free reduction gets rid of backtracking --- parts of the walk where the edges are traversed in one direction and immediately in the opposite direction.

Now, we choose a vertex $v^*$ and consider all closed walks originating at $v^*$.
We say that two closed walks, $\gamma$ and $\gamma'$, belong to the same \emph{equivalence class} (denoted $[\gamma]_{v^*} \equiv [\gamma']_{v^*}$) if $\red(\gamma) = \red(\gamma')$. We can view the graph $\Gamma$ as a topological space (a CW complex consisting of 1-cells and 0-cells), and view a closed walk as a 1-dimensional loop originating at $v^*$ in that space. Then, all closed walks from the equivalence class $[\gamma]_{v^*}$ are homotopic to each other, while closed walks from distinct equivalence classes are homotopically nonequivalent. From homotopy theory, we know that the equivalence classes under homotopy of loops form a group, known as the fundamental group, $\pi_1(\Gamma, v^*)$, of graph $\Gamma$. For a graph, the fundamental group is always a free group,
\eq{
\pi_1(\Gamma, v^*) \cong \zz^{*(\abs{E} - \abs{V} + 1)} ,
}
where $\abs{E}$ and $\abs{V}$ is the number of edges and vertices of graph $\Gamma$, respectively. 
The group multiplication (denoted as ``$\circ$'') for equivalence classes of closed walks originating at $v^*$ is inferred from homotopy theory and simply constitutes the concatenation of the edge and vertex sequences of the two representative walks.

We want to emphasize that despite closed walks having PBC, we essentially view them as ``open'' walks that have a specific origin. This helps us establish a one-to-one correspondence between equivalence classes $[\gamma]_{v^*}$ and group elements of $\pi_1(\gamma, v^*) \cong \zz^{*(\abs{E} - \abs{V} + 1)}$. If instead closed walks were viewed as having no origin, then they would correspond to the conjugacy classes of $\pi_1(\Gamma, v^*)$, rather than to its group elements. We will keep this fact in mind and apply the logical leap from the group elements of $\pi_1(\Gamma, v^*)$ to its conjugacy classes at the very end.

\subsection{Structure of intrinsic Krylov sectors}

\subsubsection{General construction}

Next, we define a $G$-group model on the graph $\Gamma$. Similarly to 1D, 2D, and 3D group models discussed in the previous sections, degrees of freedom are situated on the edges, and the local Hilbert space is $\mathcal{H} = \Span\{\ket{s}\}$, with $s \in \mathcal{A} \equiv \mathcal{S} \cup \mathcal{S}^{-1} \cup \{\tte\}$, where $\mathcal{S}$ is the generating set of group $G$. A walk $\gamma = (e_1, e_2 \dots, e_n)$ on the graph $\Gamma$ then reads a word $w_\gamma = s_{e_1}^{q_1} s_{e_2}^{q_2} \cdots s_{e_n}^{q_n}$, where $s_{e_i}$ is the character on edge $e_i$, and $q_i = +1$ ($-1$) if the walk traverses edge $e_i$ along (against) its direction.

On a 2D lattice, we deemed it natural to impose the flatness condition on elementary plaquettes, which in turn led to the flatness condition holding true on every contractible loop. There is no natural notion of ``elementary plaquettes'' on a generic graph, and we allow ourselves to choose any set of closed walks on which to impose the flatness condition. We therefore ask the following question: if the flatness condition is enforced on closed walks $\gamma_1, \dots, \gamma_k$, then what is the full set of closed walks on graph $\Gamma$ where the flatness condition holds? 

\begin{figure}[t]
\centering
  \includegraphics[scale=0.85]{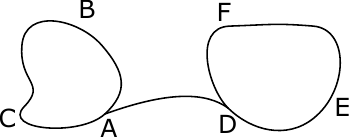}
  \caption{Imposing flatness on walks $\gamma_\mathrm{DABCAD}$ and $\gamma_\mathrm{DEFD}$ imposes it on every walk from $[\gamma_\mathrm{DABCAD}]^{\pm 1}$, $[\gamma_\mathrm{DEFD}]^{\pm 1}$, $[\gamma_\mathrm{DABCAD}] \circ [\gamma_\mathrm{DEFD}] = [\gamma_\mathrm{DABCADEFD}]$, as well as on the conjugacy classes of $[\gamma_\mathrm{DABCAD}]$ and $[\gamma_\mathrm{DEFD}]$, i.e., on any walk from $[\gamma'] \circ [\gamma_\mathrm{DABCAD}] \circ [\gamma']^{-1}$ and $[\gamma'] \circ [\gamma_\mathrm{DEFD}] \circ [\gamma']^{-1}$ for any $\gamma'$ originating at vertex D. 
  }
  \label{fig:walk_concat}
\end{figure}

To answer this question, first note that the backtracking parts of a walk do not contribute to the group element corresponding to $w_\gamma$, i.e., $\varphi(w_\gamma) = \varphi(w_{\red(\gamma)})$. This means that imposing flatness on walk $\gamma$ automatically imposes it on every walk $\gamma'$ with $\red(\gamma') = \red(\gamma)$, and hence on the entire equivalence class $[\gamma]$. Therefore, without loss of generality, we can assume that all walks $\gamma_1, \dots, \gamma_k$ originate at some fixed vertex $v^*$, since on a connected graph any walk $\gamma$ can be deformed to have an origin at any fixed vertex $v^*$ without changing $\red(\gamma)$.\footnote{This can be done by ``growing a finger,'' i.e., creating a backtracking path that connects $v^*$ with any point on the original walk. This is exemplified in Fig.~\ref{fig:walk_concat}: if we want all walks to originate at vertex D, we can deform the walk $\gamma_\mathrm{ABCA}$ into $\gamma_\mathrm{DABCAD}$.} Therefore, from now on, we will implicitly assume that all walks originate at the same vertex and we will drop the subscript $v^*$ from the notation for equivalence classes. Changing $v^*$ does not change the group structure, since, up to isomorphisms, the fundamental group $\pi_1(\Gamma, v^*)$ does not depend on $v^*$ [and hence, we will denote it simply as $\pi_1(\Gamma)$]. 

Next, note that:
\begin{enumerate}
    \item Imposing flatness on $[\gamma]$ imposes it on $[ \gamma^{-1} ] \equiv [\gamma]^{-1}$, since $w_{\gamma^{-1}} = w^{-1}_\gamma$, and $\varphi(w) = \tte$ implies $\varphi(w^{-1}) = \tte$.

    \item Imposing flatness on $[\gamma]$ imposes it on the whole conjugacy class of $[\gamma]$ in $\pi_1(\Gamma)$, i.e., on every walk from $[\gamma'] \circ [\gamma] \circ [\gamma']^{-1}$, for every $[\gamma'] \in \pi_1(\Gamma)$, since if $\varphi(w_\gamma) = \tte$, then $\varphi(w_{\gamma'} \cdot w_{\gamma} \cdot w_{(\gamma')^{-1}}) = \varphi(w_{\gamma'})\varphi(w_{\gamma})\varphi(w_{\gamma'})^{-1} = \varphi(w_{\gamma'})\varphi(w_{\gamma'})^{-1} = \tte$, for any $\gamma'$.

    \item Imposing flatness on $[ \gamma ]$ and $[ \gamma' ]$ ensures that the flatness holds on their group product, $[ \gamma ] \circ [ \gamma' ]$, since if $\varphi(w_\gamma) = \varphi(w_{\gamma'}) = \tte$, then $\varphi(w_\gamma \cdot w_{\gamma'}) = \varphi(w_\gamma) \cdot \varphi(w_{\gamma'}) = \tte$.
\end{enumerate}
These three implications are exemplified pictorially in Fig.~\ref{fig:walk_concat} (where all walks are assumed to originate at vertex D).

Consequently, imposing flatness on walks $\gamma_1, \dots, \gamma_k$ leads to flatness holding true on the normal closure\footnote{\emph{Normal closure} of a set $S$ in group $G$ is defined as the set generated by all possible conjugations of $S$ by the elements of $G$. Clearly, it is a normal subgroup of $G$.} of set $\{[\gamma_1], \dots, [\gamma_k]\}$ in group $\pi_1(\Gamma) \cong \zz^{*(\abs{E} - \abs{V} + 1)}$, which is the smallest normal subgroup of $\pi_1(\Gamma)$ containing $\{[\gamma_1], \dots, [\gamma_k]\}$ (call this subgroup $N \vartriangleleft \pi_1(\Gamma)$).\footnote{As a corollary, we immediately get that one must enforce flatness condition on at least $\abs{E} - \abs{V} + 1$ closed walks in order to impose it on \emph{every} closed walk on $\Gamma$. In that case, $N = \pi_1(\Gamma)$.} Using this fact, we can straightforwardly deduce the number of intrinsic Krylov sector labels and all constraints between them. All this information is contained in the quotient group $\pi_1(\Gamma) / N$. Indeed, $N$ splits $\pi_1(\Gamma)$ into cosets. The equivalence classes $[\gamma]$ and $[\gamma']$ belong to the same coset if they are related by an element of $N$, i.e., if $[\gamma'] = [\gamma] \circ [\tilde{\gamma}]$ for some $[\tilde{\gamma}] \in N$. And since $\varphi(\tilde{\gamma}) = \tte$, the words on $\gamma$ and $\gamma'$ must multiply to the same group elements of $G$, i.e., $\varphi(w_\gamma) = \varphi(w_{\gamma'})$.

Now, assume that group $\pi_1(\Gamma) / N$ is generated by $\left\{ [\lambda_1], \dots, [\lambda_m] \right\}$, where, in a slight abuse of notation, $[\lambda]$ now defines the walk $\lambda$ not only up to free reduction but also up to multiplication by group elements of $N$. Then, there will exist $m$ intrinsic Krylov labels, labeled by group elements of $G$, $\{ g_1 \equiv \varphi( w_{\lambda_1}), \dots, g_m \equiv \varphi(w_{\lambda_m}) \}$, subject to additional constraints defined by the group relations of $\pi_1(\Gamma) / N$.

Finally, we remind the reader that, so far, we assumed that all walks are ``open walks that start and end at the same vertex,'' as well as assuming that all walks originate at a common vertex. As we explained at the end of Sec.~\ref{sec:graph_basics}, changing the origin of a closed walk $\lambda_i$ changes $\varphi(w_{\lambda_i})$ within its conjugacy class in $G$. Therefore, the full procedure for analyzing the structure of intrinsic Krylov sectors can be summarized as follows. Assume that the flatness condition is imposed on (not necessarily overlapping) closed walks $\tilde{\gamma}_1, \dots, \tilde{\gamma}_k$. Then
\begin{enumerate}
    \item For any fixed vertex $v^*$, deform each $\tilde{\gamma}_i$ into $\gamma_i$, such that $\red(\tilde{\gamma}_i) = \red(\gamma_i)$, and $\gamma_i$ passes through $v^*$. View $\gamma_i$ as an ``open'' walk that starts at $v^*$ and ends at $v^*$.

    \item Compute the normal closure $N$ of set $\left\{[\gamma_1], \dots, [\gamma_k] \right\}$ in group $\pi_1(\Gamma) \cong \zz^{*(\abs{E} - \abs{V} + 1)}$. For this, it is convenient to know the full generating set of $\pi_1(\Gamma)$, consisting of $\abs{E} - \abs{V} + 1$ closed walks (all originating at $v^*$).

    \item Compute the quotient $\pi_1(\Gamma) / N$. It is convenient to perform this operation by presenting the fundamental group as $\pi_1(\Gamma) = \braket{[\gamma_1], \dots, [\gamma_k], [\lambda_1], \dots, [\lambda_m]\,}{\, R}$, where $\left\{ [\lambda_1], \dots, [\lambda_m] \right\}$ are the elements that complement $\left\{ [\gamma_1], \dots, [\gamma_k] \right\}$ to a full generating set of $\pi_1(\Gamma)$, and $R$ are the relations between the generators. The quotient $\pi_1(\Gamma) / N$ is then easily computed by setting $[\gamma_1], \dots, [\gamma_k]$ to identities in the above presentation, resulting in $\pi_1(\Gamma) / N = \braket{[\lambda_1], \dots, [\lambda_m]\,}{\, \tilde{R}}$. Intrinsic Krylov sectors are labeled by $m$ labels, $g_1, \dots, g_m$, where each $g_i$ is a group element of $G$. On top of that, $g_i$'s must obey constraints specified in $\tilde{R}$.

    \item Deform $\{ \lambda_1, \dots, \lambda_m \}$ to any convenient set of (not necessarily overlapping) closed walks, $\{\tilde{\lambda}_1, \dots, \tilde{\lambda}_m \}$, where the allowed deformations are (i) multiplication by elements of $N$, and (ii) removing/adding backtracking paths. Remember that such deformations (and the choice of the origin for each walk) will change the labels $g_1, \dots, g_m$ within their conjugacy classes in $G$.
\end{enumerate}

\subsubsection{Lattice constraints}
Above, we assumed that $g_1, \dots, g_m$ can assume any value in $G$, as long as they satisfy relations in $\tilde{R}$. However, on a finite graph, the number of accessible values $g_i$ will be bounded by the length $\abs{\lambda_i}$ of the corresponding walk $\lambda_i$, and will be equal to the number of distinct group elements of $G$ that can be written with words of length at most $\abs{\lambda_i}$ using characters from $\mathcal{A}$, which is the growth rate of group $G$ at radius $\abs{\lambda_i}$, $\growth(\abs{\lambda_i})$.

\subsubsection{Example: square lattice on a 2-torus}

Let us demonstrate that this formalism indeed gives the correct result when the graph $\Gamma$ is a square lattice of size $L_x \times L_y$ with PBC in both the $x$- and $y$-directions. The total number of vertices and edges is $\abs{V} = L_x L_y$ and $\abs{E} = 2L_x L_y$, respectively. Therefore, the fundamental group is $\pi_1(\Gamma) = \zz^{*(L_x L_y + 1)}$. We denote the (say, clockwise) walks around elementary plaquettes as $\gamma_1, \dots, \gamma_{L_x L_y}$, and any two non-contractible walks around the $x$- and $y$-directions as $\gamma_x$ and $\gamma_y$, respectively (for brevity, we assume that all walks have already been deformed to originate at the same vertex). Then, the set $\{[\gamma_1], \dots, [\gamma_{L_x L_y}], [\gamma_x], [\gamma_y] \}$ constitutes a valid generating set for $\pi_1(\Gamma)$. However, it is not minimal, since 
\eq{
[\gamma_1] \circ \dots \circ[\gamma_{L_x L_y}] = [\gamma_x] \circ [\gamma_y] \circ [\gamma_x]^{-1} \circ [\gamma_y]^{-1},
\label{eqn:torus-relations}
}
and we can express, say, $[\gamma_1]$ through other generators and exclude it from the generating set. This is equivalent to saying that the group $\pi_1(\Gamma)$ admits presentations
\eq{ \label{eq:presentation_pi1}
\pi_1(\Gamma) & = \braket{[\gamma_1], \dots, [\gamma_{L_x L_y}], [\gamma_x], [\gamma_y] \,}{\, \eqref{eqn:torus-relations}} \\
& = \braket{[\gamma_2], \dots, [\gamma_{L_x L_y}], [\gamma_x], [\gamma_y] \,}{\,\,\,}.
}
Imposing the flatness on elementary plaquettes $\gamma_1, \dots, \gamma_{L_x L_y}$ means taking the quotient $\pi_1(\Gamma) / N$, where $N$ is the normal closure of set $\{[\gamma_1], \dots, [\gamma_{L_x L_y}] \}$ in $\pi_1(\Gamma)$ ($N$ consists of all contractible closed walks). This quotient is easy to compute by simply replacing $[\gamma_1], \dots, [\gamma_{L_x L_y}]$ with identities in the first line of Eq.~\eqref{eq:presentation_pi1}, which results in
\eq{
\pi_1(\Gamma) / N = \braket{[\gamma_x], [\gamma_y] \,}{\,  [\gamma_x] \circ [\gamma_y] = [\gamma_y] \circ [\gamma_x]} \cong \zz^2.
}
This result coincides with the results of Sec.~\ref{subsec:intrinsic_krylov} --- intrinsic Krylov sectors are labeled by two objects, $g_x, g_y \in G$, subject to the constraint $g_x g_y = g_y g_x$ (when the corresponding closed walks originate at the same vertex). We remind that when one shifts the origin for the walk $\gamma_x$ ($\gamma_y$), the label $g_x$ ($g_y$) changes to an element of $G$ from the same conjugacy class. The total number of intrinsic Krylov sectors is upper bounded by $\growth (L_x) \growth (L_y)$, since the shortest $\gamma_x, \gamma_y$ are the straight lines crossing the system in $x$- and $y$-directions.

\subsubsection{Example: arbitrary cellulation of a 2-torus}
Similar results hold for an arbitrary cellulation of the 2-torus. By ``cellulation,'' we mean a graph whose edges split the torus into regions (also called faces) and where no face (with its boundaries excluded) supports a non-contractible loop \footnote{This definition encompasses not only graphs where each edge has two distinct faces on either side but also graphs like $K_5$ and $K_{3,3}$, where some edges are adjacent to the same face on both sides. Such a face then spans the entire torus in at least one direction, yet it still cannot support a non-contractible loop that avoids crossing any edges.}. 

Since the Euler characteristic of a torus is zero, 
\eq{
\label{eq:euler_characteristic}
\abs{V} - \abs{E} + F = 0 ,
}
where $F$ is the total number of aforementioned faces,
the fundamental group for such a graph $\Gamma$ is
\eq{
\pi_1(\Gamma) = \zz^{*\left(\abs{E} -\abs{V} + 1\right)} = \zz^{*\left(F + 1\right)} .
}
Now, similarly to the square lattice example, we denote the closed walks around every face as $\gamma_1, \dots, \gamma_F$ and the walks around the two cycles of the torus as $\gamma_x$ and $\gamma_y$.\footnote{If $\Gamma$ was not a cellulation, at least one of the walks $\gamma_x, \gamma_y$ would not be well defined.}
The set $\{[\gamma_1], \dots, [\gamma_{F}], [\gamma_x], [\gamma_y] \}$ generates $\pi_1(\Gamma)$, but, as before, there is a constraint,
\eq{
[\gamma_1] \circ \dots \circ[\gamma_{F}] = [\gamma_x] \circ [\gamma_y] \circ [\gamma_x]^{-1} \circ [\gamma_y]^{-1} .
\label{eqn:generic-torus-relation}
}
The fundamental group is then presented as
\eq{ \label{eq:presentation_pi1_general}
\pi_1(\Gamma) & = \braket{[\gamma_1], \dots, [\gamma_{F}], [\gamma_x], [\gamma_y] \,}{\, \eqref{eqn:generic-torus-relation}}
\\
& = \braket{[\gamma_2], \dots, [\gamma_{F}], [\gamma_x], [\gamma_y] \,}{\,\,\,},
}
the normal closure of $\{[\gamma_1], \dots, [\gamma_{F}]\}$ in $\pi_1(\Gamma)$ consists of all contractible closed walks on the torus, and the quotient $\pi_1(\Gamma)/N$ is, again, $\pi_1(\Gamma)/N \cong \zz^2$, meaning that for any such graph, there are two intrinsic Krylov labels, $g_x, g_y \in G$, with a constraint between them, $g_x g_y = g_y g_x$ (again, assuming that $\gamma_x$ and $\gamma_y$ originate at the same vertex).

\subsubsection{Example: cubic lattice on a 3-torus}
We now repeat the same exercise for a cubic lattice of size $L_x \times L_y \times L_z$ with PBC in all three directions. The total number of vertices is $\abs{V} = L_x L_y L_z$, while the total number of edges is $\abs{E} = 3L_x L_y L_z$, leading to the fundamental group $\pi_1(\Gamma) = \zz^{*(\abs{E} - \abs{V} + 1)} = \zz^{*(2L_x L_y L_z + 1)}$. A unit cell with coordinates $(i,j,k)$, $i=1,\dots,L_x$, $j=1,\dots,L_y$, $k=1,\dots,L_z$, consists of three elementary closed walks, $\gamma_{(i,j,k)}^{(xy)}$, $\gamma_{(i,j,k)}^{(xz)}$, and $\gamma_{(i,j,k)}^{(yz)}$, parallel to the $(xy)$-, $(xz)$-, and $(yz)$-plane, respectively. The orientations of the elementary closed walks are chosen as shown in Eq.~\eqref{eq:3d_loops_concat}. We denote the three non-contractible closed walks as $\gamma_x, \gamma_y$, and $\gamma_z$. 

Since the minimal generating set for $\pi_1(\Gamma)$ contains only $2L_xL_yL_z + 1$ generators, there must be some constraints between the $3L_x L_y L_z + 3$ closed walks introduced above. One such constraint is between the six closed walks around an elementary cube:
\begin{multline}
    [\gamma_{(i,j,k)}^{(xy)}] \circ [\gamma_{(i,j,k)}^{(xz)}] \circ [\gamma_{(i,j,k)}^{(yz)}] \circ [\gamma_{(i,j,k+1)}^{(xy)}]^{-1} \circ \\
    [\gamma_{(i,j+1,k)}^{(xz)}]^{-1} \circ [\gamma_{(i+1,j,k)}^{(yz)}]^{-1} = \tte.
    \label{eq:cube_constraint}
\end{multline}
Using Eq.~\eqref{eq:3d_loops_concat} as a pictorial aid, one can verify that concatenating the walks around the six faces results in a homotopically trivial walk. The other type of constraint is similar to the one we had in 2D:
\eq{
\label{eq:plane_constraint}
\mathop{\bigcirc}\limits_{i,j=1}^{L_x, L_y} [\gamma_{(i,j,k)}^{(xy)}] & = [\gamma_x] \circ [\gamma_y] \circ [\gamma_x]^{-1} \circ [\gamma_y]^{-1},
\\
\mathop{\bigcirc}\limits_{i,k=1}^{L_x, L_z} [\gamma_{(i,j,k)}^{(xz)}] & = [\gamma_x] \circ [\gamma_z] \circ [\gamma_x]^{-1} \circ [\gamma_z]^{-1},
\\
\mathop{\bigcirc}\limits_{j,k=1}^{L_y, L_z} [\gamma_{(i,j,k)}^{(yz)}] & = [\gamma_y] \circ [\gamma_z] \circ [\gamma_y]^{-1} \circ [\gamma_z]^{-1}.
}
where the free indices $i, j, k$ run from one to $L_x$, $L_y$, $L_z$, respectively.

Not all of the constraints are independent. In particular, from Eq.~\eqref{eq:cube_constraint}, we can derive that multiplying elementary walks on faces of any closed surface results in an identity. However, on a 3-torus, multiplying the left-hand sides of Eq.~\eqref{eq:cube_constraint} for all but one cube results in the same expression as the left side of Eq.~\eqref{eq:cube_constraint} for that remaining cube. Therefore, there are only $L_x L_y L_z - 1$ independent constraints of type \eqref{eq:cube_constraint}. Next, notice that constraints in Eq.~\eqref{eq:plane_constraint} for any two parallel planes can be related through multiplication by cube constraints from Eq.~\eqref{eq:cube_constraint}. This leaves only 3 independent constraints of type \eqref{eq:plane_constraint}. Every independent constraint allows us to remove one of the generators from the presentation of $\pi_1(\Gamma)$, and thus, we are left with $(3L_x L_y L_z + 3) - (L_x L_y L_z - 1) - 3 = 2L_x L_y L_z + 1$ generators, which is the minimal generating set for $\pi_1(\Gamma) = \zz^{*(2L_x L_y L_z + 1)}$.\footnote{This counting is reminiscent of counting the code-space dimension in stabilizer codes, which is the same as the ground-state degeneracy of stabilizer models. A particular counting for the 3D toric code can be found in Ref.~\cite{castelnovo2008topological}.}

Finally, we impose the flatness condition on every elementary closed walk. To deduce the structure of intrinsic Krylov sectors, we calculate the quotient $\pi_1(\Gamma) / N$ by simply setting the left sides in Eq.~\eqref{eq:cube_constraint}--\eqref{eq:plane_constraint} to identities.
\begin{gather}
\pi_1(\Gamma) / N = \big\langle [\gamma_x], [\gamma_y], [\gamma_z]\, \big| \, R \big\rangle \cong \zz^3,
\\
R = \left\{ [\gamma_\alpha] \circ [\gamma_\beta]  = [\gamma_\beta] \circ [\gamma_\alpha], \alpha \neq \beta, \alpha, \beta = x,y,z \right\}
\end{gather}
which means that there are three intrinsic Krylov sector labels, $g_x, g_y, g_z \in G$, subject to the constraint that they all must commute (again, we assume that $\gamma_x, \gamma_y, \gamma_z$ originate at the same vertex).

\subsection{Allowed product states}

In this subsection, we discuss the structure of the product states satisfying the constraints. For simplicity, here we assume that the flatness condition has been imposed on $\abs{E} - \abs{V} + 1$ independent closed walks and hence on every closed walk of the graph $\Gamma$.

\begin{figure}[t]
\centering
  \includegraphics[scale=0.6]{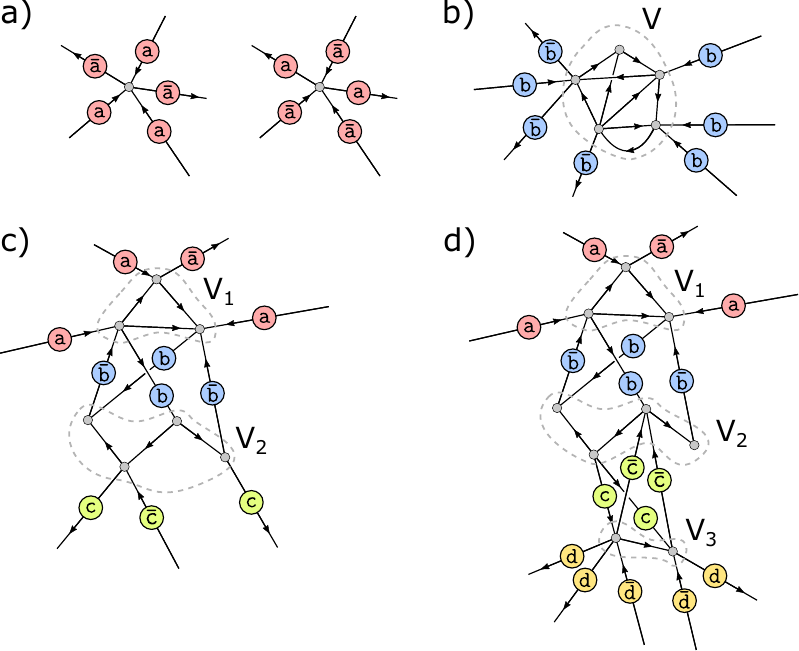}
  \caption{(a) Minimal ``loops'' on an arbitrary graph. The generators $\tta$ ($\tta^{-1}$) are situated on every edge incident to a vertex. Edges directed in and out of the vertex must host opposite generators. (b) A larger ``loop'' consists of generators placed on edges that connect a set of vertices $V$ with $\Gamma \setminus V$, while edges with both endpoints in $V$ remain in the original state. (c) A ``net'' for the relation $\tta\ttb\ttc = \tte$, which consists of two non-overlapping sets of vertices, $V_1$ and $V_2$, and the $\tta$, $\ttb$, $\ttc$ characters (or their inverses) situated on the edges connecting $V_1$ with $\Gamma \setminus (V_1 \cup V_2)$, $V_1$ with $V_2$, $V_2$ with $\Gamma \setminus (V_1 \cup V_2)$, respectively. (d) A ``net'' for the relation $\tta\ttb\ttc\ttd = \tte$.}
  \label{fig:graphs}
\end{figure}

Clearly, the ``vacuum'' --- the state with $\tte$ on every edge --- satisfies the constraints. On top of the vacuum, the simplest allowed structure one could nucleate contains a generator or its inverse on every edge adjacent to a vertex, shown in Fig.~\ref{fig:graphs}(a). This is an analog of an elementary colored closed loop in 2D (Fig\,\ref{fig:loops}(a)) and an elementary colored closed membrane in 3D (Fig.~\ref{fig:3D-states}(a)) ---  we will refer to such structures as ``loops'' for simplicity. Edges pointing in and out of the vertex must host opposite generators, such that any walk traversing through the vertex would pick up $\tta \tta^{-1}$ or $\tta^{-1} \tta$ on the two traversed edges. A larger loop can be constructed by considering a set of vertices $V$ and placing a generator or its inverse on every edge connecting $V$ with the ``outside,'' $\Gamma \setminus V$, while leaving all edges within $V$ in the original state. This is illustrated in Fig.~\ref{fig:graphs}(b) (and is analogous to Fig.~\ref{fig:loops}(b)).

Arbitrary graphs can also support analogs of nets in 2D (or sponges in 3D), shown in Figs.~\ref{fig:graphs}(c) and ~\ref{fig:graphs}(d), which we will simply call ``nets.'' To create the simplest net corresponding to the relation $\tta\ttb\ttc = \tte$, choose two non-overlapping sets of vertices, $V_1$ and $V_2$, such that there is at least one edge connecting $V_1$ and $V_2$. Then, place $\tta$ or $\tta^{-1}$ characters on the edges connecting $V_1$ with $\Gamma \setminus (V_1 \cup V_2)$, $\ttb$ or $\ttb^{-1}$ characters on the edges connecting $V_1$ with $V_2$ and $\ttc$ or $\ttc^{-1}$ characters on the edges connecting $V_2$ with $\Gamma \setminus (V_1 \cup V_2)$, while leaving edges fully within $V_1$ or $V_2$ intact. The net is illustrated in Fig.~\ref{fig:graphs}(c).

For a more complicated net, corresponding to a relator $r = \ttr_1 \ttr_2 \cdots \ttr_n$ with $\abs{r} = n > 3$, we must choose $n-1$ non-overlapping sets of vertices, $V_1, \dots, V_{n-1}$, such that there is at least one edge connecting $V_i$ and $V_{i+1}$, $i=1,\dots, n-2$. Then, we place $\ttr_1^{\pm 1}$ ($\ttr_n^{\pm 1}$) characters on the edges connecting $V_1$ ($V_{n-1}$) with $\Gamma \setminus (V_1 \cup \dots \cup V_{n-1})$, and $\ttr_i^{\pm 1}$ characters on edges connecting $V_{i-1}$ with $V_i$, $i=2,\dots,n-1$. An example for $n=4$ is shown in Fig.~\ref{fig:graphs}(d). In the simplest case, such nets would contain no edges connecting $V_i$ with $\Gamma \setminus (V_1 \cup \dots \cup V_{n-1})$ for all $i=2,\dots,n-2$ (e.g., in Fig.~\ref{fig:graphs}(d), $V_2$ is not connected to the ``outside'' of the graph). If such edges are present for $V_i$, they must support a character, call it $\tth_i$, that obeys $\ttr_1 \cdots \ttr_i \tth_i = \tth_i^{-1} \ttr_{i+1} \cdots \ttr_n = \tte$. This is analogous to the merging of sponges in 3D, shown in Fig.~\ref{fig:baseline-split}. Similarly, we assumed that there are no edges connecting $V_i$ with $V_j$ for $i<j$ with $\abs{i-j} > 1$. However, if such edges are present, they must support a character $\ttg_{ij}$ that obeys $\ttr_1 \cdots \ttr_{i} \ttg_{ij} \ttr_{j+1} \cdots \ttr_n = \ttg_{ij}^{-1} \ttr_{i+1} \cdots \ttr_{j} = \tte$.

\subsection{Dynamics}

Generic dynamics preserving the group element on every closed walk of a graph (identity or not) straightforwardly generalizes from Sec.~\ref{sub:dynamics}. Operators $T_\pm^h$ and $L_\pm^g$ are defined identically to Eqs.~\eqref{eqn:T_op} and \eqref{eqn:L_op}. Then, the operator $\tilde{A}_v(g)$ is defined in the same way as in Eq.~\eqref{eqn:generalized-Av}, but now $\partial^\dagger v$ stands for all edges adjacent to vertex $v$. It is then clear that $\tilde{A}_v(g)$ preserves the group element on any walk passing through vertex $v$ --- this can be easily seen in Eq.~\eqref{eq:action_of_Av}.\footnote{Strictly speaking, if the walk originates at vertex $v$, then $\tilde{A}_v(g)$ only preserves the conjugacy class of the group element on that walk.}

Applying $\tilde{A}_v(g)$ to a vacuum state nucleates an elementary loop around vertex $v$ (like the ones in Fig.~\ref{fig:graphs}(a)). By sequentially applying $\tilde{A}_v(g)$ operators on neighboring vertices, one can nucleate larger loops (Fig.~\ref{fig:graphs}(b)), as well as nets with 3 or 4 characters in the relator (Figs.~\ref{fig:graphs}(c) and \ref{fig:graphs}(d)), similarly to how it was done in 2D in Eqs.~\eqref{eq:abc_net} and \eqref{eq:4_letter_net_nucleation}. In order to nucleate nets with 5 or more characters, one must introduce generalized operators $\tilde{A} (\{v_i\}, \{g_i\})$ similarly to Eq.~\eqref{eqn:larger-Av}. These operators do not factorize into a product of $\tilde{A}_v(g)$.

\section{Conclusion}
\label{sec:conclusion}

We have presented a general construction yielding fragmented quantum dynamics on arbitrary lattices. The starting point is the one-dimensional group dynamics picture of Ref.~\cite{balasubramanian2023glassy}. This picture may be ``lifted'' to higher dimensions by ``promoting'' the global constraints of the one-dimensional model to local constraints on closed loops, which in particular promotes global symmetries in 1D to one-form symmetries in 2D and two-form symmetries in 3D. The constraints yield a Hilbert space generically spanned by ``string-net'' configurations in 2D and ``sponge'' configurations in 3D, where sponges are collections of merging membranes. We have explicitly demonstrated how this lifting may be accomplished on arbitrary two- and three-dimensional crystalline lattices, and have flagged new phenomena arising from lattice effects. Our construction subsumes the previously known examples of topologically robust fragmentation with ``loop-soup'' states on square and cubic lattices \cite{stephen2022ergodicity, stahl2023topologically}, and reveals them to be special cases of a far broader class of string-net and sponge models on arbitrary crystalline lattices. We have also connected our constructions to gauge theories.  
Finally, we have demonstrated how our constructions may be generalized to arbitrary graphs, without any notion of translation invariance, and perhaps without even a notion of geometric locality. Additionally, we have briefly introduced a slightly different class of 3D models, where instead of constraints on 1D loops one has constraints on 2D membranes, and instead of words one has two-dimensional ``grids'' of characters. Configurations in these models are comprised of non-intersecting strings and lead to a novel type of fragmentation due to knotting and linking of the strings.

This paper provides a roadmap to exploration of an exciting new frontier, involving constrained models with fragmented quantum dynamics on arbitrary lattices and, indeed, arbitrary graphs. Using this roadmap it should be possible to systematically investigate new types of models of fragmented dynamics. We emphasize several directions of exploration that strike us as particularly interesting. In 1D, it is straightforward to generalize group dynamics from an underlying group to a \emph{semigroup}, which might have no notion of inverse or identity. In fact, the presentation of 1D group models in Ref.~\cite{balasubramanian2023glassy} is mostly in terms of semigroups. The weaker structure of semigroups seems to be a roadblock for our robust constructions because we use inverse elements when reading words on directed paths and we use the identity elements to define flatness. Nevertheless, Ref.~\cite{balasubramanian2023Entanglement} constructs a 2D Hamiltonian whose emergent degrees of freedom are loops with semigroup dynamics, although the robustness of the dynamics is not clear. The Hamiltonian in Ref.~\cite{balasubramanian2023Entanglement} becomes particularly interesting at a generalized Rokhsar-Kivelson (RK) point, where it becomes a sum of projectors. At this point in parameter space, the ground state displays topological entanglement entropy with volume-law entanglement. Generalized RK points of \emph{group} models also host interesting physics in many settings, such as the quantum double Hamiltonians in Sec.~\ref{sub:quantum-double}. Another example is Ref.~\cite{Zhang2024Bicolor}, which in our language studies the RK point of a $G = \zz_2 * \zz_2$ model and finds interesting modifications to area laws in the topological entanglement entropy. It also might be possible to access quantum fragmentation (wherein the fragmentation is in an entangled rather than a product-state basis) by tuning a ``classically fragmented'' model to its RK point, much as tuning the pair-flip model (classically fragmented) to its RK point yields the Temperley-Lieb model, which is quantum fragmented~\cite{moudgalya2022hilbert}.

Other directions for generalization are orthogonal to the richness of generalized RK points. 
One direction involves exploring known gauge theories and seeking examples thereof that realize fragmented quantum dynamics (whether topologically robust or otherwise).
A second direction concerns expanding the class of models presented in Sec.~\ref{subsec:knots} to generic models where constraints are imposed on 2D membranes (or even on $n$-dimensional manifolds). For group-valued degrees of freedom, this is perhaps possible only for Abelian groups, however, one might ponder if a different mathematical structure (other than a group) is more suitable for this task.
Finally, the third direction includes searching for exactly solvable group models that are outside of the class of quantum doubles, which might shed light into the low-energy properties of such models.
These several lines of investigation should be greatly facilitated by this work, and we look forward to future explorations of each.

\section*{Acknowledgments}
AK thanks Shankar Balasubramanian, Ethan Lake, and Sarang Gopalakrishnan for prior collaboration and valuable insights, Claudio Chamon and Andrei Ruckenstein for ongoing related work, as well as Carolyn Zhang, Sal Pace, Guilherme Delfino, and Hongji Yu for fruitful discussions.
OH and CS thank Marvin Qi and Evan Wickenden for useful discussions, and OH and RN thank David T.~Stephen for prior collaboration. This work was supported (CS, OH and RN) by the Air Force Office of Scientific Research under Award No. FA9550-20-1-0222. AK was supported in part by DOE Grant DE-FG02-06ER46316 and the Quantum Convergence Focused Research Program, funded by the Rafik B. Hariri Institute at Boston University.

\begin{appendix}

\section{Fragmentation in 1D pair-flip model from the group-model perspective} \label{app:pair-flip}

In this Appendix, we discuss the pair-flip model in the language of group dynamics. This both provides a ``worked example'' of how to translate the quantum dynamics into the group-model language, and also illustrates two subtleties of the construction: Excluding the identity from the onsite Hilbert space leads to novel behavior, and dynamics generated by insufficiently long-ranged operators can end up implementing the ``wrong'' group.

The usual presentation of the pair-flip model~\cite{CahaNagaj, moudgalya2022hilbert} (with three colors) starts with a local Hilbert space with three states per site: $\ttr, \ttg$, and $\ttb$, called red, green, and blue. There is a $\U{1}$ symmetry that preserves the number of red spins on the even sublattice minus the number of red spins on the odd sublattice, and similarly for green and blue. Naively there are three $\U{1}$ symmetries, but actually the three charges are constrained so that the symmetry is $\U{1}^2$. The nearest-neighbor dynamics consistent with this symmetry are
\begin{equation}
\ket*{\ttr \ttr} \leftrightarrow \ket*{\ttg \ttg} \leftrightarrow \ket*{\ttb \ttb},
\end{equation}
while the next-nearest-neighbor dynamics are
\begin{equation}
\ket*{\ttr \ttg\ttb} \leftrightarrow \ket*{\ttb \ttg \ttr},
\end{equation}
along with other moves related by changing colors.
If all such dynamics are included, then there are $\sim L^2$ Krylov sectors, labeled by the $\U{1}^2$ symmetry numbers. However, if we restrict to nearest-neighbor dynamics, there are instead $\sim 2^L$ Krylov sectors~\cite{moudgalya2022hilbert}, which cannot be labeled by the symmetry numbers and instead are a signature of fragmentation. Let us see why, in the language of group models. 

From the given dynamics, we are implementing the group
\begin{equation}
G = \braket{ \ttr , \ttg , \ttb  \, }{ \,
\begin{array}{c} 
\ttr^2 = \ttg^2 = \ttb^2 = \tte ,\\
\ttr \ttg \ttb = \ttb \ttg \ttr ,\, \ttg \ttr \ttb = \ttb \ttr \ttg ,\, \ttr \ttb \ttg = \ttg \ttb \ttr 
\end{array}
},
\end{equation}
which is not obviously recognizable. However, by making the replacement $\ttr \to \ttx \tta$, $\ttg \to \tty \tta$, and $\ttb \to \ttx \tty \tta$, we recognize $G$ as 
\begin{align}
G &= \braket{ \ttx, \tty, \tta  \, }{ \,
\begin{array}{c}
\tta^2 = \tte, \, \ttx \tty = \tty \ttx,\\
\ttx \tta = \tta \ttx^{-1},\, \tty \tta = \tta \tty^{-1} 
\end{array}
}
= \zz^2 \rtimes \zz_2,
\end{align}
where the $\zz_2$ generator $\tta$ acts by inverting the two generators of $\zz^2$, $\ttx$ and $\tty$.

To define a group model in the standard way, we would consider a local Hilbert space consisting of $\ket*{\tte}$, $\ket*{\ttr}$,  $\ket*{\ttg}$, and $\ket*{\ttb}$, and allow the dynamics given by the relations. As shown in Appendix~\ref{app:general_charges}, the global symmetry is given by the Abelianization of $G$, which is $\zz_2^3$, so there are a finite number of symmetry sectors. Krylov sectors are labeled by group elements, of which there are $\sim L^2$, so the model possesses Krylov sectors that cannot be resolved by Abelian global symmetries.

Instead, we consider a modified onsite Hilbert space that only contains $\ket*{\ttr}$, $\ket*{\ttg}$, and $\ket*{\ttb}$, but not $\ket*{\tte}$. Although this is nonstandard from the perspective of Sec.~\ref{sec:1D_stuff}, the group dynamics, which includes transitions like $\ket*{\ttb \ttb} \leftrightarrow \ket*{\ttg \ttg}$ and $\ket*{\ttr \ttg\ttb} \leftrightarrow \ket*{\ttb \ttg \ttr}$, still supports the same intrinsic and fragile fragmentation structure. Even without an onsite ${\tte}$, there are still words like ${\ttr\ttr\ttg\ttg}$ that map to the identity group element. 
Interestingly, without an onsite $\tte$, the global symmetry is not given by the Abelianization of $G$. Instead, it is $\text{U}(1)^2$, showing why the requirement of an onsite $\tte$ is necessary in Appendix~\ref{app:general_charges}. This global symmetry has $\sim L^2$ symmetry sectors, so in this particular model every Krylov sector is labeled by the quantum numbers of a global symmetry, and any two states with the same symmetry numbers are connected by the dynamics~\cite{moudgalya2022hilbert}. 
This gives us dynamics based on a non-Abelian group where all Krylov sectors can be distinguished by global symmetry charges, which is a counterexample to the claim in Appendix D of Ref.~\cite{balasubramanian2023glassy}, whose proof relies on the existence of an onsite $\tte$ character.

The final step is to only consider dynamics that act on nearest-neighbor spins, which cannot implement relations like $\ttr\ttg\ttb = \ttb\ttg\ttr$. The resulting group is 
\begin{equation}
\zz_2^{*3} \equiv \zz_2 * \zz_2 * \zz_2 = \braket{ \ttr , \ttg , \ttb  \, }{ \, \ttr^2 = \ttg^2 = \ttb^2 = \tte },
\label{eqn:pair-flip-noident}
\end{equation}
where we again choose not to include the state $\ket{\tte}$ in the onsite Hilbert space. Whereas the group $\zz^2 \rtimes \zz_2$ has only $\sim L^2$ group elements corresponding to length-$L$ words, the group~\eqref{eqn:pair-flip-noident} has $\sim 2^L$ such group elements. However, the global symmetry is unchanged. This overabundance of group elements, and therefore Krylov sectors, is the source of the Hilbert space fragmentation in the pair-flip model. 

\section{Global symmetries label all intrinsic Krylov sectors for Abelian group models} \label{app:abelian_charges}
If $G$ is Abelian, then all intrinsic Krylov sectors $K_{g,L}$ can be labeled by quantum numbers of a global symmetry. In this Appendix, we show how to construct these quantum numbers for a generic Abelian $G$.\footnote{Note that we are not concerned with proving that $G$ is Abelian from a given presentation of $G$, since this question is in general undecidable. We simply assume that it is known that $G$ is Abelian.} For illustrative purposes, we will start with some examples.

Let us start with the simplest example, that of a free Abelian group,
\eq{
G = \braket{ \ttx, \tty \,}{\, \ttx\tty\ttx^{-1}\tty^{-1} = \tte }.
}
In this case, since $\ttx$ and $\tty$ commute, $G \cong \zz \times \zz \equiv \zz^2$ and it is straightforward to see that the associated group model has a $\U{1} \times \U{1}$ global symmetry, characterized by quantum numbers
\eq{
n_\ttx & = \sum_i \Big( \ket*{\ttx}\bra*{\ttx}_i - \ket*{\ttx^{-1}}\bra*{\ttx^{-1}}_i \Big),
\\
n_\tty & = \sum_i \Big( \ket*{\tty}\bra*{\tty}_i - \ket*{\tty^{-1}}\bra*{\tty^{-1}}_i \Big).
\label{eq:charges}
}
We call these numbers ``$\ttx$-charge'' and ``$\tty$-charge,'' respectively. They assume integer values, $n_\ttx, n_\tty \in \zz$, and fully determine the group element $g \in G$, and therefore the intrinsic Krylov sector $K_{g,L}$.

What if we had a free Abelian group, but with a different presentation --- the one that does not directly contain commutators between generators (a \emph{commutator} between group elements is defined as $\comm{g}{h} \equiv ghg^{-1}h^{-1}$)? For example, we could consider the following presentation of $\zz^2$,
\eq{
G = \zz^2 = \braket{\ttx, \tty, \ttz \,}{\, \ttx^2 \tty \ttz^{-1} = \ttz\ttx^{-1}\tty^{-1}\ttx^{-1} = \tte}.
}
One can see that we simply have $\ttz = \ttx^2\tty = \ttx\tty\ttx$, from which we can infer that $\ttx\tty = \tty\ttx$. The generator $\ttz$ is ``redundant,'' since it can be expressed through the ``essential'' generators $\ttx$ and $\tty$. With such a presentation, we cannot simply use quantum numbers \eqref{eq:charges}, since the words can also contain $\ttz$. It is, however, easy to modify \eqref{eq:charges} to account for the redundant generator:
\eq{
n_\ttx & = \sum_i  \ketbra{\ttx}{\ttx}_i - \ketbra{\ttx^{-1}}{\ttx^{-1}}_i + 2\ketbra{\ttz}{\ttz}_i - 2\ketbra{\ttz^{-1}}{\ttz^{-1}}_i  ,
\\
n_\tty & = \sum_i  \ketbra{\tty}{\tty}_i - \ketbra{\tty^{-1}}{\tty^{-1}}_i + \ketbra{\ttz}{\ttz}_i - \ketbra{\ttz^{-1}}{\ttz^{-1}}_i .
\label{eq:charges_with_redundant}
}
Simply put, since $\ttz = \ttx^2\tty$, every $\ttz$ character in the word contributes $+2$ to the $\ttx$-charge and $+1$ to the $\tty$-charge, while $\ttz^{-1}$ contributes $-2$ and $-1$ to the $\ttx$- and $\tty$-charges, respectively.

Another example is that of the following presentation of the group $\zz$:
\eq{
G = \zz = \braket{\tta, \ttb \,}{\, \tta^3 = \ttb^2, \tta\ttb = \ttb\tta} \, .
}
Here, one could introduce a new ``essential'' generator $\ttx = \tta^{-1} \ttb$, to get to a more familiar presentation, $\zz = \braket{\ttx \,}{\,\,}$. The modified presentation includes the relations $\tta = \ttx^2$ and $\ttb = \ttx^3$. Thus, there is only one conserved $\U{1}$ charge,
\begin{equation*}
n_\ttx = \sum_i 
2\ketbra{\tta}{\tta}_i - 2\ketbra{\tta^{-1}}{\tta^{-1}}_i + 3\ketbra{\ttb}{\ttb}_i - 3\ketbra{\ttb^{-1}}{\ttb^{-1}}_i 
\end{equation*}
even when acting on the system with the original presentation without the $\ttx$ generator.

In a similar fashion, we can construct conserved charges for an arbitrary presentation of a free Abelian group, $G = \braket{h_1, \dots, h_m\,}{\, R}$, with some relations $R$. Such a presentation can be always rewritten in the form
\begin{subequations}
\begin{gather}
G = \braket{ s_1, \dots, s_n, h_1, \dots, h_m \,}{\, C \cup E } = \braket{ s_1, \dots, s_n \,}{\, C },  \\
 C = \left\{ \comm*{s_j}{s_k} = \tte \right\}_{j,k=1}^n, \quad E = \left\{ h_j = s_1^{d_{j1}} s_2^{d_{j2}} \dots s_n^{d_{jn}} \right\}_{j=1}^m
\end{gather}
\end{subequations}
where $h_1, \dots, h_m$ are the original generators present in the model, $s_1, \dots, s_n$ are the new ``essential'' generators, $C$ is a set of commutators of all pairs of essential generators, and $E$ is a set of relations that explicitly express each original generator through the essential ones. The global $\U{1}$ symmetry charges will then be associated with the essential generators:
\begin{equation}
n_k = \sum_i \sum_{j=1}^m d_{jk} \Big( \ket*{h_j}\bra*{h_j}_i - \ket*{h_j^{-1}}\bra*{h_j^{-1}}_i \Big), 
\end{equation}
for $k = 1, \dots, n$, which can be rewritten as
\begin{equation}
n_k = \sum_{i=1}^L \sum_{g \in \mathcal{A}_h} d_{g,k} \ket{g} \bra{g}_i, 
\label{eq:charges_with_redundant_general}
\end{equation}
for $k = 1, \dots, n$,
where $\mathcal{A}_h = \{h_j\} \cup \{h_j^{-1}\} \cup \{\tte\}$ only consists of the original alphabet and we have introduced the modified weights
\begin{equation}
d_{g,k} = \begin{cases}
d_{jk},  & g = h_j \\
-d_{jk}, & g = h_j^{-1} \\
0,       & g = \tte.
\end{cases} \label{eqn:weights}
\end{equation}
for later convenience.

We now move on to generic Abelian groups. The fundamental theorem of finitely generated Abelian groups (also called ``The Basis Theorem,'' see, e.g., section 5.5 of Ref. \cite{johnson2002symmetries}) states that every finitely generated Abelian group is isomorphic to a direct product of cyclic groups, with the order of each cyclic group being a power of a prime, or possibly infinite (in which case it is the $\zz$ group). Suppose this decomposition contains $l$ infinite cyclic groups,
\eq{
G = \zz_{p_1} \times \dots \times \zz_{p_{n-l}} \times \zz^l,
}
with each $p_j$ being a power of a prime (not necessarily distinct from others). Then any finite presentation of such a group can be recast into the form
\eq{
G = \braket{ s_1, \dots, s_n, h_1, \dots, h_m \,}{\, C \cup P \cup E },
\label{eq:abelian_group_decomposition}
}
where, as before, we have $n$ essential generators $\{s_j\}$, $m$ original generators $\{h_j\}$ whose characters span the local Hilbert space, $C$ are commutators between $s_j$'s, $E$ are relations expressing $h_j$'s through $s_j$'s, while $P$ are relations of the form $P = \{ s_1^{p_1} = \dots = s_{n-l}^{p_{n-l}} = \tte \}$, signifying the cyclic nature of the first $n-l$ essential generators. The charge operators $n_k$ retain the same definition as in~\eqref{eq:charges_with_redundant_general},
but are now only conserved modulo $p_k$ for $k = 1, \dots, n-l$ while still being conserved exactly for $k=n-l+1, \dots, n$. These charges $n_k$ generate a global $\zz_{p_1}\times \cdots \times \zz_{p_{n-l}} \times \text{U}(1)^l$ symmetry.

We want to emphasize that figuring out whether a given presentation gives rise to an Abelian group is in general an undecidably hard task. However, it is easy to compute the Abelianization of a group given one of its presentations (this is done by computing the Smith normal form of the relation matrix --- see, e.g., section 5.5 of Ref.~\cite{johnson2002symmetries}). Therefore, since Abelianization of an Abelian group $G$ is the group $G$ itself, factors $p_1, \dots, p_{n-l}, l$ in decomposition \eqref{eq:abelian_group_decomposition} can be easily obtained for an arbitrary finite presentation of an Abelian group.

\section{Maximal set of global symmetries for a group model}
\label{app:general_charges}
In this Appendix, we explicitly show which intrinsic Krylov sectors can be labeled by global symmetries and which ones cannot, for a generic group $G$. In this way, we identify the ``maximal'' set of global symmetries responsible for intrinsic Hilbert space fragmentation in an arbitrary group model. Discerning intrinsic Krylov sectors that cannot be distinguished by this maximal set of global symmetries necessarily requires non-local conserved quantities.

Consider a generic (possibly non-Abelian) group $G$. The \emph{commutator subgroup} of $G$, denoted as $\comm{G}{G}$, is a subgroup generated by commutators $\comm{g}{h}$ for all $g,h \in G$. This subgroup is known to be the smallest normal subgroup such that the quotient group $G/\comm{G}{G}$ is Abelian. Such a quotient is called the \emph{Abelianization} of $G$, $G^\mathrm{ab} \equiv G/\comm{G}{G}$. The commutator subgroup $\comm{G}{G}$ splits $G$ into cosets, which form the group $G^\mathrm{ab}$. We claim that (i) intrinsic Krylov sectors $K_{g_1,L}$ and $K_{g_2,L}$ with $g_1, g_2$ from distinct cosets can always be distinguished by global symmetries, and (ii) $K_{g_1,L}$ and $K_{g_2,L}$ with $g_1, g_2$ from the same coset can never be distinguished using global symmetries.

 Claim (i) follows directly from Appendix \ref{app:abelian_charges}. For a generic group and presentation, $G = \braket{ h_1,\dots, h_m \,}{\, R }$, we first construct the Abelianization $G^\text{ab} = \braket{ h_1, \dots, h_m \,}{\, R \cup C_h }$, where 
\begin{equation}
C_h = \left\{ [h_j, h_k] = \tte \right\}_{j,k=1}^m
\end{equation}
contains commutators for all characters. Then, following the previous Appendix, we can construct conserved charges in \eqref{eq:charges_with_redundant_general}, which are conserved mod $p_k$ for $k\le n-l$ and conserved exactly otherwise. These charges are associated with the global symmetry $\zz_{p_1} \times \dots \times \zz_{p_{n-l}} \times \text{U}(1)^l$. By the first isomorphism theorem, there is a homomorphism from $G$ to $G^\text{ab}$ that maps all elements in a coset of $\comm{G}{G}$ in $G$ to a single unique element of $G^\text{ab}$. Therefore, $g_1$ and $g_2$ from distinct cosets map to distinct elements of $G^\text{ab}$, and hence $K_{g_1,L}$ and $K_{g_2,L}$ can be distinguished by the charges \eqref{eq:charges_with_redundant_general}.

Claim (ii) is similar to the proof in Appendix~D of Ref.~\cite{balasubramanian2023glassy}. We want to prove that, if $g_1$ and $g_2$ belong to the same coset of $\comm{G}{G}$ in $G$, then there does not exist a set of global symmetry operators that distinguish every $\ket{w_1}$ with $\varphi(w_1) = g_1$ from every $\ket{w_2}$ with $\varphi(w_2) = g_2$. 

To do so, we first prove that for any $g_1, g_2$ in the same coset of $\comm{G}{G}$, there exist two words, $w_1$ and $w_2$, such that $\varphi(w_1) = g_1, \varphi(w_2) = g_2$ and the characters in $w_1$ can be permuted to yield $w_2$. To prove this, first note that $g_1$ and $g_2$ can be expressed as
\eq{
g_1 & = c_1 c_2 \cdots c_n g,
\\
g_2 & = \tilde{c}_1 \tilde{c}_2 \cdots \tilde{c}_{\tilde{n}} g,
}
where $g \in G$, and each $c_i$, $i=1,\dots,n$ and $\tilde{c}_i$, $i=1,\dots \tilde{n}$ is a commutator of two group elements of $G$, i.e., $c_i = h_i p_i h_i^{-1} p_i^{-1}$, $\tilde{c}_i = \tilde{h}_i \tilde{p}_i \tilde{h}_i^{-1} \tilde{p}_i^{-1}$ for some $h_i, p_i, \tilde{h}_i, \tilde{p}_i \in G$.
Since the coset of $[G,G]$ in $G$ that includes identity is the subgroup $[G,G]$ itself (which is a subgroup generated by all commutators between group elements of $G$), $c_1 c_2 \cdots c_n$ and $\tilde{c}_1 \tilde{c}_2 \cdots \tilde{c}_{\tilde{n}}$ belong to $\comm{G}{G}$, and correspondingly, $g_1, g_2$ belong to the coset $\comm{G}{G}g$. Now, consider words $w_{h_i}, w_{p_i}, w_{\tilde{h}_i}, w_{\tilde{p}_i}$ whose characters multiply to $h_i, p_i, \tilde{h}_i, \tilde{p}_i$, respectively. 
Then construct words $w_{c_i} \equiv w_{h_i} \cdot w_{p_i} \cdot w_{h_i}^{-1} \cdot w_{p_i}^{-1}$, $i=1,\dots,n$ and $w_{\tilde{c}_i} \equiv w_{\tilde{h}_i} \cdot w_{\tilde{p}_i} \cdot w_{\tilde{h}_i}^{-1} \cdot w_{\tilde{p}_i}^{-1}$, $i=1,\dots,\tilde{n}$, such that $\varphi(w_{c_i}) = c_i$ and $\varphi(w_{\tilde{c}_i}) = \tilde{c}_i$. Note that the characters in every $w_{c_i}$ and every $w_{\tilde{c}_i}$ can be easily rearranged to yield an identity, since in any of these words every character appears as many times as its inverse. Finally, denoting $w_c \equiv w_{c_1}  \cdots  w_{c_n}$ and $w_{\tilde{c}} \equiv w_{\tilde{c}_1}  \cdots  w_{\tilde{c}_{\tilde{n}}}$, the words $w_1$ and $w_2$ can be constructed as follows:
\eq{
w_1 &= w_c \cdot w_g \cdot w_g^{-1} \cdot w_{\tilde{c}}^{-1} \cdot w_{\tilde{c}} \cdot w_g,
\\
w_2 &= w'_c \cdot w_g \cdot w_g^{-1} \cdot \left(w'_{\tilde{c}}\right)^{-1} \cdot w_{\tilde{c}} \cdot w_g,
}
where $w_g$ is a word with $\varphi(w_g) = g$, while $w'_c$ and $w'_{\tilde{c}}$ are the words with $\varphi(w'_c) = \varphi(w'_{\tilde{c}}) = \tte$ obtained from $w_c$ and $w_{\tilde{c}}$, respectively, through a permutation of characters.
Indeed, we have that $w_1$ and $w_2$ consist of the same characters up to a permutation, and $\varphi(w_1) = \varphi(w_c \cdot w_g)\varphi(w_g^{-1} \cdot w_{\tilde{c}}^{-1})\varphi(w_{\tilde{c}} \cdot w_g) = g_1 g_2^{-1} g_2 = g_1$ and $\varphi(w_2) = \varphi(w'_c) \varphi(w_g) \varphi(w_g^{-1}) \varphi(\left(w'_{\tilde{c}}\right)^{-1})\varphi( w_{\tilde{c}} \cdot w_g) = \tte g g^{-1} \tte g_2 = g_2$.

Finally, we construct the states
\begin{align}
\ket{\tilde{w}_1 } &= \ket*{\tte^{l_1}  w_c  \tte^{l_2}  w_g  \tte^{l_3}  w_g^{-1}  \tte^{l_4}  w_{\tilde{c}}^{-1}  \tte^{l_5}  w_{\tilde{c}}  \tte^{l_6}  w_g  \tte^{l_7} }, 
\\ 
\ket{\tilde{w}_2 } &= \ket*{\tte^{l_1}  w'_c  \tte^{l_2}  w_g  \tte^{l_3}  w_g^{-1}  \tte^{l_4}  \left(w'_{\tilde{c}}\right)^{-1}  \tte^{l_5}  w_{\tilde{c}}  \tte^{l_6}  w_g  \tte^{l_7} },
\end{align}
where we omitted the symbol for concatenation of characters for convenience of presentation, and
where $l_1,\dots,l_7$ are extensive in the system size. We can then use the states $\ket{\tilde{w}_1}$, $\ket{\tilde{w}_2}$ in a proof in the style of Appendix~D of Ref.~\cite{balasubramanian2023glassy}, which shows that the two intrinsic Krylov sectors containing such states cannot be distinguished by locality-preserving unitaries.

\end{appendix}

\bibliography{refs}

\end{document}